\documentclass[fleqn,usenatbib]{mnras}

\usepackage{newtxtext,newtxmath}

\usepackage[T1]{fontenc}
\usepackage{ae,aecompl}

\usepackage{graphicx}	
\usepackage{amsmath}	
\usepackage{bm}
\usepackage{adjustbox}

\usepackage{hyperref}

\def\prl{Phys.~Rev.~Lett.}

\def\apjl{ApjL}

\newcommand{\vrm}{{\rm v}}

\newcommand{\urm}{{\rm u}}

\title[CR transport theory versus data]{Reconciling cosmic-ray transport theory with phenomenological models motivated by Milky-Way data}

\author[P. Kempski and E. Quataert]{
Philipp Kempski$^{1}$\thanks{E-mail: philipp.kempski@berkeley.edu} and
Eliot Quataert$^{2}$
\\
$^{1}$Department of Astronomy and Theoretical Astrophysics Center, University of California, Berkeley, CA 94720, USA \\
$^{2}$ Department of Astrophysical Sciences, Princeton University, Princeton, NJ 08544, USA \\}
\date{Accepted XXX. Received YYY; in original form ZZZ}

\begin{document}
\label{firstpage}
\pagerange{\pageref{firstpage}--\pageref{lastpage}}
\maketitle

\begin{abstract}
Phenomenological models of cosmic-ray (CR) transport in the Milky Way (MW) can reproduce a wide range of observations assuming that CRs scatter off of magnetic-field fluctuations with spectrum $\propto k^{-\delta}$ and $\delta \sim [1.4,1.67]$. We study the extent to which such models can be reconciled with current microphysical theories of CR transport, specifically self-confinement due to the streaming instability and/or extrinsic turbulence due to a cascade of MHD fast modes. We first review why it is that on their own neither theory is compatible with observations. We then highlight that CR transport is a strong function of local plasma conditions in the multi-phase interstellar medium (ISM), and may be diffusive due to turbulence in some regions and streaming due to self-confinement in others. A multi-phase combination of scattering mechanisms can in principle reproduce the main trends in the proton spectrum and the boron-to-carbon ratio (B/C). However, models with a combination of scattering by self-excited waves and fast-mode turbulence require significant fine-tuning due to fast-mode damping, unlike phenomenological models that assume undamped Kolmogorov turbulence. The assumption that fast modes follow a weak cascade is also not well justified theoretically, as the weak cascade is suppressed by wave steepening and weak-shock dissipation even in subsonic turbulence. These issues suggest that there may be a significant theoretical gap in our understanding of MHD turbulence. We discuss a few topics at the frontier of MHD turbulence theory that bear on this (possible) gap and that may be relevant for CR scattering.
\end{abstract}


\vspace{-55pt}

\section{Introduction}

Cosmic rays may play an important role in the evolution of galaxies and diffuse gas in galaxy halos (see \citealt{zweibel2017_wind} for a recent review). However, the impact that CRs have on their host environment is a strong function of the adopted transport model (e.g., \citealt{ruszkowski17}; \citealt{farber18}; \citealt{hopkins2020_whatabout}; \citealt{qtj_2021_diff}; \citealt{qtj_2021_streaming}). As a result, the uncertainties in CR feedback are primarily driven by uncertainties in CR transport. 

The CR lifetime in galaxies is much longer than the light-crossing time. It is widely accepted that the long CR confinement time is due to (resonant) scattering by small-scale electromagnetic fluctuations. Progress in understanding this scattering has occurred on both observational and theoretical fronts. There are now detailed measurements of CR spectra in the solar neighbourhood (e.g., \citealt{stone_voyager}; \citealt{aguilar_2015}; \citealt{aguilar_bc}; \citealt{cummings_2016}), which put strong constraints on CR propagation models. On the theoretical side, there is growing understanding of how  CRs are scattered in pre-existing MHD turbulence (\citealt{chandran_scattering}; \citealt{yan_lazarian_2004}; \citealt{yan_lazarian_2008}; \citealt{xu_lazarian_ttd}; \citealt{lazarian_xu_mirror}; \citealt{fornieri_2021}) and/or by self-excited waves (\citealt{kp69}; \citealt{skilling71}; \citealt{felice_kulsrud}; \citealt{farmer_goldreich};  \citealt{bai_mhd_pic}; \citealt{squire_dust_2021}; \citealt{bai_2021}).

In the self-excitation scenario, waves are generated by the CR streaming instability (\citealt{kp69}): cosmic rays excite Alfv\'en waves if they collectively drift down their pressure gradient at speeds exceeding the Alfv\'en speed. The excited waves pitch-angle scatter cosmic rays towards isotropy in the wave frame. In the absence of damping of the self-excited waves, this limits the CR drift speed to the local Alfv\'en speed. In this limit, all CRs stream at the Alfv\'en speed, and so CR transport and the lifetime of CRs in the galaxy is energy-independent. In the presence of wave damping, the CRs are no longer fully isotropic in the frame of the self-excited Alfv\'en waves and CR transport exceeds the Alfv\'en speed by an amount that depends on the damping strength (e.g., \citealt{skilling71}; \citealt{wiener2013}). Importantly, the transport correction due to damping introduces energy dependence. However, due to the peculiar form of the correction term introduced by damping, which is neither truly diffusive nor streaming in nature,\footnote{For example, for linear damping mechanisms the term is independent of the CR distribution function. The term has been often interpreted, not entirely correctly, as super-Alfv\'enic streaming, with a streaming speed correction that is inversely proportional to the CR distribution function.} its consequences for CR transport remain somewhat unclear.

The alternative to self-confinement is that CRs are scattered by a pre-existing turbulent MHD cascade. Phenomenological models of CR transport often assume that the cascade is isotropic, undamped and follows the Kolmogorov $k^{-5/3}$ (or close to Kolmogorov, e.g. $\propto k^{-3/2}$) spectral scaling, as in hydrodynamics.  The resulting CR diffusion coefficient $\kappa_{\rm turb} \propto E^{\delta}$ with $\delta \sim 0.3-0.6$ turns out to match CR observables in the Milky Way remarkably well  (e.g., \citealt{trotta_2011}; \citealt{gaggero_2014}; \citealt{hopkins_cr_pheno}); this includes the spectra of secondary-to-primary CRs (e.g. the B/C ratio), which under the assumption of diffusive CR transport directly probe CR transport independent of injection physics (this is not true for streaming transport, as we show in Section \ref{sec:sc+et}). A subset of the phenomenological literature uses a combination of isotropic undamped Kolmogorov turbulence and waves excited by the CR streaming instability, which are assumed to cascade in $k_\parallel$ just like the background turbulence, to model the break in CR spectra around a few hundred GeV (e.g., \citealt{blasi12}; \citealt{aloisio_blasi_2013} \citealt{aloisio_2015}). However, these phenomenological models are not justified theoretically, as MHD turbulence is known to be very different from hydrodynamic Kolmogorov-like turbulence. In MHD, the turbulent cascade of Alfv\'en and slow waves does have a Kolmogorov spectrum, but only in directions perpendicular to the magnetic field (the spectrum may be slightly shallower than Kolmogorov, \citealt{boldyrev_2006}). The spectrum  along the local magnetic-field direction, which is relevant for scattering, is $k_\parallel^{-2}$ and thus steeper than Kolmogorov. Moreover, on small scales the cascade is highly anisotropic with $k_\perp \gg k _\parallel$ (\citealt{gs95}), which is very inefficient at scattering cosmic rays (\citealt{chandran_scattering}). As a result, existing models of Alfv\'enic turbulence predict negligible CR confinement. The compressible fast-mode cascade may be isotropic (\citealt{cho_lazarian}), and may have a spectral slope that is not too far from Kolmogorov, $\propto k^{-3/2}$ (\citealt{zakharov_sagdeev_1970}; \citealt{cho_lazarian}). But fast modes are subject to strong damping and the physics of their cascade (and thus spectral slope) remains uncertain.  As a result, reconciling CR observables with existing theories of MHD turbulence remains an open problem (see \citealt{fornieri_2021} for a recent attempt that uses only fast-mode turbulence).

In this paper we attempt to reconcile microscopic CR transport theory with phenomenological models based on MW data and highlight some of the main issues that existing theories face. We prioritise building physical intuition rather than deriving exact results. For this reason, we use simplified CR transport models to calculate order-of-magnitude estimates of MW  observables. We first give a pedagogical review of the issue that existing theories of self-confinement and scattering by ambient turbulence cannot, on their own, reproduce observed CR spectra. Instead, a combination of self-confinement and turbulence may be needed to explain CR observables. We then argue that the phase structure of the ISM is important because CR transport depends on the local plasma conditions.  In particular, the streaming instability is suppressed by strong damping unless the plasma is well ionized.   And how efficiently turbulence can cascade to small scales to scatter cosmic rays depends on the local plasma conditions. The interstellar medium of star-forming galaxies is multi-phase with most of the volume near the mid plane dominated by the warm and hot ISM, while a significant fraction of the mass is in denser phases (e.g., \citealt{mckee_ostriker_1977}).  At larger heights above the midplane, the hot ISM increasingly becomes the dominant ISM component by volume and perhaps by mass (e.g., \citealt{kim_ostriker_2017}).  We show that as a result of the dramatically changing conditions throughout the ISM, CR transport may be either diffusive due to turbulence or streaming due to self-confinement, depending on the ISM phase.

The paper is structured as follows. We first give an overview of CR self-confinement theory and derive associated steady-state solutions  in Section \ref{sec:sc}. We then review CR scattering by MHD turbulence, specifically the cascade of MHD fast modes, in Section \ref{sec:turb}. Many of the results in Sections \ref{sec:sc} and \ref{sec:turb}, e.g. the conclusion that neither self-confinement nor extrinsic-turbulence theory can on their own explain the CR data in the Milky Way, are not new (e.g., \citealt{kc71}; \citealt{farmer_goldreich}; \citealt{blasi12}; \citealt{fornieri_2021}), but we repeat them for pedagogical purposes (for self-confinement our approach is also quite different from what is usually done in the literature). In Section \ref{sec:sc+et} we consider CR scattering by a combination of self-excited Alfv\'en waves and fast-mode turbulence. In Section \ref{sec:two_phase_mw} we  show that CR transport may be multi-phase, i.e. streaming or diffusion depending on the ISM phase. We consider a particular multi-phase transport model and compare it to observations in Section \ref{sec:observations}. Our calculations show that  a multi-phase combination of scattering by self-excited waves and a weak fast-mode cascade can in principle reproduce the main trends in the proton spectrum and the boron-to-carbon ratio (B/C), but that requires a significant amount of fine-tuning. We note that there already exists literature that tries to combine streaming and turbulence to explain CR observables (e.g., \citealt{blasi12}; \citealt{aloisio_blasi_2013}; \citealt{aloisio_2015}). However, these models assume an undamped Kolmogorov cascade. We instead focus on the theoretically better motivated interplay of streaming instability and fast-mode turbulence. The key difference is that fast modes are damped on scales $\sim$ the Larmor radius of $\sim 100-1000$ GeV particles. This makes their effect on CR transport and their interaction with CR-streaming-unstable Alfv\'en waves very different from what previous work concluded, which focused on undamped Kolmogorov fluctuations.  We discuss uncertainties in the physics of MHD fast-mode turbulence in Section \ref{sec:fast_mode_uncertainties}, which have significant implications for CR transport, but have not been taken into account in previous work (e.g. \citealt{yan_lazarian_2004}; \citealt{yan_lazarian_2008}; \citealt{fornieri_2021}). In Section \ref{sec:discussion} we speculate about additional uncertainties and ongoing developments from the field of MHD turbulence that may be relevant for CR transport. We summarise our results in Section \ref{sec:summary}.

\section{Self-confinement theory and steady-state solutions} \label{sec:sc}
As a simple model of CR transport in galaxies, we consider 1D propagation away from the galactic disk. CRs are assumed to be injected in the disk by supernovae at a rate $2Q(p)$. We consider a uniform vertical magnetic field, $\bm{B} = B \bm{\hat{z}}$. In a steady state, ($\partial f / \partial t = 0$) the CR  distribution function as a function of momentum $p$ satisfies
\begin{gather}
\begin{aligned} \label{eq:adv_diff_turb}
         (\urm + {\rm v_{st}})  \frac{\partial f}{\partial z}= & \frac{1}{3} p \frac{\partial f}{\partial p} \frac{\partial}{\partial z}(\urm + {\rm v_{st}}) + \frac{\partial}{\partial z} \Big( \kappa \frac{\partial f}{\partial z}  \Big)  \\ & + 2 Q \delta(z) - 2\frac{f}{\tau_{\rm loss}(p)} h \delta(z).
\end{aligned}
\end{gather}
The terms in the first line describe CR transport by advection and diffusion. The first term in the second line is the source function, while the second term represents losses due to interactions with ISM material in the dense galactic mid-plane of half-thickness $h$. We ignore hadronic pion-producing collisions, as measurements in the Milky Way show that the Galaxy loses its cosmic rays through escape and not hadronic losses (\citealt{strong_2010}; \citealt{lacki_2011}). We use the expressions in \cite{schlickeiser2002} to evaluate $\tau_{\rm loss}(p)$  for ionisation and Coulomb losses.  For both loss mechanisms, $\tau_{\rm loss}(p) \propto p^{3}$ for sub-relativistic CRs and $\tau_{\rm loss}(p) \propto p$ for super-relativistic CRs (see also eq. \ref{eq:tau_ion}). Due to the strong dependence on momentum,  the loss term is negligible for ultra-relativistic CRs ($E \gg$GeV) and so we ignore it for much of the discussion in Sections \ref{sec:sc} and \ref{sec:turb}. However, the loss term is important for trans- and sub-relativistic CRs. We note that the loss term is more correctly described by a flux in momentum space. For the CR protons and the purposes of this paper, however, the approximate form in \eqref{eq:adv_diff_turb} is a reasonable approximation and we use it to compare our model predictions to observed proton spectra at low energies.  By contrast, for boron nuclei it is important to use the flux form of the energy loss term, as we discuss in Appendix \ref{app:bc_loss}. For this reason, including energy losses in the calculations of B/C spectra is beyond the scope of this paper.  

 \begin{figure}
  \centering
  \begin{minipage}[b]{\textwidth}
    \includegraphics[width=0.45\textwidth]{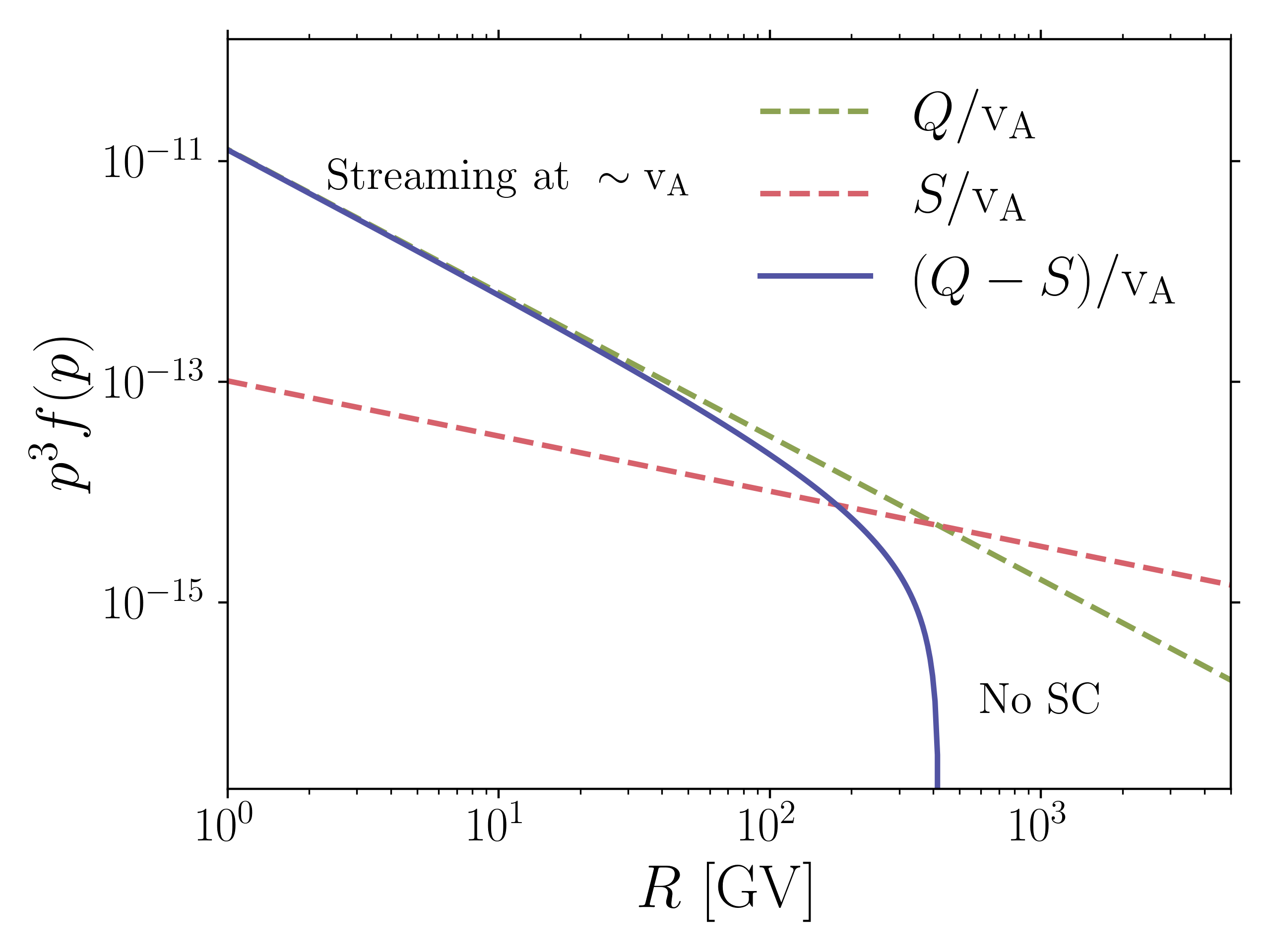}
    \end{minipage}
      \begin{minipage}[b]{\textwidth}
      \includegraphics[width=0.45\textwidth]{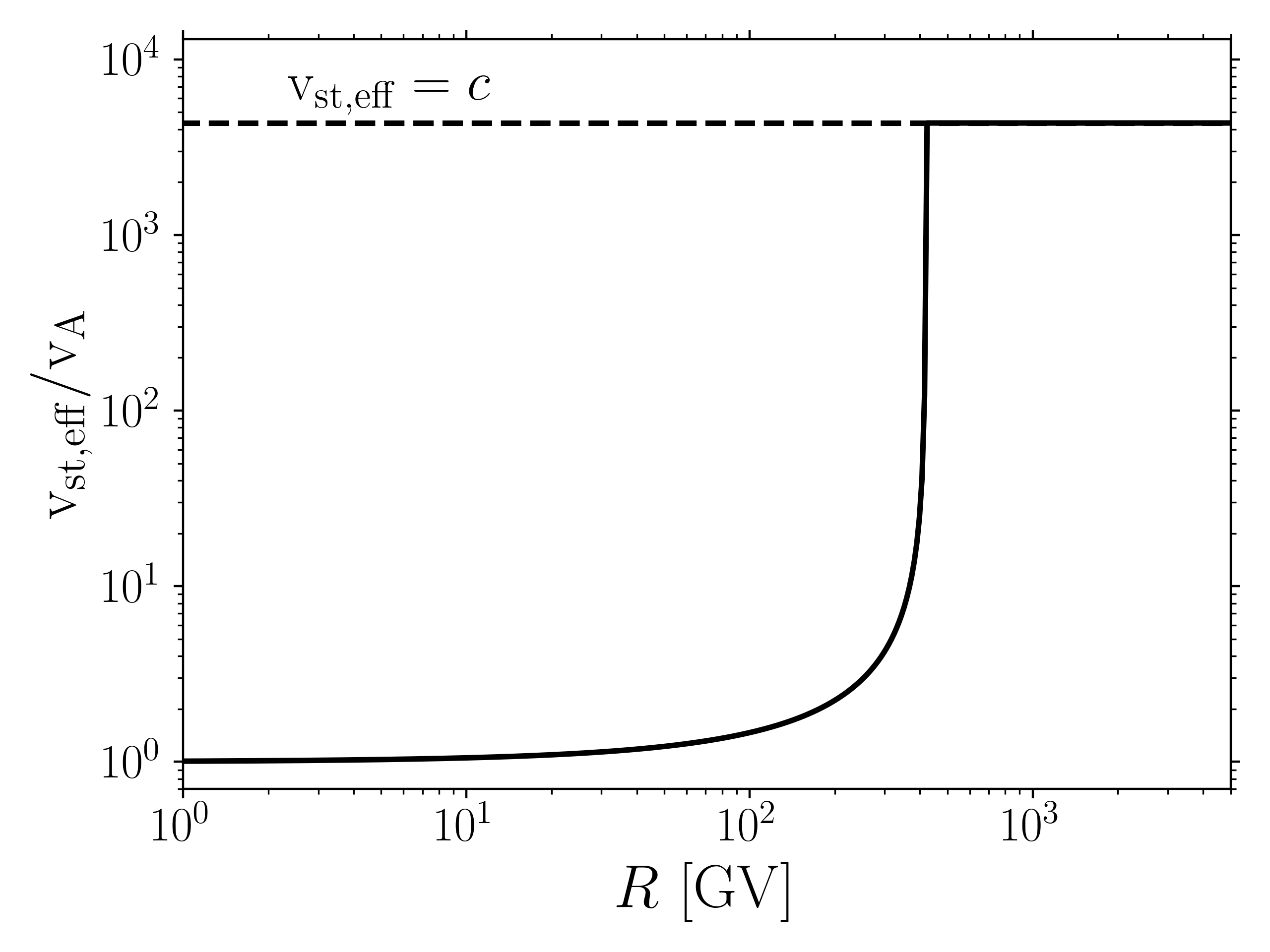}
    \end{minipage}    
  \caption{Top: steady-state solution in self-confinement theory with linear damping mechanisms (eq. \ref{eq:stead_state_lin_damp}). Here we use damping by Alfv\'enic turbulence with $\Gamma \propto k^{1/2}$. For a wide range of rigidities the solution (blue solid line) agrees well with the solution absent any corrections to Alfv\'enic streaming (green dashed line). This is followed by a sharp cutoff and at higher energies CRs are not self-confined. Bottom: effective streaming speed associated with the solution in the top panel (eq. \ref{eq:vsteff}). To reasonable approximation, for all linear damping mechanisms self-confinement theory predicts that CRs stream at either the Alfv\'en speed or the speed of light.  \label{fig:lin_damping}}
\end{figure}

In this section, we consider CR scattering by self-excited waves generated by the streaming instability (\citealt{kp69}). We consider CR scattering by extrinsic turbulence in Section \ref{sec:turb}. For self-confinement, ${\rm v_{st}}={\rm v_A sgn}(z)$ and the diffusive correction to streaming, $\kappa$, is calculated by equating the Alfv\'en wave damping rate and the growth rate due to the streaming instability (see, e.g., \citealt{skilling71}). In this work, we ignore the motion of the thermal gas, i.e. we set $\urm=0$. However, our results for ${\rm v_{st}} \neq 0$ easily generalise to the case with finite $\urm$, so our streaming results also mimic gas advection in, e.g., supernova-heated flows. For a constant ${\rm v_{A}}$ and ignoring the energy loss term, we can integrate \eqref{eq:adv_diff_turb} to find that $f$ satisfies,
\begin{equation}\label{eq:steady_master}
    {\rm v_{A}} f(z) - \kappa \frac{\partial f}{\partial z} = {\rm const} \approx Q.
\end{equation}
The interpretation of equation \ref{eq:steady_master} is simple. The rate of escape of CRs (i.e. the flux) balances the injection rate at the central source. The conserved flux on the right hand side is exactly $Q$ for scattering by turbulence and $\approx Q$ for self-confinement, with a fractional error that does not qualitatively affect the results in this paper. For a spatially varying Alfv\'en speed, \eqref{eq:steady_master} is valid for $z \lesssim H_{A}$, where $H_A$ is the $\sim $Alfv\'en scale height.\footnote{For a CR distribution function that is a power-law in momentum, $f \sim p^{-\alpha}$, the steady-state solution of \eqref{eq:adv_diff_turb} for pure Alfv\'enic streaming is,
\begin{equation} \label{eq:var_alf}
    f(z,p) = \Big[\frac{{\rm v_A}(0)}{{\rm v_A}(z)}\Big]^{\alpha / 3} \frac{3 Q}{\alpha {\rm v_A}(0)},
\end{equation}
and so the CR scale height is comparable to the Alfv\'en scale height. For a constant Alfv\'en speed, the exact conserved flux in  \eqref{eq:steady_master} is $[Q- {\rm v_A}f(0)(\alpha-3)/3]$, which gives $f=3Q/\alpha {\rm v_A}$ instead of the $f=Q/{\rm v_A}$ implied by \eqref{eq:steady_master}. The correction is therefore at most order unity (as $\alpha \approx 4.5$) and is essentially degenerate with the choice of ${\rm v_A}$. Because $\alpha$ is an a priori unknown function of CR momentum, using the exact conserved flux unnecessarily obscures the essential physics of the calculation by making it more mathematically complicated (especially when diffusive corrections are included, e.g. in Section \ref{sec:two_phase_mw}). For pedagogical purposes, we therefore use the approximation in \eqref{eq:steady_master}, which does not affect our results qualitatively.   
}

\subsection{Single-phase steady-state solutions} \label{sec:sc_single_phase}

 We assume that the diffusion coefficient $\kappa = \kappa(f_p)$ is a function of the CR \textit{proton} distribution function only, as protons are the most abundant CR species. $\kappa$ in self-confinement theory is calculated by balancing the damping rate of Alfv\'en waves and the growth rate due to the CR streaming instability. As a result, the ``diffusive" correction to Alfv\'enic streaming in \eqref{eq:steady_master} strongly depends on the mechanisms that damp the excited Alfv\'en waves, which in turn depend on the ISM phase.

\subsubsection{Linear damping mechanisms} \label{sec:lin_damp}
For linear damping mechanisms, the resulting diffusion coefficient has the form (\citealt{skilling71}),
\begin{equation}  \label{eq:kappa_lin_damp}
    \kappa_{\rm sc} = S |\bm{\hat{b} \cdot \nabla }f_p|^{-1},
\end{equation}
where $S$ depends on the properties of the plasma and the damping. Because $\kappa_{\rm sc} \propto  |\bm{\hat{b} \cdot \nabla }f_p|^{-1}$, the ``diffusion" term ends up not being  diffusive at all and is in fact independent of the CR distribution function. This peculiar result implies that linear damping mechanisms lead to a transport correction that in a steady state effectively acts like a sink of CRs, as can be seen from \eqref{eq:steady_master} combined with \eqref{eq:kappa_lin_damp},
\begin{equation} \label{eq:stead_state_lin_damp}
    f \approx \frac{Q - S}{{\rm v_A}}.
\end{equation}
 The two terms in the numerator are functions of CR momentum, $Q \sim Q_0 (p/p_0)^{-\gamma_{\rm inj}}$ with $\gamma_{\rm inj} \gtrsim 4$ and $S \propto p^{-3-a}$ for a damping $\Gamma \propto k^a$ (\citealt{skilling71}). Thus, the two terms in the numerator have a different power-law dependence on CR momentum, and for known linear damping mechanisms, $S$ will generally have a harder spectrum than Q. For example, for linear damping by ambient Alfv\'enic turbulence with $\Gamma \sim k^{1/2}$ (\citealt{farmer_goldreich}), $S \sim p^{-3.5}$, for damping by charged interstellar dust grains with $\Gamma \sim k^{3/4}$ (\citealt{squire_dust_2021}), $S \sim p^{-3.75}$, and for ion-neutral damping with $\Gamma \sim k^0$, $S \sim p^{-3}$. We can therefore define a cutoff momentum,
\begin{equation} \label{eq:sc_cutoff}
    Q(p_c) = S(p_c).
\end{equation}
For $p > p_c$, $S > Q$ which gives a negative solution for $f$: CRs are unable to confine themselves. On the other hand, for almost all $p<p_c$ we have $Q \gg S$ and 
\begin{equation} \label{eq:steady_lin_sol}
    f \approx \frac{Q}{\rm v_A} \qquad p \ll p_c.
\end{equation}
Almost all self-confined CRs therefore stream at essentially the Alfv\'en speed. Thus, the primary role of linear damping in self-confinement theory is to set the energy range of CRs that are able to self-confine. An example of this is provided in Figure \ref{fig:lin_damping}. In the top panel, we plot the steady-state solution from equation \ref{eq:stead_state_lin_damp} (blue line). For the CR source term ($=2Q$, see eq. \ref{eq:adv_diff_turb}) we use a supernova rate of 1 per 100 years, each supernova injecting $10^{50}$ ergs in cosmic rays with a spectrum $Q \propto p^{-4.3}$. For $S$, we use $\Gamma \propto k^{1/2}$ (e.g., damping by ambient Alf\'enic turbulence) with $\Gamma =  10^{-11} \ {\rm s^{-1}}$ at scales resonant with $1$ GeV CRs. We use a $1 \mu $G magnetic field and thermal-gas density $n = 0.001 \ {\rm cm^{-3}}$. We see that for a wide range of energies the solution agrees well with the solution absent any corrections to Alfv\'enic streaming (green dashed line). This is followed by a sharp cutoff and at higher energies CRs are not self-confined. We can define an effective streaming speed,
\begin{equation} \label{eq:vsteff}
    \frac{Q}{\rm v_{st,eff}} \equiv \frac{Q-S}{\rm v_A},
\end{equation}
which we plot in the bottom panel of Figure \ref{fig:lin_damping}. The effective streaming speed is essentially the Alfv\'en speed for a wide range of energies. It sharply transitions to propagation at the speed of light at the cutoff energy for self-confinement. The above analysis and Figure \ref{fig:lin_damping} show that linear damping corrections to Alfv\'enic streaming cannot produce energy-dependent transport like that needed in phenomenological models. Instead, the linear damping of Alfv\'en waves just sets a maximum CR energy above which the streaming instability ceases to operate. 

The above calculation can be easily extended to include the proton energy-loss term from eq. \ref{eq:adv_diff_turb} by modifying the RHS of eq. \ref{eq:steady_master} to $Q - f(0)h/\tau_{\rm loss} (p)$. One can then show that \begin{equation} \label{eq:steady_lin_sol_loss}
    f(0) = \frac{Q-S}{\rm v_A} \Big(1 + \frac{h}{\rm v_A \tau_{loss}(p)} \Big)^{-1}.
\end{equation}
At high energies $h \ll {\rm v_A \tau_{loss}}$ and we recover eq. \ref{eq:steady_lin_sol}, i.e. the steady state is set by CR escape from the Galaxy. At low energies $h \gg {\rm v_A \tau_{loss}}$ so that $f(0) = [Q-S(0)] \tau_{\rm loss} / h$ and CRs are in the loss-dominated regime. 

\subsubsection{Nonlinear damping mechanisms} \label{sec:nonlin_damp}
If instead nonlinear Landau damping (NLLD; $\Gamma \sim k {\rm v_{th}}  (\delta B / B)^2$; \citealt{lee_volk_1973}, \citealt{kulsrud_book}) is the dominant damping mechanism for self-excited waves, the diffusion coefficient in self-confinement theory can be calculated analogously to the linear damping case in \cite{skilling71}. We find,
\begin{equation}\label{eq:sc_kappa_nlld}
    \kappa_{\rm sc} \approx X |\bm{\hat{b} \cdot \nabla }f_p|^{-1/2},
\end{equation}
with $X$ given by,
\begin{equation}\label{eq:X}
    X \sim \frac{{\rm v_p}^2}{2 \pi^2 \Omega_0} \Big( \frac{8 \vrm_{\rm th} \Omega_0 B^2}{m_{\rm p} \vrm_{\rm A} {\rm v_p}^3 p^3} \Big)^{1/2},
\end{equation}
where ${\rm v_p} \approx c$ is the speed of individual CR protons, $\Omega_0$ is the non-relativistic gyro-frequency, and the other symbols have the usual meaning.

Using the diffusion coefficient due to NLLD in \eqref{eq:sc_kappa_nlld} and assuming $\partial f / \partial z < 0$ above the galactic plane, equation \ref{eq:steady_master} for CR protons becomes,
\begin{equation}\label{eq:steady_nlld}
    {\rm v_{A}} f_p + X \Big( - \frac{\partial f_p}{\partial z}\Big)^{1/2}  \approx Q.
\end{equation}
This differential equation is separable with solution, 
\begin{equation}\label{eq:steady_nlld_sol}
    f_p(z) = \frac{Q}{\rm v_A} + \frac{X^2}{{\rm v_A}^2 (z + C)}.
\end{equation}
$C$ is a yet undetermined integration constant, for which we need to specify a boundary condition. In phenomenological models of CR transport one traditionally specifies a ``CR halo size", such that $f_p(z_H)=0$. It may seem unphysical to introduce an ad hoc cutoff, especially if it is at a small distance above the disk (e.g. a few kpc, which is difficult to reconcile with more extended CR synchrotron emission). This boundary condition also still requires specifying a value for $z_H$, which is a commonly encountered ambiguity in the CR literature. In particular, there turns out to be a degeneracy between the CR diffusion coefficient and $z_H$ when trying to infer the CR diffusion coefficient from local measurements. This degeneracy is lifted once $z_H \gtrsim r \sim 10$kpc, i.e. of order the CR injection length scale in the disk (see, e.g., Figure 10 in \citealt{Linden2010}). The physical reason is that for $z \gtrsim r$, CRs start propagating spherically away from the galaxy and have a smaller chance of making it back to the disk. Whether the same remains true for streaming is unclear. However, given the lack of unambiguously better alternatives, we adopt $f_p(z=r)=0$ as our boundary condition here. We then have,
\begin{equation}\label{eq:steady_nlld_sol_haloC}
    f_p(z) =\frac{Q}{\rm v_A} + \frac{X^2}{{\rm v_A}^2 (z -r -X^2 / (Q{\rm v_A}) ) }.
\end{equation}
For small $X$ (negligible diffusion set by NLLD), we have escape via Alfv\'enic streaming and $f_p \approx Q / {\rm v_A} \propto p^{-\gamma_{\rm inj}}$. At higher energies ($X^2 \gg Q {\rm v_A} r$), the NLLD term becomes dominant and $f_p$ is given by $f_p(z) \approx (r - z) Q^2/X^2 \propto p^{-5.4}$, for $\gamma_{\rm inj} = 4.2$. In this regime, the CR escape time from the Galaxy therefore has a strong energy dependence, $\tau_{\rm esc} \propto p^{-\gamma_{\rm inj} + 3} \propto p^{-1.2}$(see also \citealt{ptuskin_1997}). We note that in a steady state with CR injection balancing escape, the energy dependence of CR transport due to nonlinear Landau damping is significantly stronger than the scaling usually quoted in the literature, where the measured MW CR spectrum is used to calculate the CR diffusion coefficient (e.g., $\tau_{\rm esc} \propto p^{0.75}$ in \citealt{kulsrud_book} assuming $f_p \propto p^{-4.5}$; $\tau_{\rm esc} \propto p^{0.85}$ in \citealt{blasi_2019} assuming $f_p \propto p^{-4.7}$). 

We stress that eq. \ref{eq:steady_nlld_sol} is only valid when $\Gamma_{\rm lin} < \Gamma_{\rm NLLD}$. The nonlinear Landau damping rate can be approximately expressed as,
\begin{equation} \label{eq:nlld_heuristic}
    \Gamma_{\rm NLLD} \sim \Big( \frac{n_{\rm CR}}{n_{\rm th}} \frac{\rm v_{th}}{\rm v_{A}} \Omega_0 \frac{c}{L_{\rm CR}}  \Big)^{1/2} R_{\rm GV}^{-(\alpha - 3)/2},
\end{equation}
where $n_{\rm CR}$ is the total CR number density, $R_{\rm GV}$ is the CR rigidity in GV and $\alpha$ is the slope of the CR spectrum. For $\alpha\approx{4.7}$, $\Gamma_{\rm NLLD }\sim R^{-0.85}$, which is similar to the energy dependence for damping by charged interstellar dust grains and stronger than the energy dependence for damping by ambient Alfv\'enic turbulence ($\Gamma \sim R^{-0.5}$). Thus, even if NLLD is the dominant damping mechanism at low energies, linear damping likely becomes dominant above a rigidity $R$ for which $\Gamma_{\rm lin}(R) \sim \Gamma_{\rm NLLD}(R)$.

We can again include the proton energy-loss term from \eqref{eq:adv_diff_turb} in our solution through the transformation $Q \rightarrow Q-f_p(0) h / {\rm \tau_{loss}}$ in \eqref{eq:steady_nlld_sol_haloC}. Evaluating the result at $z=0$ then involves a quadratic equation for $f_p(0)$. However, because the term associated with NLLD (second term on the RHS of eq. \ref{eq:steady_nlld_sol_haloC}) becomes important at ultra-relativistic energies where losses are unimportant, to a very good approximation it is sufficient to carry out the transformation only on the term associated with Alfv\'enic streaming (first term on the RHS of eq. \ref{eq:steady_nlld_sol_haloC}). As in \eqref{eq:steady_lin_sol_loss}, to include losses we therefore should multiply the solution absent losses (eq. \ref{eq:steady_nlld_sol_haloC}) by $(1 + h / {\rm v_A \tau_{loss}})^{-1}$:
\begin{equation}\label{eq:steady_nlld_sol_haloC_loss}
    f_p(0) \approx \Big[ \frac{Q}{\rm v_A} - \frac{X^2}{{\rm v_A}^2 (r + X^2 / (Q{\rm v_A}) ) } \Big]  \Big(1 + \frac{h}{\rm v_A \tau_{loss}(p)} \Big)^{-1}.
\end{equation}

\subsection{Self-confinement in a stratified galaxy} \label{sec:strat_lin}

 \begin{figure}
  \centering
    \includegraphics[width=0.47\textwidth]{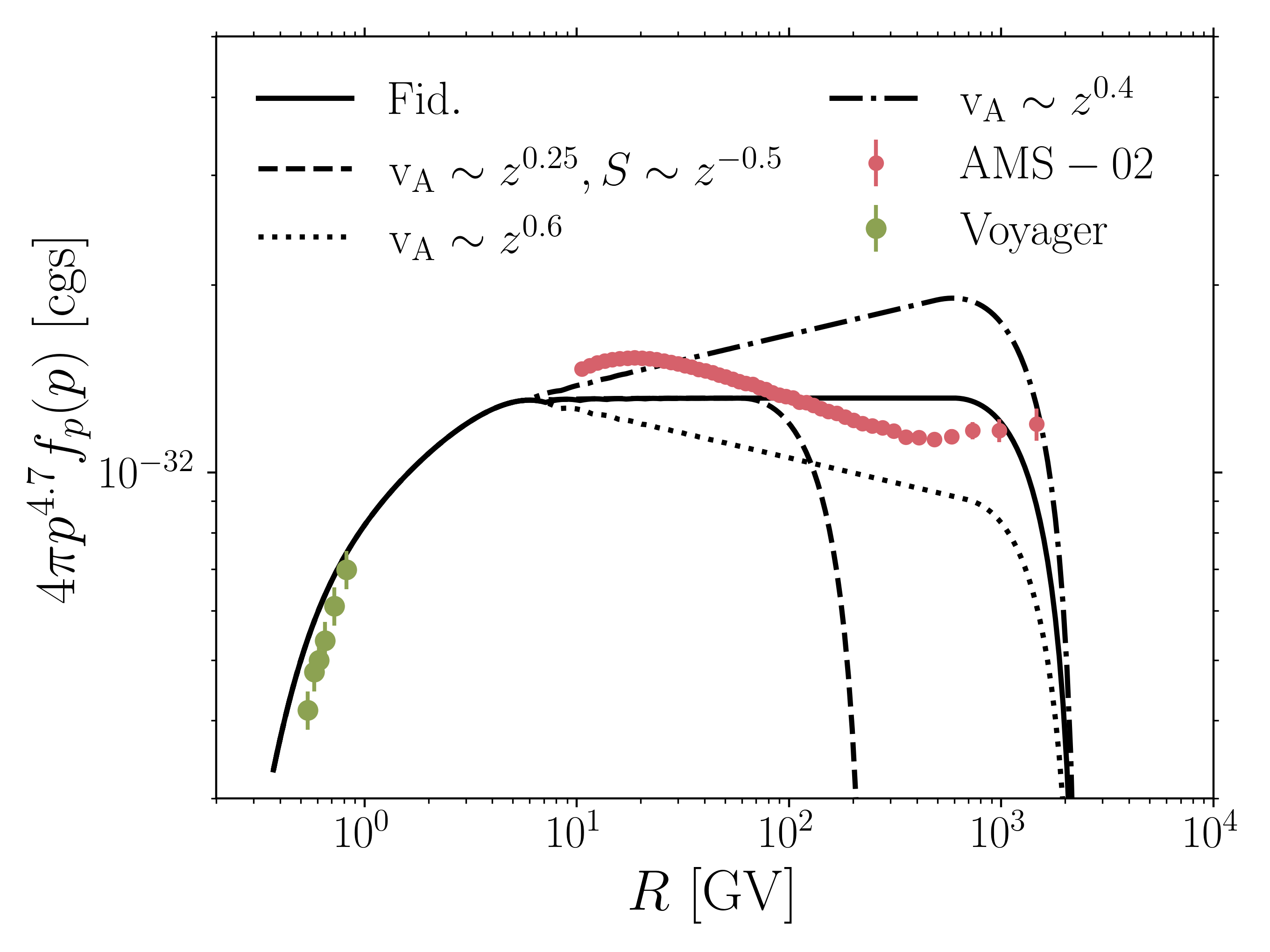}
 \caption{Comparison of self-confinement proton spectra in a  stratified galaxy with AMS-02 and Voyager data (\citealt{aguilar_2015}; \citealt{cummings_2016}). In this calculation, waves excited by the streaming instability are damped linearly by the ambient Alfv\'enic turbulence. The CR spectrum is steepened by transport in a stratified medium, in which the  Alfv\'en speed and the damping of self-excited Alfv\'en waves depend on the distance from the disk, $z$. The spectrum is steepened relative to the injection spectrum despite roughly Alfv\'enic streaming in most of the volume where CRs are confined.  The spectral slope depends sensitively on the spatial profiles of the plasma properties that affect the Alfv\'en speed and the damping strength. We show how variations around the fiducial scalings (black solid line; ${\rm v_A} \sim z^{0.5}$, $ S \sim z^{-1}$, where $S$ characterises the spatial variation of the linear damping) affect the spectrum.  See section \ref{sec:strat_lin} for details of the calculation. The  galactic phase structure has significant implications for the interpretation of CR observables. \label{fig:strat_halo_fp} }
\end{figure}

In Section \ref{sec:sc_single_phase} we showed that predictions from self-confinement theory are not compatible with phenomenological models of CR transport based on MW data. In particular, for both linear and nonlinear damping mechanisms, the energy dependence of CR transport is very different from the empirically derived $\kappa \sim E^{\delta}$ with $\delta \sim 0.3-0.7$ (see, e.g., \citealt{blasi12}, or \citealt{hopkins_cr_pheno} who favour $\delta \sim 0.5-0.6$). The discrepancy between theoretical predictions and observations is particularly strong for linear damping mechanisms: based on Figure \ref{fig:lin_damping} it may seem hopeless to reconcile the theoretical transport prediction with empirical scalings.

However, before we completely discard transport corrections due to linear damping mechanisms, we note that an energy dependence in CR observables can also be introduced in rather unconventional ways. As an example, in this section we will show how roughly Alfv\'enic streaming, a process usually considered to be energy independent, in a stratified galaxy with linear damping of self-excited Alfv\'en waves can mimic energy-dependent transport with the appropriate scaling.  While our calculation is not much more than a toy model, it illustrates the importance of the ISM phase structure, and its spatial variation, for interpreting CR observables. We note that CR transport in a stratified galaxy has been studied in recent years by numerous authors (see, e.g., \citealt{ptuskin_1997}, \citealt{recchia_2016}, \citealt{evoli_2018} for CR transport with advection by winds or Alfv\'en waves, and \citealt{tomassetti_2012} for pure diffusion with vertical variation), who showed that features in CR spectra (e.g. hardening) can be the product of vertical variations in CR transport. However, our calculation is significantly different from their models. In particular, the energy dependence in CR transport that we derive is a product of  the spatial variation of the linear damping rate of self-excited Alfv\'en waves, which can shut off the streaming instability (not considered in the works referenced above), and the simultaneous spatial variation of the Alfv\'en speed. Our calculation is thus similar to the models considered in \cite{holmes_1974} and \cite{holmes_1975}, who considered exponentially varying ion-neutral damping rates and Alfv\'en speeds. The calculation presented here is valid for arbitrary linear damping mechanisms, including damping due ambient Alfv\'enic turbulence, which unlike ion-neutral damping operates in the hot ionized phases of the ISM. These occupy most of the galactic volume and are most important for setting CR observables.

In reality, galaxies and their halos are vertically stratified. In particular, the Alfv\'en speed likely increases, while the damping strength of self-excited Alfv\'en waves decreases, with increasing distance from the galactic disk. As a result of the decreasing damping, the range of CR energies that are self-confined changes with distance from the disk. And because of the simultaneous increase in Alfv\'en speed, CRs of different energies sample different effective escape speeds from the galaxy, even though their transport is approximately Alfv\'enic as discussed in Section \ref{sec:lin_damp}. In a stratified medium, the distribution function for self-confined CR protons roughly satisfies (cf. equation \ref{eq:stead_state_lin_damp}),
\begin{equation} \label{eq:crp_stratified}
    f(p,z) \sim \frac{Q(p)- S(p,z)}{{\rm v_{A}}(z)} \qquad z \gtrsim z_{\rm sc},
\end{equation}
where we have introduced the ``self-confinement height" $z_{\rm sc}(p)$, the height beyond which CRs of a given momentum are self-confined, $Q(p)=S(p, z_{\rm sc})$, and $S$ is the correction term that comes from an arbitrary linear damping mechanism of self-excited Alfv\'en waves. For $z< z_{\rm sc}$ CRs are not self-confined and the distribution function is approximately given by equations \ref{eq:f>p_*} and \ref{eq:f>p_*_approx} below. We shall consider cases in which the spatial variation of $S$ is stronger than the spatial variation of ${\rm v_A}$. We note that \eqref{eq:crp_stratified} is an approximation, as we used \eqref{eq:stead_state_lin_damp}, which assumes a constant ${\rm v_A}$. In a vertically stratified galaxy, with ${\rm v_A} \neq {\rm const}$, the Alfv\'enic loss term on the RHS of \eqref{eq:adv_diff_turb} is not zero and so \eqref{eq:steady_master}, used to derive \eqref{eq:stead_state_lin_damp}, is not correct. Equation \ref{eq:crp_stratified} is, however, a reasonable approximation in the region $(z-z_{\rm sc})/H_A \lesssim 1$, where $H_A$ is the Alfv\'en scale height.

The approximate solution at some small height above the disk, $z=z_0$, for low-energy CRs with $Q \gg S(z_0)$  is found by directly evaluating \eqref{eq:crp_stratified}, which yields the very weak energy dependence of CR transport  described in Section \ref{sec:lin_damp} with $f \sim Q/{\rm v_A}(z_0)$. The more interesting energy dependence comes from CR momenta $p > p_*$ which are unable to self-confine at small $z$ due to large damping rates, $S(p>p_*, z=z_0) > Q(p>p_*)$. While the injection rate $Q$ is fixed, $S(z)$ is plausibly a decreasing function of $z$, and so for sufficiently low-energy CRs there may be a $z_{\rm sc}(p)$ where $Q(p)=S(p, z_{\rm sc})$ and CRs start to self-confine. We stress that $z_{\rm sc}(p)$ is a function of CR momentum, which will give rise to energy dependence in the CR distribution function. 

What are the consequences for CR observables close to the disk? Because close to the disk CRs with $p>p_*$ are not self-confined, one might first guess the free-streaming solution $f(p>p_*) \sim Q/c$ at small $z$. However, if CRs do confine themselves above some height and $f(z \sim z_{\rm sc}) \sim Q/ {\rm v_A}$, the free-streaming solution is unphysical as it corresponds to CRs streaming up their pressure gradient. Instead, the solution relaxes to a steady state with a flat spatial profile in regions where self-confinement does not operate,
\begin{equation} \label{eq:f>p_*}
    f(p>p_*, z_0 \leq z \leq z_{\rm sc}) \sim {\rm max}\Big[ \frac{Q(p>p_*) - S(p>p_*,z)}{{\rm v_A}(z)} \Big]_z,
\end{equation}
i.e. $f$ is independent of $z$ for $z<z_{\rm sc}$. To get a sense of the energy scaling implied by  \eqref{eq:f>p_*}, it is useful to further approximate the above by,
\begin{equation}\label{eq:f>p_*_approx}
    f(p>p_*, z_0 \leq z \leq z_{\rm sc}) \sim \frac{Q(p>p_*)}{{\rm v_A}(z_{\rm sc}(p))} \sim p^{-\gamma_{\rm inj}  - \lambda_1 \lambda_3  },
\end{equation}
where we used our assumption that the spatial variation of $S$ is stronger than the spatial variation of ${\rm v_A}$ and in the last step we assumed that the spatial variations are well described by simple power laws, $ {\rm v_A} \sim z^{\lambda_1}$, $S \sim z^{- \lambda_2}$, and as a result $z_{\rm sc} \sim p^{\lambda_3}$. For $\lambda_1 > 0$ and $\lambda_3 >0$, the CR spectrum is steepened. For a linear damping $\Gamma \propto k^a$, $S \sim z^{- \lambda_2} p^{-3 - a}$ and one can show that the self-confinement height $z_{\rm sc}$ is given by,
\begin{equation}\label{eq:zsc}
    z_{\rm sc} \sim z_0 p ^{\lambda_3} \sim z_0 p^{\frac{\gamma_{\rm inj} - (3+a)}{\lambda_2}},
\end{equation}
and \eqref{eq:f>p_*_approx} becomes,
\begin{equation}\label{eq:f>p_*_approx2}
    f(p>p_*, z_0 \leq z \leq z_{\rm sc}) \sim p^{-\gamma_{\rm inj}  - \lambda_1 ( \gamma_{\rm inj} -3 -a)/\lambda_2 }.
\end{equation}
For $\lambda_1 >0$ (${\rm v_A}$ increases with increasing $z$), $\lambda_2>0$ ($S$ decreases with increasing $z$) and $a < \gamma_{\rm inj} -3$ (true for all known linear damping mechanisms except ion-neutral damping at long wavelengths), the spectrum is steepened relative to the injection spectrum. 

So far we have kept the discussion fairly general and considered arbitrary linear damping mechanisms. Let us now provide a more concrete solution for the steady-state $f$. We consider CR propagation in a turbulent inner galactic halo, so that the linear damping is due to the shearing of self-excited waves by the ambient Alfv\'enic turbulence. Thus, $\Gamma \propto k^{0.5}$ and for turbulence injected with ${\rm Ma} \sim 1$ on scales $L$ (\citealt{skilling71}; \citealt{farmer_goldreich}), 
\begin{equation}
   S \sim p^{-3.5} \frac{L^{-0.5} B_{\rm tot} B_z}{4 \pi^3 (m \Omega_0)^{0.5} } \propto B_{\rm tot}^{0.5} B_z L^{-0.5} \propto z^{- \lambda_2},
\end{equation}
where $B_{\rm tot}$ is the total magnetic field (as opposed to just the vertical component). Suppose we take ${\rm v_A} \sim z^{0.5}$ i.e. $\lambda_1 = 0.5$ (e.g. $\rho \sim z^{-1}$ and constant $B_z$). Let us further assume $S \sim z^{-1}$, i.e. $\lambda_2=1$. For $\gamma_{\rm inj} = 4.3$, $z_{\rm sc} \sim p^{0.8}$ per equation \ref{eq:zsc}, i.e. $\lambda_3 = 0.8$. Then, $f(p>p_*, z \leq z_{\rm sc})$  per equations \ref{eq:f>p_*_approx} or \ref{eq:f>p_*_approx2} scales as $f \sim p^{-4.3 - 0.4} \sim p^{-4.7}$, approximately consistent with observations. This is despite the fact that CRs are advected at $\approx {\rm v_A}$ in most of the volume in which they are self-confined.  Roughly Alfv\'enic streaming in a stratified medium with the right spatial variations can therefore mimic energy-dependent streaming $\propto E^{0.3-0.7}$.

In Figure \ref{fig:strat_halo_fp} we show example proton spectra from the stratified-halo calculation and compare them to AMS-02 and Voyager data (\citealt{aguilar_2015}; \citealt{cummings_2016}; for AMS-02 we only consider $E \gtrsim 10$GeV, as measurements of lower-energy CR protons are strongly affected by solar modulation). We take $z_0=500$pc as our base height. For the CR source term ($=2Q$, see eq. \ref{eq:adv_diff_turb}) we take a supernova rate of 1 per 100 years (every supernova injecting $10^{50}$ ergs of CR energy) with a spectral slope $\gamma_{\rm inj}=4.3$. We assume that the spatial variation of $f$ below $z_0$ is small, so that $f(z_0) \approx f(0)$. At the base we take $n_{\rm th}=0.005{\rm cm}^{-3}$ and $B = 1 \mu$G. The domain of our calculation extends from $z_0$ to $z_{\rm max}=20$ kpc. We include ionisation losses as described in Section \ref{sec:sc_single_phase} and eq. \ref{eq:steady_lin_sol_loss}, which we assume occur in a disk of neutral-gas density $1.25 \ {\rm cm^{-3}}$ and half-thickness 200 pc.  Turbulence at the base is injected on scales $L=z_0=500$pc. As our fiducial spatial scalings we take ${\rm v_A} \sim z^{0.5}$ ($\lambda_1 = 0.5$) and $S \sim z^{-1}$ ($\lambda_2 = 1$), as described in the paragraph above. We show the proton spectrum for the fiducial parameters and scalings as the solid line. We also show how variations around the fiducial scalings affect the spectrum.  Interestingly, according to this mechanism of spectral steepening, small changes in the spectral slope as a function of CR energy in MW data are the result of changes in the spatial profiles with increasing distance from the galaxy. 

The cutoff in the spectra in Figure \ref{fig:strat_halo_fp} is set by $z_{\rm max}=20$kpc. We could in principle shift the cutoff to higher energies by increasing $z_{\rm max}$ (or $r_{\rm max}$ in spherical geometry), but our adopted scalings would not hold indefinitely and would likely be different in the CGM relative to the near-disk environment. Thus, in the stratified-halo model it remains true that self-confinement cannot explain CR observations at high energies and there remains a need for additional scattering.

\subsection{The need for extrinsic waves} \label{sec:need_ext}
Sections \ref{sec:sc_single_phase} and \ref{sec:strat_lin} showed the well-known result that higher-energy CRs are unable to self-confine due to the linear damping of waves excited by the streaming instability. Moreover, even in the limit of negligible linear damping, nonlinear Landau damping predicts a proton spectrum that is asymptotically significantly steeper than the observed spectrum in the solar neighbourhood (Section \ref{sec:nonlin_damp}). 
Thus, it transpires that self-confinement alone cannot explain observations of CRs with energies $\gtrsim$TeV. There must exist a different scattering mechanism that is necessarily dominant at high energies, and potentially dominant at low energies. A natural way of generating waves over a wide range of spatial scales, that can scatter CRs over a wide range of energies, is via a turbulent cascade. Currently, the most likely candidate for efficient CR scattering is the cascade of fast modes (\citealt{yan_lazarian_2004}; \citealt{yan_lazarian_2008}). 

\section{Scattering by turbulence} \label{sec:turb}

 \begin{figure}
  \centering
    \includegraphics[width=0.47\textwidth]{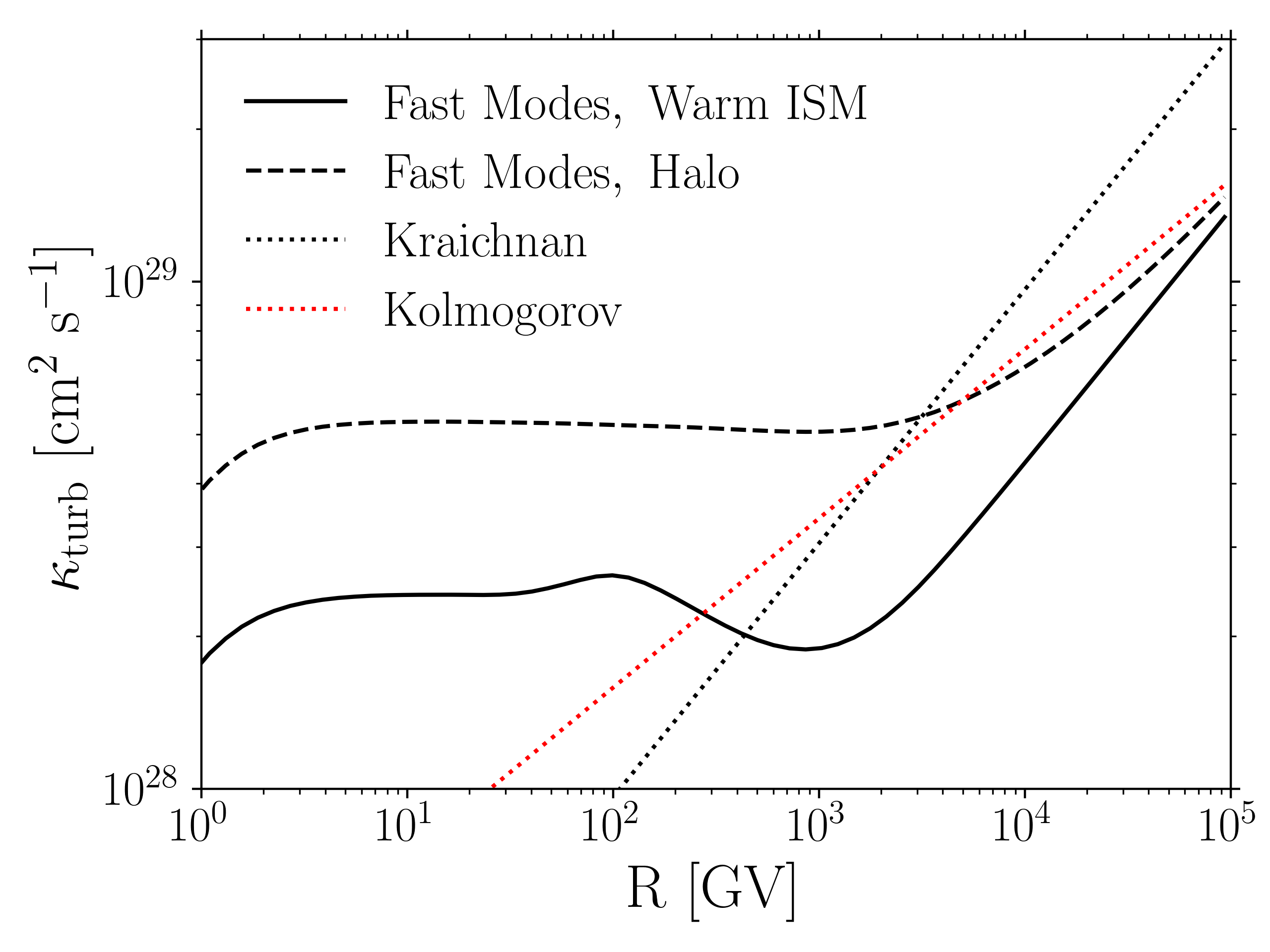}
  \caption{Diffusion coefficients due to the weak-turbulence MHD fast-mode cascade in a low-$\beta$ warm ISM (black solid line) and a low-$\beta$ halo (black dashed line; see main text in Section \ref{sec:turb} for parameters). In the halo, the diffusion coefficient has a very weak energy dependence due to the collisionless damping of the cascade. In the warm ISM, the thermal mean free path is small, and the fast-mode cascade is isotropic (with $\kappa_{\rm turb} \propto p^{0.5}$) until viscous damping becomes important on scales resonant with $\sim 10^3$ GeV CRs. For comparison, we show the scaling of $\kappa_{\rm turb}$ for undamped Kolmogorov ($\propto R^{1/3}$) and Kraichnan-like ($\propto R^{1/2}$) cascades (often assumed in phenomenological models), which have a very different energy dependence from the MHD fast-mode calculation at low energies.   \label{fig:kappa_turb}}

\end{figure}

When CR scattering is due to an extrinsic (balanced) turbulent cascade, ${\rm v_{st}} =0$ in \eqref{eq:adv_diff_turb} and \eqref{eq:steady_master}, and $\kappa=\kappa_{\rm turb}$ is specified by the properties of the turbulent cascade. For $\kappa_{\rm turb}$ independent of position, the solution to \eqref{eq:steady_master} is particularly simple,
\begin{equation} \label{eq:sol_turb}
    f(z) = \frac{Q}{\kappa_{\rm turb}} (H - z) + f(H), \qquad z < H,
\end{equation}
where we used $f=f(H)$ at $z=H$ as an outer boundary condition.

The main theoretical challenge with \eqref{eq:sol_turb}  lies in specifying $\kappa_{\rm turb}$. The MHD turbulent cascades of Alfv\'en and slow magnetosonic waves are believed to be inefficient at scattering CRs due to the fact that eddies are highly elongated along the local magnetic field (\citealt{chandran_scattering}). For this reason, \cite{yan_lazarian_2004} proposed that CRs are scattered by the MHD fast-mode cascade, which they took to be isotropic and obey a weak-turbulence $\propto k^{-3/2}$ spectrum based on the theory put forward by \cite{zakharov_sagdeev_1970} and the numerical work by \cite{cho_lazarian}. Following \cite{yan_lazarian_2004} and \cite{yan_lazarian_2008}, in this section we review the calculation of $\kappa_{\rm turb}$ using the weak-turbulence formalism for fast modes. In Section \ref{sec:fast_mode_uncertainties} we discuss some of the significant uncertainties in this calculation.

We include both collisionless damping (\citealt{ginzburg1961}) on scales smaller than the thermal-particle mean free path and anisotropic viscous damping (\citealt{br65})  on scales larger than the mean free path ($\sim 10^{13} \ {\rm  cm}$ in the warm ISM, comparable to the Larmor radius of $\sim 10$ GeV CRs). At low $\beta$, where $\beta=8 \pi p_g / B^2$  is the ratio of thermal to magnetic pressure, the damping makes the fast-mode cascade highly anisotropic below the viscous scale because parallel propagating modes are damped least efficiently. In particular, there is a scale-dependent critical wave pitch angle $\theta_c(k)$ for which the cascade rate,
\begin{equation} \label{eq:tau_fast_isotropic}
    \tau_{\rm casc}^{-1} \sim \frac{k \delta \vrm ^2}{\vrm_{\rm ph}} \sim \Big(\frac{k}{L} \Big)^{1/2} \frac{\delta V^2}{\vrm_{\rm ph}}
\end{equation}
 ($\delta \vrm$ denotes the amplitude at scale $k$, while $\delta V$ is the amplitude at the injection scale) is equal to the wave damping rate. For viscous damping of quasi-parallel modes at low $\beta$ the damping rate is approximately (\citealt{br65}),
 \begin{equation} \label{eq:gamma_collisional}
         \Gamma (k_\parallel l_{\rm mfp} {\rm v_A / v_{th}} \ll 1) \sim \frac{\nu_{\rm B} k^2}{6} \theta^2 \qquad \theta \ll 1,
 \end{equation}
 and for collisionless damping of quasi-parallel modes at low $\beta$ it is approximately (\citealt{ginzburg1961}),
 \begin{equation}\label{eq:gamma_collisionless}
         \Gamma (k_\parallel l_{\rm mfp} {\rm v_A / v_{th}} \gg 1)  \sim \frac{\sqrt{\pi \beta} \theta^2  }{4} \Big( \frac{m_e}{m_i} \Big)^{1/2} k {\rm v_A} \qquad \theta \ll 1.
 \end{equation}
We provide more general expressions for the damping rates in Appendix \ref{app:diff_fast_mode} (eq. \ref{eq:Gamma_brag} for collisional and eq. \ref{eq:Gamma_ginz} for collisionless damping, respectively). Equating the cascade rate (eq. \ref{eq:tau_fast_isotropic}) and the angle-dependent damping rates (eq. \ref{eq:Gamma_brag} and \ref{eq:Gamma_ginz}) gives the scale-dependent critical wave pitch angle $\theta_c(k)$. Modes with $\theta < \theta_c$ are not strongly damped.  \citealt{yan_lazarian_2004} and \citealt{yan_lazarian_2008} assume that fast modes with $\theta < \theta_c$  continue cascading to smaller scales unaffected by the damping, while the remaining modes are fully damped. Here we use the same assumption, but we stress that it is quite uncertain. When we evaluate the cascade and damping rates in \eqref{eq:tau_fast_isotropic} and \eqref{eq:Gamma_brag}--\eqref{eq:Gamma_ginz}, we take the average of the cascade/damping rate in the interval $(\theta- \delta \theta, \theta + \delta \theta)$, where $\delta \theta$ is the spread in mode pitch angle experienced by a fast mode during one cascade time due to turbulent magnetic-field-line wandering. The field-line wandering due to ambient Alfv\'enic turbulence with ${\rm Ma_{Alf}} \sim 1$ experienced by quasi-parallel fast modes with wavenumber $k$ is,
    \begin{equation} \label{eq:dtheta_alf_iso}
    \frac{\delta B}{B} \sim   \Big({\rm Ma}^2(kL)^{1/2}  \Big)^{-1/2}.
\end{equation}
Throughout this work ${\rm Ma_{Alf}}$ is the turbulent amplitude (normalised by ${\rm v_A}$) at the injection scale in the Alfv\'enic cascade, while ${\rm Ma}$ corresponds to the amplitude (normalised by ${\rm v_A}$) in the fast-mode branch. 
In our calculation, we also assume that the medium is sufficiently ionized (neutral fraction $\lesssim 1\%$) to ignore ion-neutral damping (see \citealt{xu_2016} for a recent discussion of MHD turbulence in partially ionized media). In the warm phase of the ISM, however, this assumption is likely only valid in a fraction of the volume. We provide a more complete summary of the $\kappa_{\rm turb}$ calculation in Appendix \ref{app:diff_fast_mode}.

We show example diffusion coefficients in fast-mode turbulence in Figure \ref{fig:kappa_turb}, for turbulence injected with Mach number ${\rm Ma =1}$ on scales $L=100$pc. The warm ISM ($B=6\mu$G, $n_{\rm th}=0.1 {\rm cm^{-3}}$, $T=2 \times 10^4$K) is shown as the solid line. We also show the diffusion coefficient for a low-$\beta$ halo for comparison ($B=6\mu$G, $n_{\rm th}=0.001 {\rm cm^{-3}}$, $T=2 \times 10^6$K). The diffusion coefficient in the halo has a very weak energy dependence, due to the influence of collisionless damping on the cascade. In the warm ISM with a short thermal mean free path, the fast-mode cascade is unaffected by damping (with $\kappa_{\rm turb} \propto p^{0.5}$) down to scales resonant with $\sim 10^3$GV CRs, where viscous damping becomes important. Below a few GV, $\kappa_{\rm turb}$ decreases with decreasing rigidity as CRs become trans-relativistic. For comparison we show the scaling of $\kappa_{\rm turb}$ for the undamped Kolmogorov ($\kappa_{\rm turb} \propto p^{0.33}$) and Iroshnikov-Kraichnan ($\kappa_{\rm turb} \propto p^{0.5}$) phenomenologies commonly used in the CR transport literature, which differ significantly from the fast-mode diffusion coefficients at low energies. 

We can explain the main trends in Figure \ref{fig:kappa_turb} heuristically. On large scales the cascade is isotropic and unaffected by the damping, with $(\delta B /B)^2 \sim k^{-1/2}$ and $\kappa \sim {\rm v}^2 / \nu^G_{\Gamma=0}   \sim {\rm v}^2 / ( \Omega (\delta B / B)^2) \sim R^{0.5}$, where $\nu^G_{\Gamma=0}$ is the gyroresonant scattering rate absent any damping. Recall that $R \sim k^{-1}$ for gyro-resonance. For collisionless damping, equating \eqref{eq:tau_fast_isotropic} and \eqref{eq:gamma_collisionless} gives $\theta_c^2 \sim k^{-1/2}$, i.e. surviving fast modes cover a solid angle $\sim \theta_c^{2}$ that shrinks $\propto k^{-1/2}$. Thus, the wave power available to scatter CRs decreases with decreasing spatial scale. Gyroresonant scattering then scales as $\nu^G \sim \nu^G_{\Gamma=0} \theta_c^2 \sim R^{1/2-1/2} \sim R^0$. Gyroresonant scattering is therefore rigidity independent in the collisionlessly damped regime. Since scattering by transit time damping (TTD; see Appendix \ref{app:diff_fast_mode}) is also energy independent, this results in a roughly energy independent CR diffusion coefficient, as shown by the black dashed line at small rigidities in Figure \ref{fig:kappa_turb}. The analogous calculation for viscous damping gives $\theta_c^2 \sim k^{-3/2}$ and $\nu^G \sim \nu^G_{\Gamma=0} \theta_c^2 \sim R^{1}$. Thus, gyroresonant scattering decreases with decreasing rigidity, while the TTD scattering is energy-independent. As a result, the total $\kappa_{\rm turb}$ in the warm ISM, shown as the solid line in Figure \ref{fig:kappa_turb}, increases slowly with decreasing rigidity in the viscous regime (roughly $10^2-10^3$ GV), followed by weak energy dependence in the collisionless regime (energies $\lesssim 100$ GV).

The low-energy scattering frequency associated with the diffusion coefficients in Figure \ref{fig:kappa_turb}, $\nu \gtrsim 10^{-8} {\rm s^{-1}}$, likely sufficiently isotropises the CR distribution function to suppress the excitation of the streaming instability. The CR anisotropy $\sim \vrm_{\rm D} / c$, where $\vrm_{\rm D}$ is the CR drift speed and $c$ is the speed of light, is (\citealt{skilling_1975})
\begin{equation} \label{eq:vd}
    \frac{\vrm_{\rm D}}{c} \sim \frac{\kappa / L_{\rm CR}}{c}\sim \frac{c / L_{\rm CR}}{\nu} \sim 10^{-4} \ \frac{10 \ {\rm kpc}}{L_{\rm CR}} \  \frac{10^{-8} \ {\rm s}^{-1}}{\nu},
\end{equation}
where $L_{\rm CR}$ is the CR scale height and $\nu$ is the scattering rate ($10^{-8} {\rm s}^{-1}$ corresponds to a CR mean free path of $\sim 1$pc and a diffusion coefficient of $\sim 10^{29} \ {\rm cm^{2} s^{-1}} $). If $\vrm_{\rm D} < {\rm v_A}$, the streaming instability is suppressed. At low $\beta$ with ${\rm v_A} > c_s$, $\vrm_{\rm D} \sim 3 \times 10^6 \ {\rm cm \ s^{-1}}$ is comparable to (less than) the Alfv\'en speed in the warm ISM (the halo/hot ISM). The CR anisotropy in low-$\beta$ turbulence is therefore plausibly small enough that it does not excite the streaming instability (especially if damping of Alfv\'en waves is also taken into account), consistent with external turbulence dominating the transport. Interestingly, for a plausible $L_{\rm CR} \sim 10$kpc, equation \ref{eq:vd} suggests that it is implausible that CR diffusion coefficients due to turbulence are $\gg 10^{29} \ {\rm cm^2 s^{-1}}$ for the bulk of the CRs: in such a regime, the streaming instability is excited, which increases the scattering rate on top of the turbulent scattering rate, thus lowering the effective diffusion coefficient (at higher energies the streaming instability is fully damped and so $\kappa_{\rm turb}$ can in principle be arbitrarily large).

Observations suggest a CR diffusion coefficient that approximately scales as $\kappa \propto E^{0.3-0.7}$ down to $\sim$GeV energies. Thus, the very different scaling of $\kappa_{\rm turb}$ for fast modes at low energies in Figure \ref{fig:kappa_turb} is not observed.

\section{Self-excited Waves + Fast-Mode Turbulence and Multi-phase Models of CR Transport} \label{sec:sc+et}

Sections \ref{sec:sc} and \ref{sec:turb}, and Figures \ref{fig:lin_damping}--\ref{fig:kappa_turb} suggest that existing theories of self-confinement and scattering by weak fast-mode turbulence cannot, on their own, explain the rather consistent scaling with energy of the CR diffusion coefficient inferred from observations,  $\kappa \propto E^{0.3-0.7}$ (see also \citealt{kc71}; \citealt{farmer_goldreich}; \citealt{fornieri_2021}). This suggests that 1) there is a different, yet unknown, universal process/cascade that efficiently scatters CRs with the right energy dependence, or that 2) a combination of scattering by self-excited waves and fast-mode turbulence conspires to mimic the empirically derived CR diffusion coefficient. At present, no MHD cascade satisfies the properties required for scenario 1). We present some speculative suggestions that might remedy this in Section \ref{sec:discussion}. In what follows, we consider the alternative possibility that CRs are scattered by a combination of fast modes and self-excited waves. We shall assume that the MHD-fast mode cascade is isotropic and well-described by weak-turbulence theory (Section \ref{sec:turb}), but we stress that this assumption is very uncertain, as we discuss in Sections \ref{sec:weak_shocks} and \ref{sec:weak_casc_anisotropic}. 

Our calculation is the first attempt to combine microphysical theories of CR self-confinement and scattering by MHD fast-mode turbulence. It will highlight the important issue that this theoretically motivated combination of scattering mechanisms can in principle reproduce rough trends in CR spectra, but this requires significant fine-tuning of plasma parameters. The fine-tuning issue arises primarily because the MHD fast-mode cascade is significantly damped on small scales corresponding to CR energies $\lesssim$ TeV (Figure \ref{fig:kappa_turb}). As a result, the CR diffusion coefficient due to fast-mode turbulence deviates strongly from the empirically derived $E^{0.3-0.7}$ scaling. This is not captured in phenomenological models of CR transport based on undamped isotropic ($\sim$ isotropic Alfv\'enic) turbulence.

\subsection{Two-phase model of CR transport}\label{sec:two_phase_mw}
We now calculate  CR spectra (the proton spectrum and the B/C spectrum) using a combination of self-excited Alfv\'en waves and the fast-mode turbulence model from Section \ref{sec:turb}.

\begin{figure}
  \centering
    \includegraphics[width=0.45\textwidth]{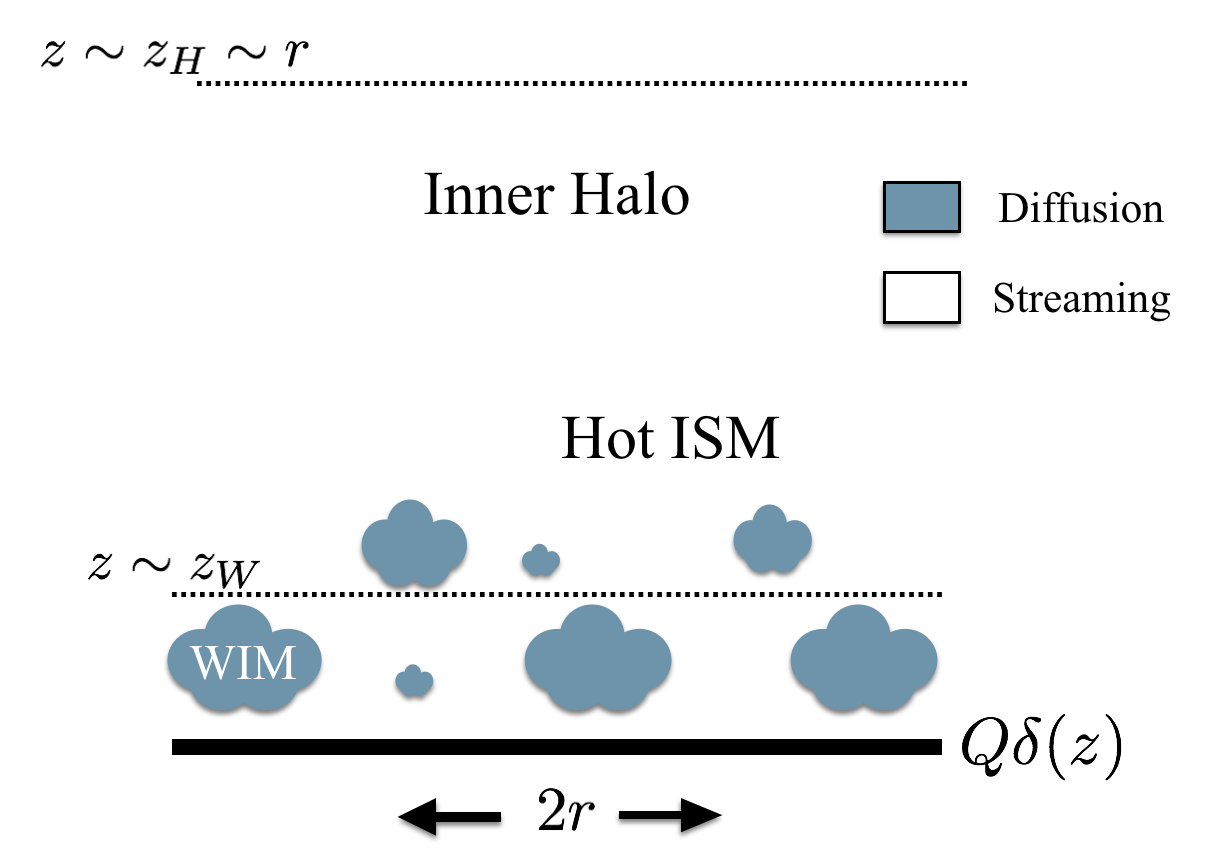}
  \caption{Example model for multi-phase CR transport in the Milky Way. Cosmic rays stream/diffuse away from the galactic disk where they are injected by supernovae and/or spallation reactions ($Q$ is the CR injection rate per unit area). In the hot ionized medium and halo, CRs are self-confined and streaming. In the warm ionized medium (WIM), where it is plausible that  $\beta \equiv 8 \pi p_g / B^2 < 1$, CRs are scattered by fast-mode turbulence. This results in diffusive transport.     \label{fig:galaxy_sketch}}
\end{figure}

Because fast modes are strongly damped in dilute high-$\beta$ environments (which likely make up a large fraction of the volume), while in low-$\beta$ regions they may be very efficient scatterers and suppress the streaming instability (assuming the weak-turbulence calculation of Figure \ref{fig:kappa_turb} correctly describes the fast-mode cascade), a possible picture emerges that CRs are scattered by only fast modes or only self-excited Alfv\'en waves, depending on the ISM phase. This motivates treating CR propagation as a multi-phase problem. We show an illustration of what multi-phase CR propagation may look like in Figure \ref{fig:galaxy_sketch}. There is observational evidence that the warm phases of the ISM in spiral galaxies are magnetically dominated (e.g., \citealt{beck2015_Bfield_spirals}), i.e. $\beta = 8 \pi p_g / B^2 \lesssim 1$. This may also be true in the inner CGM (e.g., \citealt{beck2015_Bfield_spirals}). The transport model in Figure \ref{fig:galaxy_sketch} has a low-$\beta$ warm ISM in which CRs diffuse due to scattering by fast-mode turbulence. If $\beta > 1$, however, fast modes are damped and low-energy CRs are likely self-confined. We thus assume that cosmic rays are self-confined and streaming in the remaining regions of the galaxy, i.e. the hot ISM and inner halo. We note that different ISM phase structures are also plausible and the depiction in Figure \ref{fig:galaxy_sketch} should be regarded as one particular example of a broader class of multi-phase propagation models. We focus on the particular model illustrated in Figure \ref{fig:galaxy_sketch} because we find that it can, in principle, match observations. This is important in terms of assessing whether existing theories of CR transport are at all compatible with observations.

The warm ionized medium in our 1D formalism is modelled as a thin layer in the disk, i.e. we assume that the region between $z=0$ and $z=z_W$ is entirely filled by the warm ISM (this 1D picture could, for example, approximate a more realistic multi-phase ISM where the warm ISM fills 50\% of the volume up to $z=2z_W$, as in Fig. \ref{fig:galaxy_sketch}). We treat CRs as self-confined in the coronal regions and the inner halo of the galaxy, $z> z_W$. We now consider solutions to \eqref{eq:steady_master} for the two regions with different transport mechanisms, starting with the self-confinement region.

\subsubsection{Self-confinement regions with NLLD: the proton spectrum} \label{sec:protons}
\begin{figure}
  \centering
    \includegraphics[width=0.47\textwidth]{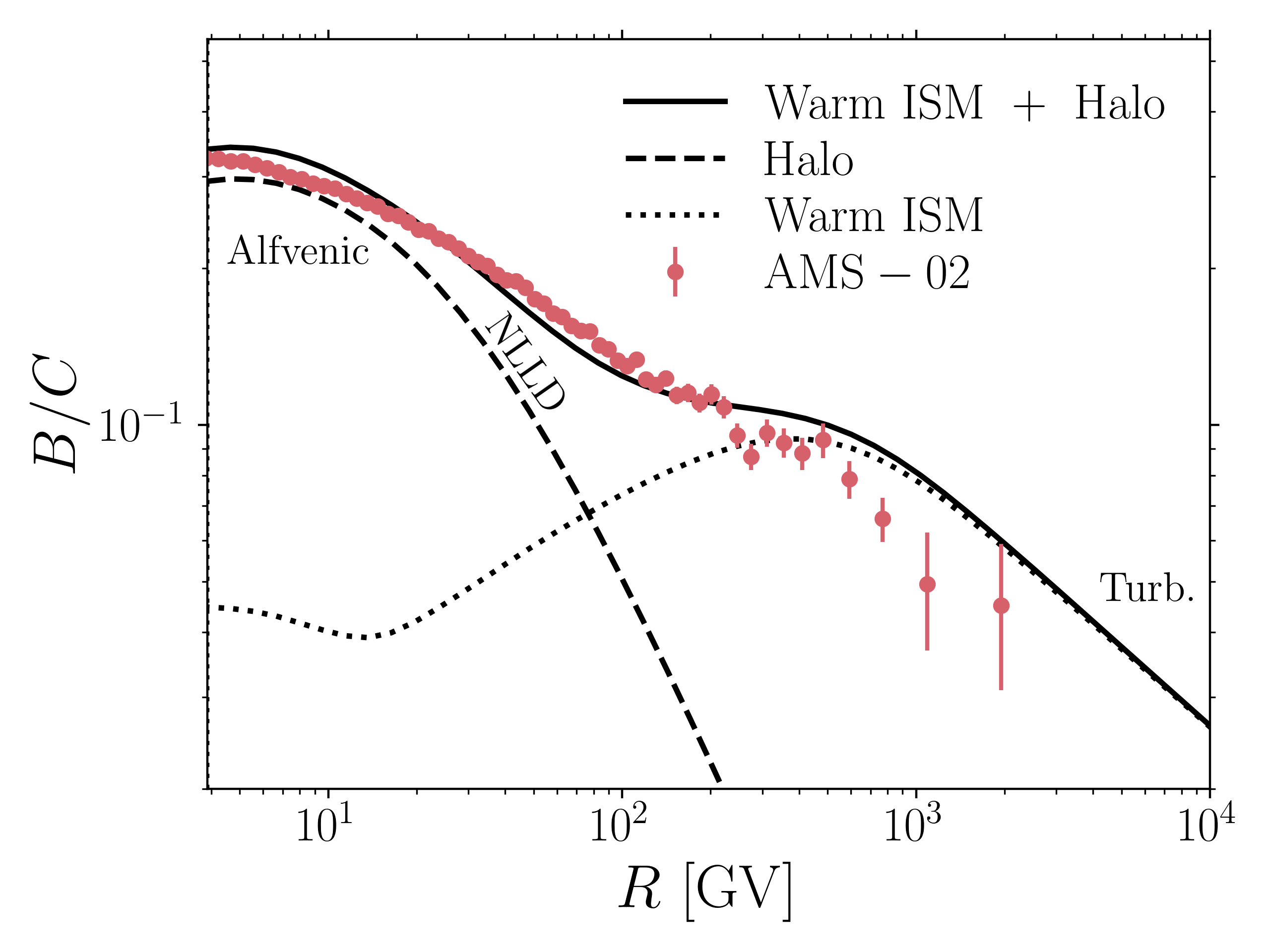}
 \caption{The boron-to-carbon ratio (B/C) for our fiducial two-phase ISM model. The red points are AMS-02 measurements (\citealt{aguilar_bc}) and the black line is the spectrum obtained using \eqref{eq:boron_steady} and \eqref{eq:sol_wism} evaluated at $z=0$.  The dotted and dashed lines show the individual contributions from the turbulent warm ISM, and the self-confined hot ISM and halo (first and second terms in eq. \ref{eq:sol_wism}, respectively).  There is good agreement between the model and the data. For $R \lesssim 100$ GV, self-confinement in the diffuse hot ISM and halo sets the local B/C. At GeV energies, CR transport has a weak energy dependence as CRs stream at the Alfv\'en speed. At intermediate energies, NLLD is important and introduces a strong dependence on energy. Above a few hundred GeV, diffusion in the turbulent warm ISM becomes the escape-rate-limiting step and sets the local ($z=0$) CR observables. We show rigidities that correspond to kinetic energies $>1$ GeV/nucleon, for which we find that escape dominates over ionisation losses, i.e. $\tau_{\rm ion} \gg h / {\rm v_A}$, consistent with our model assumptions.   } \label{fig:bc}
\end{figure}

For simplicity, we first consider a constant Alfv\'en speed throughout the self-confined coronal regions and inner halo, so that \eqref{eq:steady_master} is the adequate CR transport equation  (as we discuss below, relaxing this assumption does not change our conclusions qualitatively). Since the ``diffusive" correction due to linear damping mechanisms effectively acts like a sink of CRs, which is $\ll Q$ for most self-confined CRs and thus does not introduce energy-dependence of the right form (eq. \ref{eq:stead_state_lin_damp}; although we again point the reader to Section \ref{sec:strat_lin} for an interesting caveat), we here focus on nonlinear Landau damping. Including linear damping mechanisms in addition to nonlinear Landau damping does not significantly affect our conclusions as long as the cutoff rigidity imposed by linear damping $R_{\rm cutoff} \gg 100$GV (see Figure \ref{fig:bc}). In the hot and ionized halo, the relevant linear damping is set by the ambient Alfv\'enic turbulence (\citealt{farmer_goldreich}), so we essentially assume a weakly turbulent halo, so that NLLD is more important than linear damping (this requires fairly small amplitudes at the turbulence injection scale, $\delta {\rm V} / {\rm v_A} \sim \mathcal{O}(0.1)$).  Under these assumptions, per equations \ref{eq:adv_diff_turb} and \ref{eq:sc_kappa_nlld} the CR flux in the self-confined halo, $z>z_W$, satisfies,
\begin{equation} \label{eq:sc_flux_bc}
    {\rm v_{A}} f_p + X \Big( - \frac{\partial f_p}{\partial z}\Big)^{1/2}  = {\rm const} = - \kappa_{\rm turb} \Big[ \frac{\partial f_p}{\partial z} \Big]_{z \rightarrow z_W^-} = Q,
\end{equation} 
where we imposed flux continuity at $z=z_W$ and in the last step we used the solution in \eqref{eq:sol_turb} to evaluate the CR flux $-\kappa_{\rm turb} \partial f_p / \partial z$ in the warm ISM in which CRs are scattered by ambient fast-mode turbulence. In the self-confinement region, the CR distribution function therefore satisfies the same equation as for the single-phase medium in Section \ref{sec:nonlin_damp} (eq. \ref{eq:steady_nlld}). Thus, the steady-state CR proton distribution function for $z>z_W$ is in our model given by equation \ref{eq:steady_nlld_sol_haloC}.

In deriving eq. \ref{eq:steady_nlld_sol_haloC} we made the assumption that ${\rm v_A}$ is constant and one may wonder how much our results depend on it. For negligible NLLD and for an Alfv\'en speed that increases with $z$, the solution to \eqref{eq:adv_diff_turb} is $f_p \approx Q/{\rm v_A} \exp(-z / H_A)$, where $H_A$ is the $\sim$ Alfv\'en scale height. One can show that in this case NLLD becomes important when $X^2 \gtrsim Q {\rm v_A H_A}$. Previously, in \eqref{eq:steady_nlld_sol_haloC} we found that NLLD becomes important when $X^2 \gtrsim Q {\rm v_A r}$. At high energies (large $X$) our results are independent of ${\rm v_A}$ in both cases. So, a spatially-varying ${\rm v_A}$ merely changes the CR energy at which NLLD becomes important by order unity (since $H_A$ is expected to be quite large in the halo, $\gtrsim 5$kpc). Importantly, the exact spatial profile of ${\rm v_A}$  does not qualitatively change our results.

The solution in \eqref{eq:steady_nlld_sol_haloC} shows that the proton spectrum depends sensitively on the injection spectrum $Q$. The slope of the injection spectrum remains uncertain, which is a  limitation for putting tight constraints on CR transport using just the proton flux. In this work we will assume that the spectral slope at injection is around $\gamma_{\rm inj} \approx 4.2 \pm 0.1$, motivated by acceleration at strong shocks and FERMI data on luminous starbursts (\citealt{ackermann2012}), where gamma-ray emission likely traces the injection spectrum without any energy-dependent losses (e.g., \citealt{lacki_2011}).

\subsubsection{Self-confinement regions with NLLD: the boron-to-carbon ratio} \label{sec:bc}
Cleaner constraints on CR transport can usually be obtained using the boron-to-carbon ratio (B/C). While C nuclei are CR primaries, i.e. they are injected at the sources, B nuclei are secondaries created by spallation reactions of C nuclei in the ISM. The B/C ratio is therefore a direct probe of the average column density / grammage traversed by CRs during their lifetime in the galaxy. Importantly, for diffusion in turbulence B/C is independent of the injection spectrum. However, there is significant dependence on the injection spectrum if one considers self-confinement (see eq. \ref{eq:boron_steady} below).

In self-confinement regions ($z>z_W$), B nuclei are passively scattered by the Alfv\'en waves excited by the protons via the streaming instability. By imposing flux continuity at $z=z_W$ as in Section \ref{sec:protons} and equation \ref{eq:sc_flux_bc}, we find that B nuclei satisfy equation \ref{eq:steady_master}, with $\kappa = \kappa(f_p)$ according to \eqref{eq:sc_kappa_nlld},\footnote{For B nuclei with the same rigidity as protons, we have to multiply the proton diffusion coefficient by the speed ratio ${\rm v_B}(R) / {\rm v_p}(R)$ to get the boron diffusion coefficient at the same rigidity.} and $f_p$ given by \eqref{eq:steady_nlld_sol_haloC}. As the production mechanism for B is different, however, we need to replace the proton source term $Q$ with $f_C h \delta(z) / \tau_{\rm spall}$, where $f_C$ is the carbon distribution function, $h$ is the thickness of the dense thin disk in which boron is assumed to be produced, and $\tau_{\rm spall}$ is the spallation reaction timescale in the disk (we assume the cross section to be constant for relativistic CRs, $=61 {\rm mB}$). The differential equation is solvable, with 

\begin{equation}\label{eq:boron_steady}
    \frac{f_B(z,R)}{f_C(0, R)} = \frac{h}{\rm v_A \tau_{\rm spall}} \frac{(r - z + X^2 / (Q{\rm v_A}))^{\rm v_p / v_B} - (X^2 / (Q{\rm v_A}))^{\rm v_p / v_B} }{ (r - z + X^2 / (Q{\rm v_A}))^{\rm v_p / v_B}  },
\end{equation}
where ${\rm v_B}$ and ${\rm v_p}$ are the B and proton speeds at the same rigidity $R$ (${\rm v_p / v_B} \approx 1$ for $R \gg 1 \ {\rm GV}$), and $Q$ and $X$ are evaluated for proton momenta corresponding to rigidity $R$. We note that $f_C$ in the denominator is evaluated at $z=0$ due to the delta function that appears in the B source term. At low energies (small $X$), $f_B$ is set by Alfv\'enic escape, while at higher energies NLLD becomes important. At high energies with ${\rm v_p} \approx {\rm v_B}$ and $ r {\rm v_A} \ll X^2 / Q$, $f_B(z=0)/f_C(z=0) \propto Q/X^2 \propto p^{-\gamma_{\rm inj} + 3}$, and so B/C in self-confinement theory depends strongly on the injection spectrum.

\subsubsection{Diffusion in turbulence in the warm ISM} \label{sec:wism}
We now complete our solution for B/C and the proton spectrum by considering CR diffusion in the warm ISM layer ($z<z_W$). Because protons and B nuclei are both passively scattered by the turbulent cascade, they are described by the same equation. Assuming $\kappa_{\rm turb} = {\rm const}$, the solution to \eqref{eq:steady_master} is,
\begin{equation} \label{eq:sol_wism}
    f(z) = \frac{Q'}{\kappa_{\rm turb}} (z_W - z) + f(z_W), \qquad z < z_W,
\end{equation}
where $f$ can denote any CR species and $Q'$ is the source term for either protons or B nuclei. We imposed continuity of $f$ between the warm ISM and the coronal regions as the outer boundary condition, i.e. we use \eqref{eq:steady_nlld_sol_haloC} and \eqref{eq:boron_steady} for the protons and boron nuclei, respectively (evaluated at $z=z_W$).  Because $f(z_W)$ is set by the transport in the halo, eq. \ref{eq:sol_wism} highlights a key property of our multi-modal CR transport models: different transport mechanisms set observables at $z=0$ depending on CR energy and the ISM phase structure. In particular, the transport in the warm ISM (assumed diffusive here) sets observables in the disk if
\begin{equation} \label{eq:wism_condition}
    \frac{Q' z_W}{\kappa_{\rm turb}} \gtrsim f(z_W).
\end{equation}
In the opposite limit, it is the transport in the halo (assumed to be streaming here) that sets the local observables. We note that the above inequality is similar, though not equivalent, to the common assumption that low-energy CRs are scattered by self-excited waves, while higher-energy CRs are scattered by turbulence. In fact, in the low-$\beta$ warm ISM CRs of all energies are scattered by the fast-mode cascade (see Figure \ref{fig:kappa_turb}). For the model presented here, the more accurate interpretation is that \textit{the escape-rate-limiting step} is self-confinement for low-energy CRs and fast-mode turbulence for higher-energy CRs.

\subsubsection{Proton ionisation losses}
CRs are also subject to ionisation losses in the dense disk in which boron is produced (here we assume that the disk is mostly neutral, so that ionisation losses dominate over Coulomb losses). The effect of ionisation losses on the CR proton spectrum can be included analogously to equations \ref{eq:steady_lin_sol_loss} and \ref{eq:steady_nlld_sol_haloC_loss} by transforming the solution in \eqref{eq:sol_wism},
\begin{equation} \label{eq:multiphase_loss}
    f(0) \rightarrow f(0) \Big(1 + \frac{h}{\rm v_A \tau_{loss}} + \frac{h}{\rm \tau_{loss} } \frac{z_W}{\kappa_{\rm turb}}  \Big)^{-1}.
\end{equation}

\subsection{Two-phase model versus observations} \label{sec:observations}

\begin{figure}
  \centering
      \begin{minipage}[b]{\textwidth}
      \includegraphics[width=0.45\textwidth]{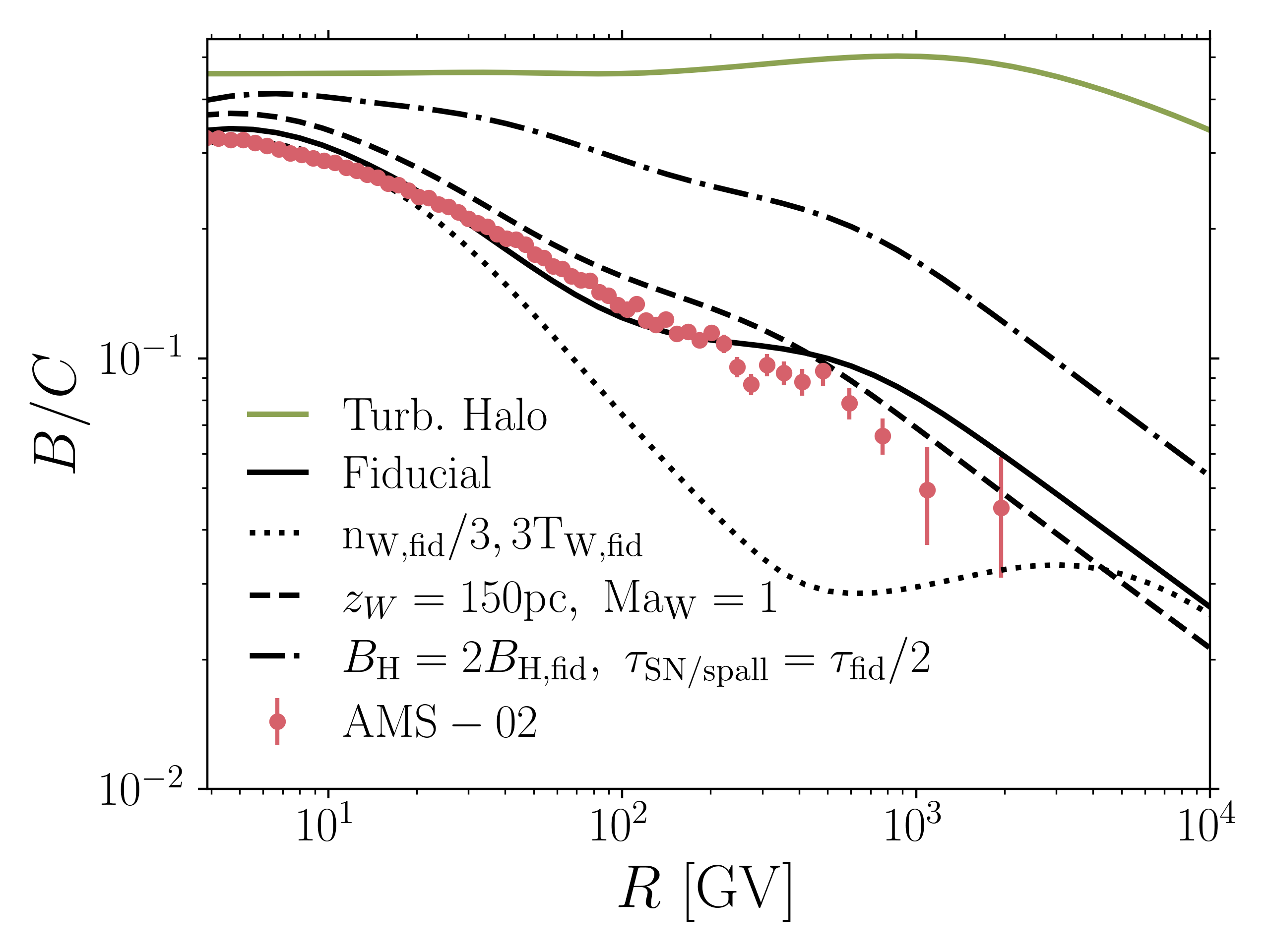}
    \end{minipage} 
      \begin{minipage}[b]{\textwidth}
      \includegraphics[width=0.45\textwidth]{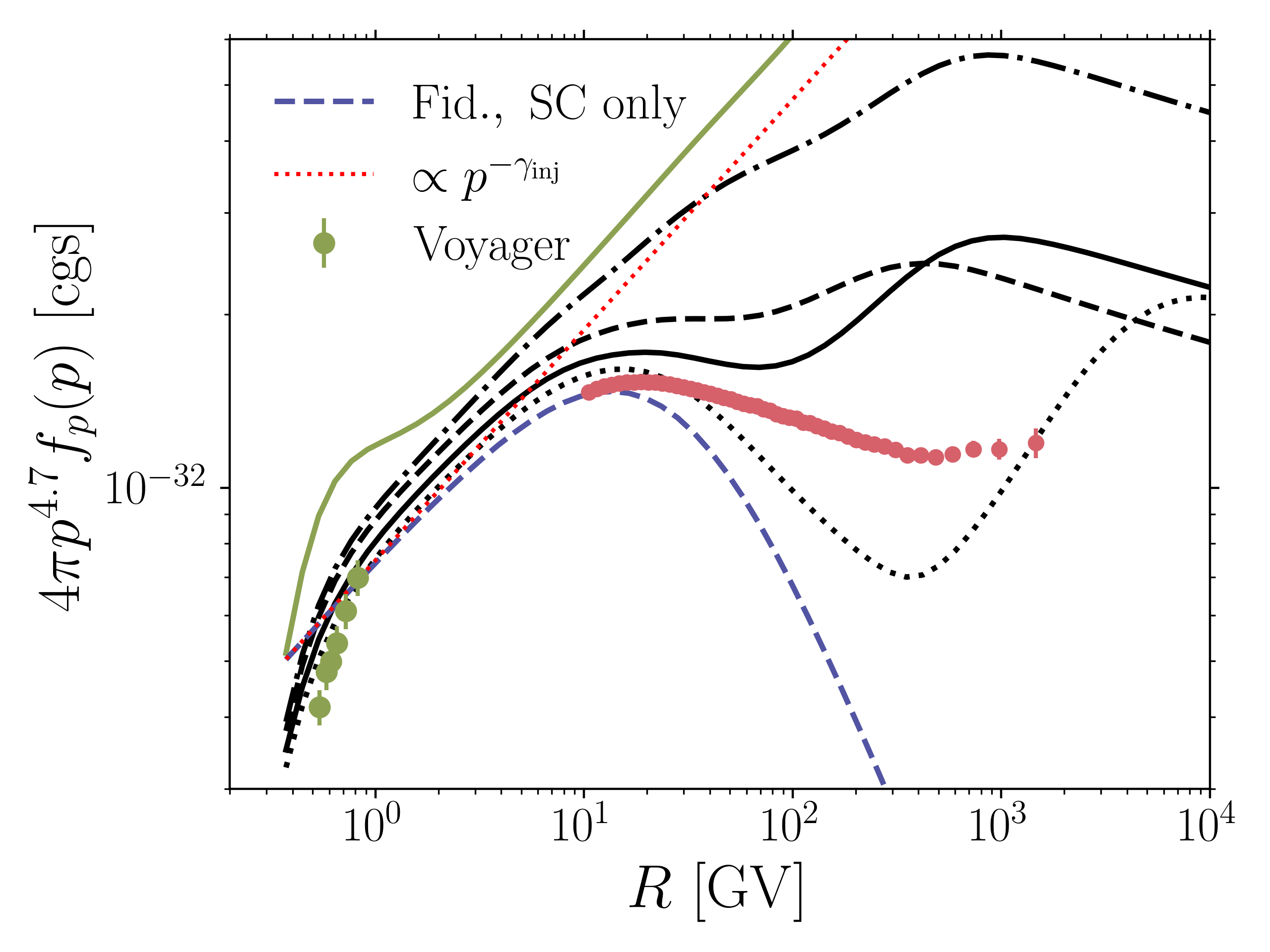}
    \end{minipage} 

  \caption{Dependence of spectra (top: B/C, bottom: protons) on our choice of parameters and propagation model. Red points are AMS-02 measurements, green points are Voyager measurements.  The black lines show the effect of variations in the parameters for our fiducial two-phase propagation model (Figure \ref{fig:galaxy_sketch}).  Modifying the parameters (black lines) changes the quality of the fit, but some qualitative trends remain the same. The proton spectrum is significantly steeper than the injection spectrum (red dotted line) for energies $\gtrsim 10$GeV. However, even small changes can introduce significant deviations from a power-law (e.g., the dotted line), which is incompatible with the observations. Due to the wide range of plasma conditions present in a realistic multi-phase galaxy, CR transport is likely some convolution of our model's results for different parameters. A multi-phase combination of scattering mechanisms can in principle reproduce rough trends in CR spectra, but this requires significant fine-tuning. The green line is a different model with scattering by fast-mode turbulence everywhere, i.e. no self-confinement; it shows qualitatively very different behaviour. Due to the weak energy dependence of $\kappa_{\rm turb}$ in the halo (Figure \ref{fig:kappa_turb}), B/C is essentially flat and the proton spectrum $\sim p^{-\gamma_{\rm inj}}$, inconsistent with observations. The blue dashed line in the bottom panel shows the spectrum calculated using the self-confined halo only (no contribution from the warm ISM). Without scattering by fast-mode turbulence, the spectrum at high energies is too steep unless the transport region is stratified as in Figure \ref{fig:strat_halo_fp}. 
  \label{fig:vars}}
\end{figure}

We now compare our multi-phase model to local CR measurements.  Finding a ``best-fit" set of parameters would be overkill given the simplicity of our model and the theoretical uncertainties. Instead, we show that there exists a plausible set of physical parameters for which we can recover the normalisation and essential trends of the proton and B/C spectra. However, our results also show that the spectra calculated using a combination of self-excited Alfv\'en waves and fast-mode turbulence are extremely sensitive to local plasma conditions. As a result, generating CR spectra with approximately constant power-law slope using a combination of self-excited waves and fast-mode turbulence requires significant fine-tuning, unlike phenomenological models of CR transport based on self-generated waves and undamped isotropic (typically Kolmogorov) turbulence.

We use the following fiducial parameters: a supernova rate of 1 per century, injecting $10^{50} \ {\rm ergs}$ of CRs per supernova with a spectrum $Q \propto p^{-4.3}$. In the hot ISM  we use $n_{\rm th} = 0.003 \ {\rm cm^{-3}}$, $T = 3 \times 10^{6} \ {\rm K}$ and $B=1 \mu$G. For the warm ISM, we use a density $n_{\rm th} = 0.2 \ {\rm cm^{-3}}$, temperature $T =  10^{4} \ {\rm K}$ and magnetic field $B=6 \ \mu$G. This corresponds to $\beta \approx 0.2 $ in the warm ISM and $\beta \approx 30$ in the hot ISM. We assume that the transition between the warm ISM and the coronal regions occurs at $z_W = 500$ pc above the disk and we use $r=10$ kpc as the characteristic CR injection length scale in the disk. We assume that fast-mode turbulence in the warm ISM is injected with ${\rm Ma } =0.6$ on scales of order $L=50$ pc. For the boron source term and ionisation loss term, we assume a thin disk of half-thickness $200$ pc and density $\approx 1.25 \ {\rm cm^{-3}}$. We stress that this is not a constrained, unique set of parameters. It is merely a set of physical parameters that matches the normalisation and main spectral trends in the observed data. We show some variation about these parameters in Figure \ref{fig:vars} discussed below.

The B/C spectrum for our fiducial  parameters is shown in Figure \ref{fig:bc}. The red points are AMS-02 measurements (\citealt{aguilar_bc}) and the black solid line is the spectrum obtained using \eqref{eq:boron_steady} and \eqref{eq:sol_wism} evaluated at $z=0$. The dotted and dashed lines show the contributions from the turbulent warm ISM and self-confined halo, respectively.  At rigidities $\lesssim 100$ GV, self-confinement in the diffuse hot ISM and halo sets the local B/C. At GeV energies, CRs stream at the Alfv\'en speed. At higher energies, but still in the self-confined limit, NLLD introduces a stronger dependence on energy. Above a few hundred GeV, diffusion in the turbulent warm ISM becomes the escape-rate-limiting step and sets the local CR observables. There is good agreement between our solution and the observations, especially given the simplicity of our model. As we do not include ionisation losses in the B/C calculation, we do not compare our model predictions to low-energy B/C data measured by Voyager (\citealt{cummings_2016}), which probe CRs in the energy-loss dominated regime. In Figure \ref{fig:bc} we instead only show rigidities that correspond to kinetic energies $>1$ GeV/nucleon, for which we find that escape dominates over ionisation losses, i.e. $\tau_{\rm ion} \gg h / {\rm v_A}$, consistent with our model assumptions (at these energies solar modulation also does not significantly affect B/C; e.g. \citealt{aloisio_2015}, \citealt{bresci_2019_reacc}).

While Figure \ref{fig:bc} suggests that there exist parameters for which there is good agreement between the model and the data, this is not generally the case. We consider variations in our parameters and the propagation model in Figure \ref{fig:vars}. We show results for B/C in the top panel, and for the proton spectrum in the bottom panel. We include proton ionisation losses using eq. \ref{eq:multiphase_loss} to enable comparison to both AMS-02 and low-energy Voyager measurements (\citealt{aguilar_2015}; \citealt{cummings_2016}). We again do not include energy losses in our calculation of the B/C spectra, as our simple implementation of the loss term in \eqref{eq:adv_diff_turb} is not appropriate for B particles (see Appendix \ref{app:bc_loss}). We therefore compare our calculated B/C only to AMS-02 data points with $E > 1$ GeV/nucleon.  The black lines all correspond to the fiducial propagation model (Figure \ref{fig:galaxy_sketch}) for different choices of parameters (for example, $n_{\rm W, fid}/3$ in the legend means that the density in the warm ISM is decreased by a factor of 3 relative to the fiducial parameter). Importantly, the proton spectrum is significantly steeper than the injection spectrum (red dotted line) for energies $\gtrsim 10$GeV. There is a break in the spectra between $10^2$ and $10^3$ GeV, which comes from the transition between $f_p(0)$ being set by self-confinement in the halo and $f_p(0)$ being set by turbulent diffusion in the warm ISM. The transition between self-confinement and extrinsic turbulence dominating the transport also gives rise to a spectral break in phenomenological models based on undamped Kolmogorov-like turbulence (e.g., \citealt{blasi12}; \citealt{aloisio_blasi_2013};  \citealt{aloisio_2015}). However, unlike these phenomenological models, modest changes in our model parameters can introduce significant deviations from a power-law (e.g., the dotted line), which is incompatible with the observations. The real ISM samples a wide range of plasma conditions and so CR transport likely is some convolution of our toy model's results for different parameters. It is unclear what the resulting CR spectrum would be in a more realistic multiphase model.

The green line in Figure \ref{fig:vars} shows the prediction for a different propagation model, in which CRs are scattered by fast-mode turbulence throughout the entire volume of the galaxy, including the halo and hot ISM (i.e. no self-confinement at all). We use the diffusion coefficients from Figure \ref{fig:kappa_turb} and a $5$kpc halo size to adjust the normalisation. The green line shows qualitatively very different behaviour from our multiphase model. Due to the weak energy dependence of $\kappa_{\rm turb}$ in the halo (Figure \ref{fig:kappa_turb}), B/C is essentially flat and the proton spectrum $\sim p^{-\gamma_{\rm inj}}$. The blue dashed line in the bottom panel of Figure \ref{fig:vars} is the proton spectrum calculated using the self-confined halo only (no warm ISM contribution; same as the dashed line in Figure \ref{fig:bc} for B/C). Without scattering by fast-mode turbulence, the spectrum at high energies is too steep (see discussion under eq. \ref{eq:steady_nlld_sol_haloC}). 

Figure \ref{fig:vars} thus shows that neither fast-mode turbulence nor self-confinement alone give reasonable agreement with the observations. The combination in principle can, but it requires a significant amount of fine-tuning of plasma conditions. This includes our rather uncertain assumption of a weakly turbulent halo, such that linear damping of self-excited Alfv\'en waves by ambient Alfv\'enic turbulence does not inhibit self-confinement at $R\sim 100$ GV. We briefly note that this fine-tuning issue is not a consequence of multi-phase CR propagation, but is also present if CRs are scattered by self-excited Alfv\'en waves and fast-mode turbulence in the same ISM phase, as we discuss in Appendix \ref{app:sc_et_same_phase}. It is also worth stressing that models that use self-confinement + damped fast-mode turbulence require significantly more  fine-tuning than phenomenological models based on self-confinement + undamped Kolmogorov turbulence. In these phenomenological models, CR injection rates and streaming speeds are constrained by observables at low energies where self-confinement dominates the transport, while  the turbulence strength is set by CR observables at high energies where self-confinement no longer operates. Combined, these two pieces of physics constrain CR transport. By contrast, the MHD fast-mode cascade is very sensitive to local plasma conditions, which strongly affect the damping rates. This is best illustrated by the solid and dotted black lines in Figure \ref{fig:vars}. The only parameters that are different between these two models are the warm ISM density and temperature. The turbulence strength at the outer scale is unchanged and yet the two models have very different predictions for the CR proton and B/C spectra. In addition to being theoretically unsatisfying, the fine-tuning required in Figures \ref{fig:bc} and \ref{fig:vars} may also face challenges explaining the relatively small spatial variations in CR spectra in the MW inferred from synchrotron and gamma-ray data (e.g., \citealt{miville_deschenes_2008}; \citealt{acero_2016}; \citealt{yang_2016}).

\section{Uncertainties in fast-mode turbulence} \label{sec:fast_mode_uncertainties}
We now discuss significant uncertainties in the physics of MHD fast-mode turbulence, which are usually not taken into account in the CR literature (e.g. \citealt{yan_lazarian_2004}; \citealt{yan_lazarian_2008}; \citealt{fornieri_2021}), but may strongly affect the CR diffusion coefficients from Figure \ref{fig:kappa_turb} and the predicted spectra in Section \ref{sec:sc+et}. These uncertainties appear to primarily \textit{increase} the discrepancy between phenomenological CR scattering models and weak fast-mode turbulence predictions, as we discuss below.

\subsection{Suppression of the weak cascade by wave steepening} \label{sec:weak_shocks}

The governing principles of the MHD fast-mode cascade are still up for debate.  It is uncertain what happens to the cascade below the viscous scale at low $\beta$, where anisotropic damping becomes important for non-parallel propagating modes.  Moreover, the calculation from \cite{yan_lazarian_2008}, used in Sections \ref{sec:turb} and \ref{sec:sc+et} and in Figures \ref{fig:kappa_turb}--\ref{fig:vars}, hinges on the assumption that MHD fast modes (or hydro sound waves) indeed follow a weak-turbulence cascade as argued by \cite{zakharov_sagdeev_1970}. However, the weak turbulence assumption is uncertain. \cite{kadomtsev1973acoustic} instead argued that for sufficiently small viscosities sound waves inevitably steepen to form (weak) shocks. Indeed, even for low-Ma turbulence the sound-wave steepening timescale is significantly shorter than the weak-turbulence nonlinear interaction (cascading) timescale,
\begin{equation} \label{eq:shock}
    \frac{\tau_{\rm steepen}^{-1}}{\tau_{\rm casc}^{-1}  } \sim \frac{k \delta \vrm}  {  k \delta \vrm^2 / \vrm_{\rm ph} } \sim \frac{\vrm_{\rm ph}}{\delta \vrm} \gg 1,
\end{equation}
where $\vrm_{\rm ph}$ is the phase speed of sound waves. In the \cite{kadomtsev1973acoustic} picture steepened fast modes produce weak shocks and then follow a $k^{-2}$ spectrum. This spectrum was indeed observed by \cite{kowal_lazarian_2010} in their compressible MHD simulations and, more recently, by \cite{makwana_yan_2020} in their sub-sonic turbulence simulations. 

After sound waves steepen and form weak shocks, they dissipate on a timescale of order the steepening timescale (\citealt{landau_lifshitz}),
\begin{equation}
    \tau_{\rm diss}^{-1} \sim k \delta {\rm v}.
\end{equation}
Both the steepening and weak-shock dissipation timescales are thus shorter than the weak-cascade timescale by a factor ${\rm \delta v / v_{ph}}$ (eq. \ref{eq:shock}). This factor is $\ll 1$ deep inside the inertial range  on small scales  resonant with $\lesssim$ PeV CRs. It is therefore likely that wave steepening suppresses the weak cascade and so the $k^{-3/2}$ spectrum.

CR scattering in a field of weak shocks is qualitatively different from standard pitch-angle diffusion in wave turbulence. In the latter case, CRs undergo frequent uncorrelated small changes in pitch angle through wave-particle interactions. By contrast, in a field of weak shocks CRs are likely scattered by single strong events at the shock discontinuities, if the width of the shock is $\ll$ the CR gyroradius. Thus, for CRs with sufficiently large gyroradii (energies), the scattering mean free path is constant and set by the separation of the shocks, which is likely of the same order of magnitude as the turbulence injection scale. The width of the shock is of the order of the ion mean free path in the  collisional case ($w \sim l_{\rm mfp} {\rm v}_{\rm th}/ \delta V $), and much smaller than the mean free path in the collisionless case ($\sim$ ion gyroradius). In the warm ionized medium the mean free path is of order $10^{13} \ {\rm cm}$, and so CRs with with energies $E\gtrsim 10$GeV can be scattered by this mechanism.  We note, however, that it is not fully clear how the steepening of sound waves in a low-collisionality plasma progresses once the spatial scales approach the ion mean free path.The perturbation might transition to a collisional shock or in some cases generate collisionless velocity space instabilities that alter the subsequent evolution from what would be predicted by fluid theory.  While this weak-shock-mediated CR transport could be tuned to have the right normalisation, the transport is energy independent, in disagreement with the empirical $\kappa \propto E^{0.3-0.7}$ (see blue dotted line in Figure \ref{fig:kappa_turb_anisotropic}). This is a problem especially at high energies where self-confinement certainly no longer operates. At lower energies, $\lesssim $TeV, there can in principle still be an energy dependence to CR transport due to scattering by a combination of self-excited waves and extrinsic weak shocks.

\subsection{Anisotropic fast-mode turbulence at low $\beta$} \label{sec:weak_casc_anisotropic}

 \begin{figure}
  \centering
        \begin{minipage}[b]{\textwidth}

    \includegraphics[width=0.47\textwidth]{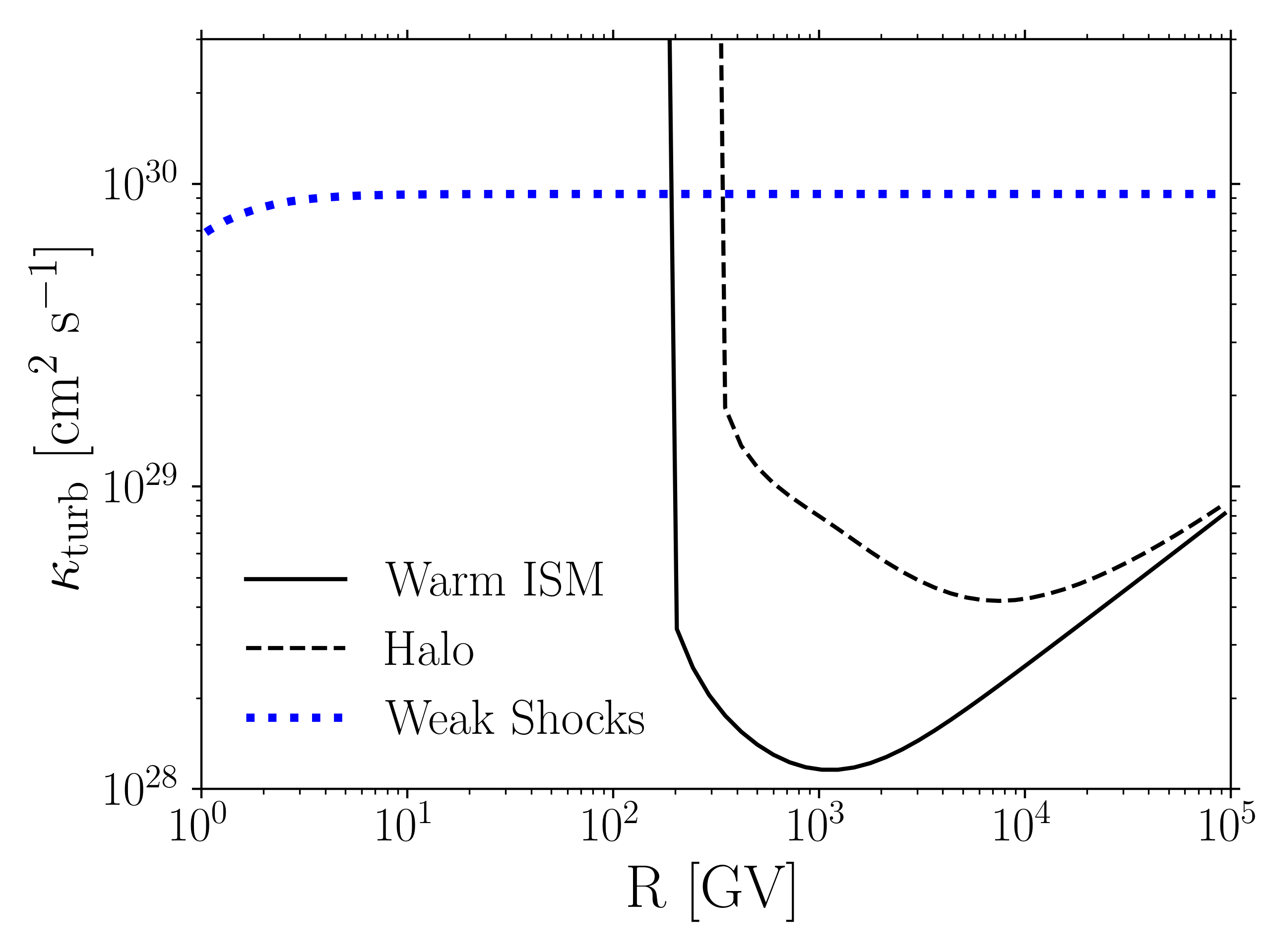}
        \end{minipage}

      \begin{minipage}[b]{\textwidth}
    \includegraphics[width=0.47\textwidth]{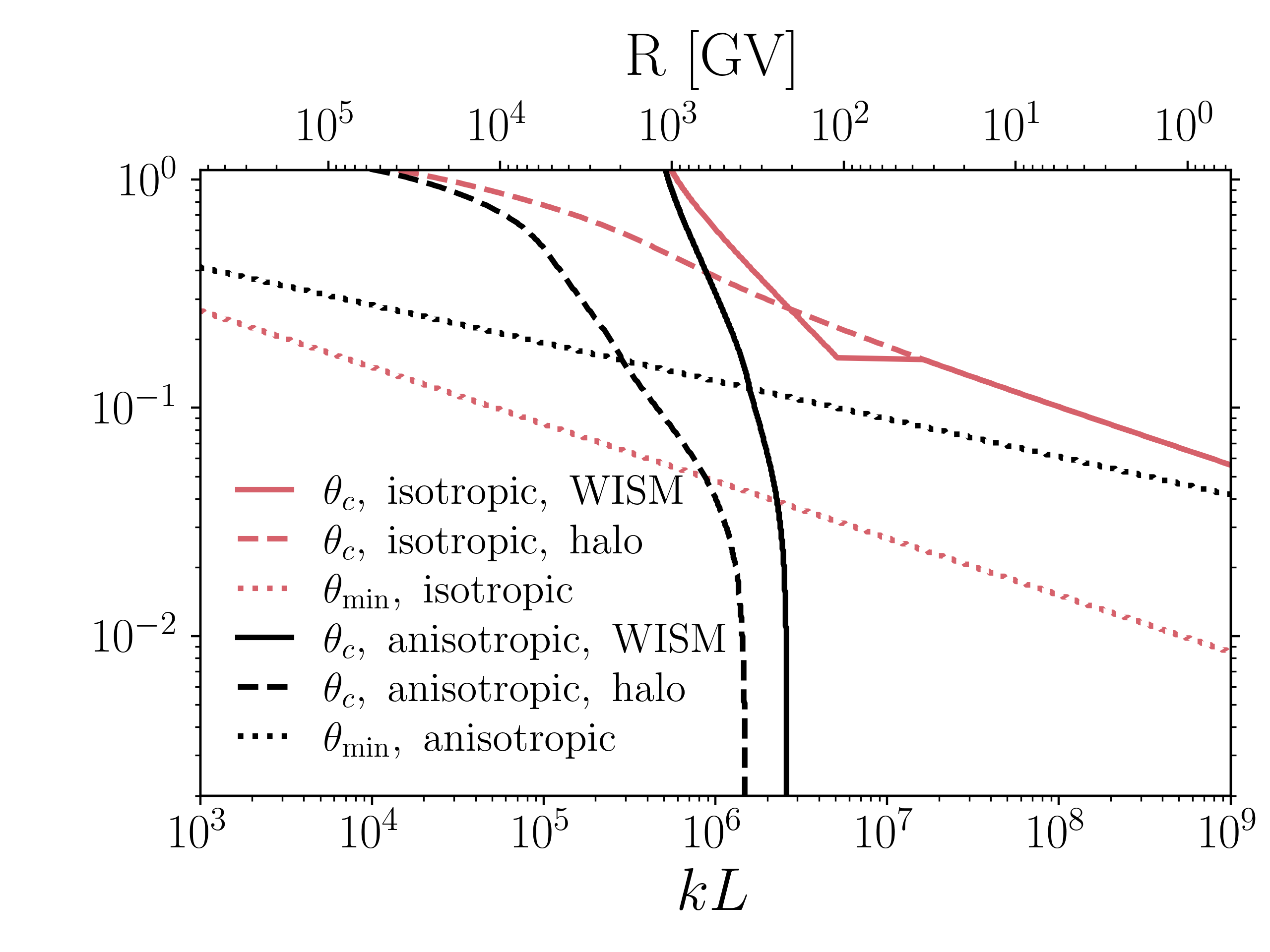}
    \end{minipage}
    
  \caption{Uncertainties in MHD fast-mode turbulence significantly affect CR transport. Top: the blue dotted line shows a roughly energy-independent CR diffusion coefficient if wave steepening suppresses the weak cascade of fast modes (Section \ref{sec:weak_shocks}). We assume $\mathcal{O}(1)$ shocks  separated by $\sim 10$ pc. The black lines are CR diffusion coefficients calculated  using the low-$\beta$ anisotropic scalings for a weak fast-mode cascade from \eqref{eq:tau_fast_anisotropic}--\eqref{eq:dtheta_alf} for the same warm ISM and halo plasma conditions as in Figure \ref{fig:kappa_turb}. Due to the slow cascade rate of quasi-parallel modes (eq. \ref{eq:tau_fast_anisotropic}) and field-line wandering (eq. \ref{eq:dtheta_alf}), CR scattering is completely suppressed for rigidities $\lesssim 100$ GV. Bottom: the critical mode pitch angle $\theta_c$ as a function of spatial scale. Fast modes with $\theta > \theta_c(k)$ are fully damped.  The red solid and dashed lines show $\theta_c$ for the isotropic fast-mode scalings in Section \ref{sec:turb} and Figure \ref{fig:kappa_turb}. The black solid and dashed lines are for the anisotropic fast-mode scalings from equations \ref{eq:tau_fast_anisotropic}--\ref{eq:dtheta_alf} (and correspond to the black solid and dashed lines in the top panel). In the anisotropic case $\theta_c$ shrinks significantly faster with increasing $k$. On scales $kL \gtrsim 10^6$ (CR rigidities $\lesssim$ few hundred GV) $\theta_c$ is smaller than $\theta_{\rm min}$, where $\theta_{\rm min}$ is the minimum well-defined pitch angle a fast mode can have over one cascade timescale due to field-line wandering. As a result, the cascade is fully damped for $kL \gtrsim 10^6$.  \label{fig:kappa_turb_anisotropic}}
\end{figure}

\cite{yan_lazarian_2004} and \cite{yan_lazarian_2008} assume that in the absence of damping the fast-mode power spectrum is isotropic, $P \propto k^{-3/2}$, which is largely based on the numerical results in  \cite{cho_lazarian}. While fast-modes are very likely isotropic at high $\beta$, where they behave essentially like hydro sound waves, this is probably not true at low $\beta$ (which is the most important regime for CR scattering). In particular, at low-$\beta$ quasi-parallel fast modes are significantly less compressive than oblique fast modes. At low $\beta$, $\delta \vrm_\parallel / \delta \vrm_\perp \sim \beta \sin \theta \cos \theta$ (where $\parallel$ and $\perp$ denote the velocity-fluctuation components parallel and perpendicular to $\bm{B}$, and $\theta$ is the angle between $\bm{k}$ and $\bm{B}$). This means that for quasi-parallel modes $\delta \vrm_\parallel / \delta \vrm_\perp \sim \beta \sin \theta \ll 1$ and $\bm{k \cdot \delta \vrm} \sim k \delta \vrm  \sin \theta  \ll k \delta \vrm$. Quasi-parallel fast modes therefore interact significantly more weakly than quasi-perpendicular modes. This plausibly generates anisotropy in the cascade. In particular, the cascade timescale becomes angle-dependent,
\begin{equation} \label{eq:tau_fast_anisotropic}
    \tau_{\rm casc}^{-1} \sim k \frac{\delta \vrm ^2}{\vrm_{\rm ph}} \sin^2\theta,
\end{equation}
(cf. equation \ref{eq:tau_fast_isotropic}). If turbulence is forced isotropically at the outer scale (the energy transfer rate $\epsilon$ is isotropic), then
\begin{equation} \label{eq:v_fast_anisotropic}
    \frac{\delta \vrm}{\vrm_{\rm ph}} \sim {\rm Ma} (kL)^{-1/4} (\sin \theta)^{-1/2},
\end{equation}
where ${\rm Ma}$ is defined as $\delta \vrm / \vrm_{\rm ph}$ at the injection scale for $\sin \theta \sim 1$. This corresponds to a power spectrum,
\begin{equation} \label{eq:power_fast_anisotropic}
    P \propto k^{-3/2} / \sin \theta
\end{equation}
(see also \citealt{chandran_2005} for a more rigorous weak-turbulence-theory derivation of this result). The power does not diverge as $\theta \rightarrow 0$ due to the effect of field-line random walk. Before it cascades, a fast mode with wavenumber $k$ travels a distance ${\rm v_{ph}} \tau_{\rm casc}(k, \theta) $. A quasi-parallel mode which cascades at a rate given by \eqref{eq:tau_fast_anisotropic} experiences field-line wandering due to ${\rm Ma_{Alf}} \sim 1$ Alfv\'enic turbulence of order,
\begin{equation} \label{eq:dtheta_alf}
    \frac{\delta B}{B} \sim \Big({\rm Ma}^2 (kL)^{1/2} \sin \theta \Big)^{-1/2}
\end{equation}
(cf. equation \ref{eq:dtheta_alf_iso}). A mode has a well-defined pitch-angle  over a cascade time only if $\theta \gtrsim \delta \theta \sim  \delta B /B$.
As in Section \ref{sec:turb}, when we evaluate the cascade (eq. \ref{eq:tau_fast_anisotropic}) and damping (eq. \ref{eq:Gamma_brag} and \ref{eq:Gamma_ginz}) rates, we take the average of the cascade/damping rate in the interval $(\theta- \delta \theta, \theta + \delta \theta)$. According to \eqref{eq:dtheta_alf}, $\delta \theta$ diverges as $\theta \rightarrow 0$. This is, however, unphysical given the presence of finite field-line wandering. We remedy this by imposing a lower limit on $\theta$ in \eqref{eq:dtheta_alf}, $\theta_{\rm min}(k)$, defined such that $\delta \theta(k, \theta_{\rm min}) = \theta_{\rm min}$. $\theta_{\rm min}$ is therefore the smallest well-defined pitch-angle a fast mode can have. For an isotropic fast-mode cascade, $\theta_{\rm min}$ is simply equal to the pitch-angle spread due to field-line wandering in \eqref{eq:dtheta_alf_iso}, $\theta_{\rm min} \sim \delta B /B$.

In the top panel of Figure \ref{fig:kappa_turb_anisotropic}, the black lines show CR diffusion coefficients calculated using the anisotropic scalings of the fast-mode cascade from \eqref{eq:tau_fast_anisotropic}--\eqref{eq:dtheta_alf} for the same warm ISM and halo plasma conditions as in Figure \ref{fig:kappa_turb}. CR scattering is completely suppressed for rigidities $\lesssim 100$ GV. The suppression of CR scattering at small energies relative to Figure \ref{fig:kappa_turb} is due to the slower cascade rate of quasi-parallel modes (eq. \ref{eq:tau_fast_anisotropic}) and increased field-line wandering experienced by quasi-parallel modes during one cascade time (eq. \ref{eq:dtheta_alf}). The solid and dashed lines in the bottom panel of Figure \ref{fig:kappa_turb_anisotropic} show how the critical mode pitch angle $\theta_c$ depends on spatial scale. Fast modes with $\theta > \theta_c(k)$ are fully damped. The solid and dashed lines therefore show how the cone of undamped fast modes shrinks with spatial scale.  The red solid and dashed lines show $\theta_c$ for the isotropic fast-mode scalings in Section \ref{sec:turb} and Figure \ref{fig:kappa_turb}. On collisionless scales $\theta_c \propto k^{-1/4}$ (see Section \ref{sec:turb}), and because $\theta_{\rm min}$ (red dotted line) has the same scaling with $k$ (eq. \ref{eq:dtheta_alf_iso}), the fast-mode cascade is not fully damped. This is in contrast to the results for the anisotropic fast-mode scalings from equations \ref{eq:tau_fast_anisotropic}--\ref{eq:dtheta_alf}. In this case $\theta_c$ (black solid and dashed lines) shrinks significantly faster with increasing $k$ than $\theta_{\rm min}$ (black dotted line). On scales $kL \gtrsim 10^6$ (corresponding to CR energies of a few hundred GeV) $\theta_c$ is smaller than $\theta_{\rm min}$ and the cascade is fully damped. 

We also note that the $\theta_{\rm min}$ lines in Figure \ref{fig:kappa_turb_anisotropic} are assumed to be due to  Alfv\'enic turbulence. There is, however, also the additional field-line wandering generated by the cascade of fast modes. This may truncate the fast-mode cascade on even larger scales. The field-line wandering experienced by a quasi-parallel fast-mode during one cascade time due to undamped fast-mode turbulence is roughly,
\begin{equation} \label{eq:dtheta_fast}
    \frac{\delta B}{B} \sim \Big(\frac{{\rm Ma}^2}{(kL)^{1/2} \sin^3 \theta} \Big)^{1/4}
\end{equation} 
(for an isotropic fast-mode cascade, the $\sin^{-3} \theta$ factor should be dropped). Due to the weak dependence on $k$, $\propto (kL)^{-1/8}$, weak fast-mode turbulence can be the dominant source of field-line wandering experienced by high-$k$ modes. The weaker dependence on $k$ relative to eq. \ref{eq:dtheta_alf} is due to the fact that the weak-turbulence fast-mode spectrum ($\propto k^{-3/2}$) is shallower than the Alfv\'enic-turbulence spectrum in the parallel direction ($\propto k_\parallel^{-2}$). Importantly, the field-line wandering experienced by high-$k$ modes is due to longer-wavelength modes on scales ${\rm v_{ph} \tau_{\rm casc}}(k) \gg k^{-1}$ which may not be affected by damping. As a result, assuming an undamped cascade in \eqref{eq:dtheta_fast} turns out to be a reasonable assumption in e.g. the warm ISM.

As we noted in Section \ref{sec:weak_shocks}, it is unclear whether weak-turbulence theory is at all applicable to the MHD fast-mode cascade given that wave steepening and weak-shock dissipation occur on a shorter timescale than the weak-turbulence cascade.\footnote{This is also true if fast-mode anisotropies at low $\beta$ are taken into account. For $\beta \ll 1$, fast modes steepen at a rate $\tau_{\rm steepen}^{-1} \sim k \delta {\rm v} \sin \theta$. At low $\beta$ quasi-parallel fast modes therefore steepen at rate that is slower than at high $\beta$, but the steepening rate is still faster than the weak-turbulence cascade rate (eq. \ref{eq:tau_fast_anisotropic}) by a factor ${\rm v_{ph} / (\delta v }\sin \theta)$.} In this section we showed that if theoretically motivated cascade anisotropies at $\beta \lesssim 1$ are taken into account, then even within the weak-turbulence theory framework  CR scattering by fast modes is completely suppressed at energies $\lesssim $100 GeV. The anisotropy of weak fast-mode turbulence at low $\beta$ has not yet been observed in simulations (e.g. \citealt{cho_lazarian}). However, this is not very surprising given that the weak cascade itself remains elusive. Moreover, close to the injection scale field-line-wandering effects are significant and act to suppress the anisotropy of the cascade.

\section{Discussion} \label{sec:discussion}
The good agreement between the multi-phase model and observations in Section \ref{sec:observations} may be viewed as a success. After all, there appears to be \textit{a} set of simplified yet plausible conditions where existing theories of CR propagation meet observations. In that sense, our results can be interpreted as a proof of concept for existing theoretical models of CR transport. However, the success of a theory depends not only on its ability to match the data, but also on its robustness. For this reason, the good fit in Section \ref{sec:observations} should not overshadow the uncertain micro-physical building blocks of our model (e.g., the significant uncertainties in the MHD fast-mode cascade discussed in Sections \ref{sec:weak_shocks} and \ref{sec:weak_casc_anisotropic}). The fine-tuning of parameters that is necessary to create smooth spectra with the right energy dependence may also, understandably, leave a bittersweet taste.  Especially since the fine-tuning carries a message of non-universality: in this model other Milky-Way-like galaxies, even ones without calorimetric CR losses, may have very different CR spectra. And one might expect rather different spectra in different regions of the Milky Way, which is naively not consistent with synchrotron and gamma-ray data (e.g., \citealt{miville_deschenes_2008}; \citealt{acero_2016}; \citealt{yang_2016}).

In this section we discuss some speculative extensions of self-confinement and MHD turbulence theory that might improve agreement between CR scattering theory and phenomenological models. We begin by noting that alternative nonlinear damping mechanisms of self-excited waves would modify the energy dependence predicted by self-confinement theory, potentially yielding better agreement with observations (Section \ref{sec:nl_sc}). A more theoretically attractive alternative is scattering of CRs by a single mechanism, which would be devoid of the fine-tuning problem discussed above, e.g., scattering by a turbulent cascade with a universal spectral index. This is in part why the phenomenological Kolomogorov-turbulence model has been so successful in explaining CR observations. However, the Kolmogorov phenomenology is not theoretically well motivated and at present no MHD cascade is known to have the desired properties over the entire range of CR energies. Still, MHD turbulence remains sufficiently uncertain to not discard this option. In Sections \ref{sec:echo}--\ref{sec:imb} we discuss a few topics at the frontier of MHD turbulence theory that may be relevant for a universal, turbulence-based theory of CR transport. In Section \ref{sec:ext_drive} we also briefly consider the possibility that the waves that scatter CRs are \textit{excited} on a wide range of scales by an external driving process (e.g., stellar feedback), instead of only being generated via a turbulent cascade from large scales.  

\subsection{Nonlinearities in self-confinement theory} \label{sec:nl_sc}
We showed in Section \ref{sec:nonlin_damp} that nonlinear damping of self-excited Alfv\'en waves gives rise to a smooth transition from Alfv\'enic streaming to energy-dependent transport with constant spectral index (in contrast to linear damping). However, the spectral index for nonlinear Landau damping is too steep relative to observations (Section \ref{sec:nonlin_damp}). This is not necessarily true for nonlinear damping mechanisms that are different from $\Gamma_{\rm NLLD} \sim k {\rm v_{th}} (\delta B/ B)^2$. For example, Alfv\'en waves on scales larger than the thermal-particle mean free path are nonlinearly damped by pressure-anisotropy effects at a rate $\Gamma \sim (k l_{\rm mfp}) \Gamma_{\rm NLLD}$ (\citealt{squire_et_al_2017}). With this extra $k$ dependence, nonlinear damping yields energy-dependent transport very close to the observationally inferred $E^{0.3-0.7}$ scaling. However, this particular physical mechanism is only valid above the thermal-particle mean free path, which is very large ($\gtrsim$pc) in the dilute and hot plasmas in which these nonlinear damping mechanisms are important. This particular damping is thus not applicable for CRs below PeV energies. Nevertheless, this example illustrates that other nonlinear damping mechanisms can in principle give rise to appropriate energy dependence. The nonlinear damping would need to be strong, and linear damping negligible, so as to not introduce a sharp cutoff as in Figure \ref{fig:lin_damping}.

\subsection{External driving on a range of scales} \label{sec:ext_drive}
It is also possible that the waves that scatter CRs over a wide range of energies are directly excited by some external driving process. This could be due to fluid instabilities acting on a range of scales, e.g. Kelvin-Helmholtz or dust-driven instabilities (\citealt{squire_hopkins_2018_rdi}). Alternatively, it is possible that stellar feedback (winds, supernovae, etc.) drives not only large-scale motions (which then cascade down to smaller scales), but also fluctuations on a wide range of smaller scales. Driving on smaller scales leads to less anisotropy in the Alfv\'enic cascade and thus more efficient CR scattering. For example, the periods of the Alfv\'en waves resonant with GeV CRs are comparable to the variability timescale of solar-type stars, and so it is in principle possible that stellar winds interacting with the magnetized ISM can contribute to some scattering of lower-energy CRs. The power required to excite waves with amplitudes consistent with phenomenological CR transport models is $\ll$ the total power from stellar feedback. In particular, the power required to excite $\delta B/B\sim 10^{-3}$ waves (rough amplitude suggested by phenomenological models) in a cylindrical galaxy with radius $R=10$kpc and height $H=1$kpc is $P_{\rm scatt} \sim 4 \times 10^{36} \ {\rm ergs /s}$, assuming $B\sim 1 \mu$G and a plausible wave damping rate $\Gamma \sim 10^{-11} {\rm s^{-1}}$. This is indeed orders of magnitude smaller than the total power injected by supernovae, $P_{\rm SN} \sim 3 \times 10^{41} \ {\rm ergs /s}$ for a supernova rate of 1 per century. However, this speculation faces a severe fine-tuning problem, as it is unclear how waves excited on a range of scales would conspire to have a spectrum that resembles an undamped turbulent cascade.

\vspace{-10pt}

\subsection{Cascade damping: role of the plasma echo} \label{sec:echo}
As shown in Figure \ref{fig:kappa_turb} and Figure \ref{fig:kappa_turb_anisotropic}, damping prevents a weak-turbulence cascade of fast modes from maintaining a universal $k^{-3/2}$ power-law at high $k$ (or $\kappa_{\rm turb} \propto E^{1/2}$ at scales resonant with low-energy CRs). The damping sets in at particularly large scales in dilute and hot plasmas like the hot ISM or the galactic halo, which constitute most of the confining volume. In these dilute systems, linear compressive waves are Landau-damped, which transfers energy from spatial fluctuations to ever-finer structures in velocity space in the thermal-particle distribution function,  a process called phase-mixing.  

Our calculation (based on the calculation in \citealt{yan_lazarian_2008}) assumes that compressive fluctuations in a turbulent cascade are damped at their linear damping rates. However, recent theoretical and numerical progress suggests that in the presence of nonlinear interactions and sufficiently small particle collisionalities (so that phase mixing can be reversible), fluctuations that have phase-mixed away to fine velocity scales can be brought back by coupling to other perturbations, a process called  the ``stochastic plasma echo", which effectively reverses phase-mixing (\citealt{scheko_2016_echo}; \citealt{meyrand_2019}). Landau damping is suppressed and the collisionless plasma behaves more fluid-like. The fluidization of plasma turbulence appears consistent with the seemingly undamped cascade of compressive fluctuations in the solar wind (in this case, slow modes; \citealt{chen_2016}; \citealt{verscharen_2017}). An analogous fluidization of fast-mode turbulence would allow $\kappa_{\rm turb}$ in the halo (Figure \ref{fig:kappa_turb}) to continue its $R^{0.5}$ scaling to lower CR energies. 

One issue with this idea is that the plasma echo can anti-phase-mix fluctuations before they are damped by collisions only if the nonlinear timescale is sufficiently short. This in turn requires large amplitudes for the turbulent fluctuations, which then leads to small CR diffusion coefficients if the desired $E^{0.5}$ scaling continues to small energies.  For example, using the results by \cite{adkins_2018}, we estimate that for the stochastic echo to operate in the hot ISM, amplitudes of order ${\rm Ma} \gtrsim$1 on the outer scale appear necessary, leading to diffusion coefficients of order $10^{27} \ {\rm cm^2 \ s^{-1}}$ at a GeV, far too small compared to observationally inferred CR diffusion coefficients. However, the theory of the stochastic echo in a fast-mode cascade has not been developed and clearly merits more work. Moreover, if the conditions to realise the stochastic echo only occur in a small fraction of the ISM volume, this would reduce the effective volume-averaged diffusion coefficient. 

\subsection{Reconnection and the anisotropy of Alfv\'enic turbulence} \label{sec:rec}
An alternative to scattering by fast-mode turbulence is that there is a different, yet unknown, regime of Alfv\'enic turbulence that can efficiently scatter CRs. Though critical balance as the main governing principle of MHD turbulence stands on firm ground, it is possible that additional physics on small scales can change the cascade physics. For example, the large aspect ratios of structures formed in strong turbulence on small scales plausibly become unstable to instabilities such as tearing and/or Kelvin-Helmholtz. In particular, Alfv\'enic turbulence may enter a new regime of tearing-mediated turbulence on scales much larger than one would naively expect, i.e. $\gg$ the resistive scale (\citealt{loureiro_boldyrev_2016}; \citealt{mallet_2017_collisionless}; \citealt{mallet_2017_tearing}). There is some numerical evidence/theory that turbulence in reconnecting layers may obey different anisotropy scalings than standard Alfv\'enic turbulence, e.g. $k_\parallel \sim k_\perp$ (\citealt{huang_bhattacharjee_2016}; \citealt{boldyrev_loureiro_2019}), or even $k_\parallel \sim k_\perp^{1.2}$ (\citealt{yang_2020}). This has the potential for more efficient CR scattering and the latter scaling could even give rise to an energy dependence of CR scattering that is roughly in the right ballpark. However, applying the theory of \cite{mallet_2017_colless_rec}, we estimate that tearing-mediated turbulence is expected to become important only on scales $\ll$ the gyroradius of GeV CRs. If this is generally true, it would not be important for CR scattering. 

\subsection{Balanced versus imbalanced turbulence}\label{sec:imb}
Numerical studies of MHD turbulence usually assume approximate equipartition between kinetic and magnetic energies, and roughly equal energies in forward and backward propagating Alfv\'en waves. While this setup is numerically convenient, it is generally not realised  in nature. Many of the cascade properties, including the spectral slopes and anisotropy, of imbalanced Alfv\'enic turbulence are quite different from the usual balanced case (e.g., \citealt{beresnyak_lazarian_2009} or Section 7 in the review by \citealt{schekochihin_mhd} for a summary). However, it is believed that the parallel spectrum remains steep, with a spectral slope $\lesssim -2$ and $k_\perp \gg k_\parallel$. This still means inefficent CR scattering. Nevertheless, imbalanced MHD turbulence remains a fairly unexplored terrain compared to its balanced counterpart, so there is still room for surprises there.

\section{Summary} \label{sec:summary}

Neither confinement of CRs by self-excited Alfv\'en waves nor scattering by an ambient fast-mode cascade can plausibly explain CR observations over the entire range of CR energies (e.g., \citealt{kc71}; \citealt{farmer_goldreich}; \citealt{fornieri_2021}). We provide a pedagogical review of this important result in Sections \ref{sec:sc} and \ref{sec:turb}. Due to the strong damping of fast modes on small scales, the energy dependence of the CR diffusion coefficient in fast-mode turbulence is incompatible with observations at CR energies  $\lesssim 10^3$ GeV (Figure \ref{fig:kappa_turb} and Figure \ref{fig:kappa_turb_anisotropic}). The discrepancy is significantly larger than in phenomenological models which assume CR scattering by an undamped isotropic turbulent cascade (e.g., \citealt{trotta_2011}; \citealt{gaggero_2014}; \citealt{hopkins_cr_pheno}). In self-confinement theory, higher-energy CRs are unable to self-confine due to the linear damping of Alfv\'en waves excited by the streaming instability.  Indeed, for self-excited waves damped by linear damping mechanisms, self-confined CRs stream, to reasonable approximation, at either the Alfv\'en speed or the speed of light (Figure \ref{fig:lin_damping}), which is naively incompatible with the smooth energy dependence measured empirically.  Non-linear damping of self-excited waves (e.g., non-linear Landau damping) introduces a smoother energy-dependence to CR transport but existing theoretical models predict an energy dependence that is stronger than that observed (see Section \ref{sec:nonlin_damp} and the dashed line in Figure \ref{fig:bc}).  

CR transport is theoretically predicted to depend sensitively on spatial variations in the plasma properties because local plasma conditions change the strength of ambient turbulence, the efficiency of fast-mode damping and the damping of self-excited Alfv\'en waves. This strong dependence of CR transport on local plasma conditions motivates our considering how such variations might affect CR observables at Earth. To start, we have shown that even pure Alfv\'enic streaming can in fact produce  energy dependent transport and energy dependent observables if both the Alfv\'en speed and the region in which CRs of a given energy are self-confined vary with height above the galactic disk (see also \citealt{holmes_1974}). These two conditions mean that CRs of different energy effectively sample different Alfv\'en speeds, leading to energy dependent transport that can in principle be similar to that observed (Section \ref{sec:strat_lin} and Figure \ref{fig:strat_halo_fp}). Even in this model, however, self-excited Alfv\'en waves are eventually fully damped at sufficiently high energies and a separate CR scattering mechanism is required.

The fact that neither self-confinement theory nor scattering by weak fast-mode turbulence can, on their own, explain CR spectra in the Milky Way suggests two possibilities. There is either a yet unidentified process that efficiently scatters CRs with the right energy dependence, or, a combination of scattering by self-excited waves and fast-mode turbulence conspires to mimic the empirically derived CR diffusion coefficient. In Section \ref{sec:sc+et}  we considered the latter possibility.  The multi-phase nature of the ISM of galaxies independently suggests that CR transport may be multi-modal, i.e. a mix of streaming and diffusion in turbulence. Self-confinement likely operates in regions where the turbulent cascade of scattering waves (e.g., fast modes) is strongly damped, whereas in regions with efficient scattering by turbulence the streaming instability is suppressed. In Section \ref{sec:two_phase_mw}, we have considered a particular model in which CRs diffuse in the warm ISM due to scattering by the MHD fast-mode cascade, and are self-confined in the coronal regions and the halo (Figure \ref{fig:galaxy_sketch}). We have assumed that turbulence in the halo is weak, so that nonlinear Landau damping is the dominant damping of Alfv\'en waves excited by the streaming instability. We stress that this particular model belongs to a broader class of possible multi-phase CR propagation models. In particular, the exact filling fractions of different phases is likely a function of a number of properties of the galaxy, such as the local gas surface density and star formation rate surface density, which set the cooling rates and supernovae heating rate.  This will in turn modify CR propagation. The purpose of the model considered in this work is to demonstrate that a relatively simple model of multi-phase CR transport can in principle explain some of the main trends observed in CR spectra in the solar neighbourhood (Figure \ref{fig:bc} and Figure \ref{fig:vars}), but that significant fine-tuning of plasma parameters is needed. These trends are difficult to explain using only self-confinement theory (Section \ref{sec:sc} and, e.g., dashed line in Figure \ref{fig:bc}) or only scattering by MHD turbulence (Section \ref{sec:turb}, Figure \ref{fig:kappa_turb} and the green line in Figure \ref{fig:vars}).

The calculation in Section \ref{sec:sc+et} is the first attempt to combine microphysical theories of CR self-confinement and scattering by MHD fast-mode turbulence. While there already exists literature that tries to combine streaming and turbulence to explain CR spectra measured in the MW (e.g. \citealt{blasi12}; \citealt{aloisio_blasi_2013}; \citealt{aloisio_2015}), these models assume undamped Kolmogorov-like ($\sim$ isotropic Alfv\'enic) turbulence. In this work, we instead considered the theoretically better motivated interplay of the streaming instability and fast-mode turbulence. Because fast modes are damped on scales $\sim$ the Larmor radius of $\lesssim$ TeV particles, their impact on CR transport and their interaction with self-excited Alfv\'en waves are very different from results based on an undamped Kolmogorov cascade (see also Appendix \ref{app:sc_et_same_phase}). In particular, Section \ref{sec:sc+et} highlights the important issue that, due to fast-mode damping, CR propagation models based on a combination of self-excited Alfv\'en waves and MHD fast-mode turbulence require a significant amount of fine-tuning of plasma parameters to recover the almost-pure-power-law CR spectra measured in the Milky Way. This issue is not captured in phenomenological models based on undamped isotropic Kolmogorov-like turbulence.

Models of CR scattering by MHD fast-mode turbulence in the literature also rely on the  uncertain assumption that fast modes follow a weak isotropic cascade in the absence of wave damping (e.g. \citealt{yan_lazarian_2004}; \citealt{yan_lazarian_2008}; \citealt{xu_lazarian_ttd}; \citealt{fornieri_2021}). As we discussed in Section \ref{sec:fast_mode_uncertainties}, weak fast-mode turbulence is theoretically expected to be anisotropic at low $\beta$ (Section \ref{sec:weak_casc_anisotropic}), and is probably completely suppressed by wave steepening (Section \ref{sec:weak_shocks}). This raises the significant possibility that fast modes are unimportant at scattering CRs.

Interpreting CR data using a combination of scattering mechanisms in different phases of the ISM is physically well-motivated, but will inevitably require some fine-tuning (especially if the turbulent cascade is strongly damped). Moreover, CR scattering by fast-mode turbulence is likely inefficient due to wave steepening.  An attractive alternative is that CR scattering is dominated by a currently unidentified source of a roughly Kolmogorov-like cascade, as is often assumed in phenomenological models. Such a cascade is not well motivated by current theoretical models of MHD turbulence. In Section \ref{sec:discussion}, we discussed some of the uncertain aspects of MHD turbulence theory that could bear on the presence of such a cascade and thus on CR scattering. In this context it is worth noting that the energetic requirements for confining CRs by scattering are very small compared to the overall energetics of interstellar turbulence (Section \ref{sec:ext_drive}). Thus, an energetically minor component of the cascade or turbulent driving by an energetically subdominant channel could nonetheless dominate CR scattering.

In this work we have focused on CR transport along field lines. There is a separate question of how the geometry of magnetic-field lines, e.g. turbulent field-line wandering, affects CR transport perpendicular to the mean magnetic field. We have ignored this aspect of CR transport in this work because the perpendicular diffusion time is expected to be  orders of magnitude longer than the parallel diffusion time (e.g., \citealt{giacalone_jokipii_1999}; \citealt{desiati_zweibel_2014}; \citealt{dundovic_2020}).  Nevertheless, given that the magnetic field is on average close to being planar in the galactic disk and that perpendicular diffusion of CRs remains quite uncertain, it is possible that perpendicular transport is important for regulating CR escape from the Galaxy and setting CR observables.
    
In the course of this work, we became aware of similar efforts to reconcile CR transport theories with MW observables by \cite{hopkins_sc_et_problems}. Using galaxy-formation simulations combined with a broad selection of CR transport models motivated by microphysics, \cite{hopkins_sc_et_problems} also conclude that self-confinement theory and “extrinsic turbulence" theory alone cannot qualitatively reproduce CR spectra observed in the solar neighbourhood. As we showed in Section \ref{sec:strat_lin} and Section \ref{sec:sc+et}, the exact ISM phase structure and spatial variations of plasma properties can significantly affect CR observables and there exist multi-phase solutions (combinations of self-confinement and extrinsic turbulence) that can, in principle, qualitatively reproduce local CR spectra (Figure \ref{fig:vars}). These solutions require a particular set of fine-tuned ISM plasma conditions, which are not realised in the simulations of \cite{hopkins_sc_et_problems}.   \cite{hopkins_sc_et_problems} instead elucidate the form of driving/damping necessary for extrinsic turbulence and/or self-confinement to be compatible with MW measurements (analogous to our Section \ref{sec:discussion}).

\section*{Data Availability}
The calculations from this article will be shared on reasonable request to the corresponding author.

\section*{Acknowledgements}
We thank Philip Hopkins and Jonathan Squire for many enlightening discussions and helpful comments on earlier drafts of this manuscript. We are grateful to the referee for very useful feedback on the paper. We also thank Susan Clark, Brandon Hensley, Chang-Goo Kim, Matthew Kunz, Eve Ostriker, Alexander Schekochihin, and Anatoly Spitkovsky for useful conversations. This research was supported  by NSF grant AST-1715070 and a Simons Investigator award from the Simons Foundation. 

\bibliographystyle{mnras}

\bibliography{cr_scatt}

\begin{thebibliography}{}
\makeatletter
\relax
\def\mn@urlcharsother{\let\do\@makeother \do\$\do\&\do\#\do\^\do\_\do\%\do\~}
\def\mn@doi{\begingroup\mn@urlcharsother \@ifnextchar [ {\mn@doi@}
  {\mn@doi@[]}}
\def\mn@doi@[#1]#2{\def\@tempa{#1}\ifx\@tempa\@empty \href
  {http://dx.doi.org/#2} {doi:#2}\else \href {http://dx.doi.org/#2} {#1}\fi
  \endgroup}
\def\mn@eprint#1#2{\mn@eprint@#1:#2::\@nil}
\def\mn@eprint@arXiv#1{\href {http://arxiv.org/abs/#1} {{\tt arXiv:#1}}}
\def\mn@eprint@dblp#1{\href {http://dblp.uni-trier.de/rec/bibtex/#1.xml}
  {dblp:#1}}
\def\mn@eprint@#1:#2:#3:#4\@nil{\def\@tempa {#1}\def\@tempb {#2}\def\@tempc
  {#3}\ifx \@tempc \@empty \let \@tempc \@tempb \let \@tempb \@tempa \fi \ifx
  \@tempb \@empty \def\@tempb {arXiv}\fi \@ifundefined
  {mn@eprint@\@tempb}{\@tempb:\@tempc}{\expandafter \expandafter \csname
  mn@eprint@\@tempb\endcsname \expandafter{\@tempc}}}

\bibitem[\protect\citeauthoryear{{Acero} et~al.,}{{Acero}
  et~al.}{2016}]{acero_2016}
{Acero} F.,  et~al., 2016, \mn@doi [\apjs] {10.3847/0067-0049/223/2/26}, \href
  {https://ui.adsabs.harvard.edu/abs/2016ApJS..223...26A} {223, 26}

\bibitem[\protect\citeauthoryear{{Ackermann} et~al.,}{{Ackermann}
  et~al.}{2012}]{ackermann2012}
{Ackermann} M.,  et~al., 2012, \mn@doi [\apj] {10.1088/0004-637X/755/2/164},
  \href {https://ui.adsabs.harvard.edu/abs/2012ApJ...755..164A} {755, 164}

\bibitem[\protect\citeauthoryear{{Adkins} \& {Schekochihin}}{{Adkins} \&
  {Schekochihin}}{2018}]{adkins_2018}
{Adkins} T.,  {Schekochihin} A.~A.,  2018, \mn@doi [Journal of Plasma Physics]
  {10.1017/S0022377818000089}, \href
  {https://ui.adsabs.harvard.edu/abs/2018JPlPh..84a9007A} {84, 905840107}

\bibitem[\protect\citeauthoryear{{Aguilar} et~al.,}{{Aguilar}
  et~al.}{2015}]{aguilar_2015}
{Aguilar} M.,  et~al., 2015, \mn@doi [\prl] {10.1103/PhysRevLett.114.171103},
  \href {https://ui.adsabs.harvard.edu/abs/2015PhRvL.114q1103A} {114, 171103}

\bibitem[\protect\citeauthoryear{{Aguilar} et~al.,}{{Aguilar}
  et~al.}{2016}]{aguilar_bc}
{Aguilar} M.,  et~al., 2016, \mn@doi [\prl] {10.1103/PhysRevLett.117.231102},
  \href {https://ui.adsabs.harvard.edu/abs/2016PhRvL.117w1102A} {117, 231102}

\bibitem[\protect\citeauthoryear{{Aloisio} \& {Blasi}}{{Aloisio} \&
  {Blasi}}{2013}]{aloisio_blasi_2013}
{Aloisio} R.,  {Blasi} P.,  2013, \mn@doi [\jcap]
  {10.1088/1475-7516/2013/07/001}, \href
  {https://ui.adsabs.harvard.edu/abs/2013JCAP...07..001A} {2013, 001}

\bibitem[\protect\citeauthoryear{{Aloisio}, {Blasi}  \& {Serpico}}{{Aloisio}
  et~al.}{2015}]{aloisio_2015}
{Aloisio} R.,  {Blasi} P.,   {Serpico} P.~D.,  2015, \mn@doi [\aap]
  {10.1051/0004-6361/201526877}, \href
  {https://ui.adsabs.harvard.edu/abs/2015A&A...583A..95A} {583, A95}

\bibitem[\protect\citeauthoryear{{Bai}}{{Bai}}{2021}]{bai_2021}
{Bai} X.-N.,  2021, arXiv e-prints, \href
  {https://ui.adsabs.harvard.edu/abs/2021arXiv211214782B} {p. arXiv:2112.14782}

\bibitem[\protect\citeauthoryear{{Bai}, {Ostriker}, {Plotnikov}  \&
  {Stone}}{{Bai} et~al.}{2019}]{bai_mhd_pic}
{Bai} X.-N.,  {Ostriker} E.~C.,  {Plotnikov} I.,   {Stone} J.~M.,  2019,
  \mn@doi [\apj] {10.3847/1538-4357/ab1648}, \href
  {https://ui.adsabs.harvard.edu/abs/2019ApJ...876...60B} {876, 60}

\bibitem[\protect\citeauthoryear{{Beck}}{{Beck}}{2015}]{beck2015_Bfield_spirals}
{Beck} R.,  2015, \mn@doi [\aapr] {10.1007/s00159-015-0084-4}, \href
  {https://ui.adsabs.harvard.edu/abs/2015A&ARv..24....4B} {24, 4}

\bibitem[\protect\citeauthoryear{{Beresnyak} \& {Lazarian}}{{Beresnyak} \&
  {Lazarian}}{2009}]{beresnyak_lazarian_2009}
{Beresnyak} A.,  {Lazarian} A.,  2009, \mn@doi [\apj]
  {10.1088/0004-637X/702/1/460}, \href
  {https://ui.adsabs.harvard.edu/abs/2009ApJ...702..460B} {702, 460}

\bibitem[\protect\citeauthoryear{{Blasi}}{{Blasi}}{2019}]{blasi_2019}
{Blasi} P.,  2019, \mn@doi [Galaxies] {10.3390/galaxies7020064}, \href
  {https://ui.adsabs.harvard.edu/abs/2019Galax...7...64B} {7, 64}

\bibitem[\protect\citeauthoryear{{Blasi}, {Amato}  \& {Serpico}}{{Blasi}
  et~al.}{2012}]{blasi12}
{Blasi} P.,  {Amato} E.,   {Serpico} P.~D.,  2012, \mn@doi [\prl]
  {10.1103/PhysRevLett.109.061101}, \href
  {https://ui.adsabs.harvard.edu/abs/2012PhRvL.109f1101B} {109, 061101}

\bibitem[\protect\citeauthoryear{{Boldyrev}}{{Boldyrev}}{2006}]{boldyrev_2006}
{Boldyrev} S.,  2006, \mn@doi [\prl] {10.1103/PhysRevLett.96.115002}, \href
  {https://ui.adsabs.harvard.edu/abs/2006PhRvL..96k5002B} {96, 115002}

\bibitem[\protect\citeauthoryear{{Boldyrev} \& {Loureiro}}{{Boldyrev} \&
  {Loureiro}}{2019}]{boldyrev_loureiro_2019}
{Boldyrev} S.,  {Loureiro} N.~F.,  2019, \mn@doi [Physical Review Research]
  {10.1103/PhysRevResearch.1.012006}, \href
  {https://ui.adsabs.harvard.edu/abs/2019PhRvR...1a2006B} {1, 012006}

\bibitem[\protect\citeauthoryear{Braginskii}{Braginskii}{1965}]{br65}
Braginskii S.~I.,  1965, Rev. Plasma Phys., 1, 205

\bibitem[\protect\citeauthoryear{{Bresci}, {Amato}, {Blasi}  \&
  {Morlino}}{{Bresci} et~al.}{2019}]{bresci_2019_reacc}
{Bresci} V.,  {Amato} E.,  {Blasi} P.,   {Morlino} G.,  2019, \mn@doi [\mnras]
  {10.1093/mnras/stz1806}, \href
  {https://ui.adsabs.harvard.edu/abs/2019MNRAS.488.2068B} {488, 2068}

\bibitem[\protect\citeauthoryear{Chandran}{Chandran}{2000}]{chandran_scattering}
Chandran B. D.~G.,  2000, \mn@doi [Phys. Rev. Lett.]
  {10.1103/PhysRevLett.85.4656}, 85, 4656

\bibitem[\protect\citeauthoryear{{Chandran}}{{Chandran}}{2005}]{chandran_2005}
{Chandran} B. D.~G.,  2005, \mn@doi [\prl] {10.1103/PhysRevLett.95.265004},
  \href {https://ui.adsabs.harvard.edu/abs/2005PhRvL..95z5004C} {95, 265004}

\bibitem[\protect\citeauthoryear{{Chen}}{{Chen}}{2016}]{chen_2016}
{Chen} C.~H.~K.,  2016, \mn@doi [Journal of Plasma Physics]
  {10.1017/S0022377816001124}, \href
  {https://ui.adsabs.harvard.edu/abs/2016JPlPh..82f5302C} {82, 535820602}

\bibitem[\protect\citeauthoryear{{Cho} \& {Lazarian}}{{Cho} \&
  {Lazarian}}{2003}]{cho_lazarian}
{Cho} J.,  {Lazarian} A.,  2003, \mn@doi [\mnras]
  {10.1046/j.1365-8711.2003.06941.x}, \href
  {https://ui.adsabs.harvard.edu/abs/2003MNRAS.345..325C} {345, 325}

\bibitem[\protect\citeauthoryear{{Cummings} et~al.,}{{Cummings}
  et~al.}{2016}]{cummings_2016}
{Cummings} A.~C.,  et~al., 2016, \mn@doi [\apj] {10.3847/0004-637X/831/1/18},
  \href {https://ui.adsabs.harvard.edu/abs/2016ApJ...831...18C} {831, 18}

\bibitem[\protect\citeauthoryear{{Desiati} \& {Zweibel}}{{Desiati} \&
  {Zweibel}}{2014}]{desiati_zweibel_2014}
{Desiati} P.,  {Zweibel} E.~G.,  2014, \mn@doi [\apj]
  {10.1088/0004-637X/791/1/51}, \href
  {https://ui.adsabs.harvard.edu/abs/2014ApJ...791...51D} {791, 51}

\bibitem[\protect\citeauthoryear{{Dundovic}, {Pezzi}, {Blasi}, {Evoli}  \&
  {Matthaeus}}{{Dundovic} et~al.}{2020}]{dundovic_2020}
{Dundovic} A.,  {Pezzi} O.,  {Blasi} P.,  {Evoli} C.,   {Matthaeus} W.~H.,
  2020, \mn@doi [\prd] {10.1103/PhysRevD.102.103016}, \href
  {https://ui.adsabs.harvard.edu/abs/2020PhRvD.102j3016D} {102, 103016}

\bibitem[\protect\citeauthoryear{{Evoli}, {Blasi}, {Morlino}  \&
  {Aloisio}}{{Evoli} et~al.}{2018}]{evoli_2018}
{Evoli} C.,  {Blasi} P.,  {Morlino} G.,   {Aloisio} R.,  2018, \mn@doi [\prl]
  {10.1103/PhysRevLett.121.021102}, \href
  {https://ui.adsabs.harvard.edu/abs/2018PhRvL.121b1102E} {121, 021102}

\bibitem[\protect\citeauthoryear{{Farber}, {Ruszkowski}, {Yang}  \&
  {Zweibel}}{{Farber} et~al.}{2018}]{farber18}
{Farber} R.,  {Ruszkowski} M.,  {Yang} H. Y.~K.,   {Zweibel} E.~G.,  2018,
  \mn@doi [\apj] {10.3847/1538-4357/aab26d}, \href
  {https://ui.adsabs.harvard.edu/abs/2018ApJ...856..112F} {856, 112}

\bibitem[\protect\citeauthoryear{{Farmer} \& {Goldreich}}{{Farmer} \&
  {Goldreich}}{2004}]{farmer_goldreich}
{Farmer} A.~J.,  {Goldreich} P.,  2004, \mn@doi [\apj] {10.1086/382040}, \href
  {https://ui.adsabs.harvard.edu/abs/2004ApJ...604..671F} {604, 671}

\bibitem[\protect\citeauthoryear{{Felice} \& {Kulsrud}}{{Felice} \&
  {Kulsrud}}{2001}]{felice_kulsrud}
{Felice} G.~M.,  {Kulsrud} R.~M.,  2001, \mn@doi [\apj] {10.1086/320651}, \href
  {https://ui.adsabs.harvard.edu/abs/2001ApJ...553..198F} {553, 198}

\bibitem[\protect\citeauthoryear{{Fornieri}, {Gaggero}, {Cerri}, {De La Torre
  Luque}  \& {Gabici}}{{Fornieri} et~al.}{2021}]{fornieri_2021}
{Fornieri} O.,  {Gaggero} D.,  {Cerri} S.~S.,  {De La Torre Luque} P.,
  {Gabici} S.,  2021, \mn@doi [\mnras] {10.1093/mnras/stab355}, \href
  {https://ui.adsabs.harvard.edu/abs/2021MNRAS.502.5821F} {502, 5821}

\bibitem[\protect\citeauthoryear{{Gaggero}, {Maccione}, {Grasso}, {Di Bernardo}
   \& {Evoli}}{{Gaggero} et~al.}{2014}]{gaggero_2014}
{Gaggero} D.,  {Maccione} L.,  {Grasso} D.,  {Di Bernardo} G.,   {Evoli} C.,
  2014, \mn@doi [\prd] {10.1103/PhysRevD.89.083007}, \href
  {https://ui.adsabs.harvard.edu/abs/2014PhRvD..89h3007G} {89, 083007}

\bibitem[\protect\citeauthoryear{{Giacalone} \& {Jokipii}}{{Giacalone} \&
  {Jokipii}}{1999}]{giacalone_jokipii_1999}
{Giacalone} J.,  {Jokipii} J.~R.,  1999, \mn@doi [\apj] {10.1086/307452}, \href
  {https://ui.adsabs.harvard.edu/abs/1999ApJ...520..204G} {520, 204}

\bibitem[\protect\citeauthoryear{{Ginzburg}}{{Ginzburg}}{1961}]{ginzburg1961}
{Ginzburg} V.~L.,  1961, {Propagation of electromagnetic waves in plasma (New
  York: Gordon \& Breach)}

\bibitem[\protect\citeauthoryear{{Goldreich} \& {Sridhar}}{{Goldreich} \&
  {Sridhar}}{1995}]{gs95}
{Goldreich} P.,  {Sridhar} S.,  1995, \mn@doi [\apj] {10.1086/175121}, \href
  {https://ui.adsabs.harvard.edu/abs/1995ApJ...438..763G} {438, 763}

\bibitem[\protect\citeauthoryear{{Holmes}}{{Holmes}}{1974}]{holmes_1974}
{Holmes} J.~A.,  1974, \mn@doi [\mnras] {10.1093/mnras/166.2.155}, \href
  {https://ui.adsabs.harvard.edu/abs/1974MNRAS.166..155H} {166, 155}

\bibitem[\protect\citeauthoryear{{Holmes}}{{Holmes}}{1975}]{holmes_1975}
{Holmes} J.~A.,  1975, \mn@doi [\mnras] {10.1093/mnras/170.2.251}, \href
  {https://ui.adsabs.harvard.edu/abs/1975MNRAS.170..251H} {170, 251}

\bibitem[\protect\citeauthoryear{{Hopkins} et~al.,}{{Hopkins}
  et~al.}{2020}]{hopkins2020_whatabout}
{Hopkins} P.~F.,  et~al., 2020, \mn@doi [\mnras] {10.1093/mnras/stz3321}, \href
  {https://ui.adsabs.harvard.edu/abs/2020MNRAS.492.3465H} {492, 3465}

\bibitem[\protect\citeauthoryear{{Hopkins}, {Butsky}, {Panopoulou}, {Ji},
  {Quataert}, {Faucher-Giguere}  \& {Keres}}{{Hopkins}
  et~al.}{2021a}]{hopkins_cr_pheno}
{Hopkins} P.~F.,  {Butsky} I.~S.,  {Panopoulou} G.~V.,  {Ji} S.,  {Quataert}
  E.,  {Faucher-Giguere} C.-A.,   {Keres} D.,  2021a, arXiv e-prints, \href
  {https://ui.adsabs.harvard.edu/abs/2021arXiv210909762H} {p. arXiv:2109.09762}

\bibitem[\protect\citeauthoryear{{Hopkins}, {Squire}, {Butsky}  \&
  {Ji}}{{Hopkins} et~al.}{2021b}]{hopkins_sc_et_problems}
{Hopkins} P.~F.,  {Squire} J.,  {Butsky} I.~S.,   {Ji} S.,  2021b, arXiv
  e-prints, \href {https://ui.adsabs.harvard.edu/abs/2021arXiv211202153H} {p.
  arXiv:2112.02153}

\bibitem[\protect\citeauthoryear{{Huang} \& {Bhattacharjee}}{{Huang} \&
  {Bhattacharjee}}{2016}]{huang_bhattacharjee_2016}
{Huang} Y.-M.,  {Bhattacharjee} A.,  2016, \mn@doi [\apj]
  {10.3847/0004-637X/818/1/20}, \href
  {https://ui.adsabs.harvard.edu/abs/2016ApJ...818...20H} {818, 20}

\bibitem[\protect\citeauthoryear{Kadomtsev \& Petviashvili}{Kadomtsev \&
  Petviashvili}{1973}]{kadomtsev1973acoustic}
Kadomtsev B.~B.,  Petviashvili V.~I.,  1973, in Doklady Akademii Nauk. pp
  794--796

\bibitem[\protect\citeauthoryear{{Kim} \& {Ostriker}}{{Kim} \&
  {Ostriker}}{2017}]{kim_ostriker_2017}
{Kim} C.-G.,  {Ostriker} E.~C.,  2017, \mn@doi [\apj]
  {10.3847/1538-4357/aa8599}, \href
  {https://ui.adsabs.harvard.edu/abs/2017ApJ...846..133K} {846, 133}

\bibitem[\protect\citeauthoryear{{Kowal} \& {Lazarian}}{{Kowal} \&
  {Lazarian}}{2010}]{kowal_lazarian_2010}
{Kowal} G.,  {Lazarian} A.,  2010, \mn@doi [\apj]
  {10.1088/0004-637X/720/1/742}, \href
  {https://ui.adsabs.harvard.edu/abs/2010ApJ...720..742K} {720, 742}

\bibitem[\protect\citeauthoryear{{Kulsrud}}{{Kulsrud}}{2005}]{kulsrud_book}
{Kulsrud} R.~M.,  2005, {Plasma physics for astrophysics}

\bibitem[\protect\citeauthoryear{{Kulsrud} \& {Cesarsky}}{{Kulsrud} \&
  {Cesarsky}}{1971}]{kc71}
{Kulsrud} R.~M.,  {Cesarsky} C.~J.,  1971, Astrophysical Letters, \href
  {https://ui.adsabs.harvard.edu/abs/1971ApL.....8..189K} {8, 189}

\bibitem[\protect\citeauthoryear{{Kulsrud} \& {Pearce}}{{Kulsrud} \&
  {Pearce}}{1969}]{kp69}
{Kulsrud} R.,  {Pearce} W.~P.,  1969, \mn@doi [\apj] {10.1086/149981}, 156, 445

\bibitem[\protect\citeauthoryear{{Lacki}, {Thompson}, {Quataert}, {Loeb}  \&
  {Waxman}}{{Lacki} et~al.}{2011}]{lacki_2011}
{Lacki} B.~C.,  {Thompson} T.~A.,  {Quataert} E.,  {Loeb} A.,   {Waxman} E.,
  2011, \mn@doi [\apj] {10.1088/0004-637X/734/2/107}, \href
  {https://ui.adsabs.harvard.edu/abs/2011ApJ...734..107L} {734, 107}

\bibitem[\protect\citeauthoryear{{Landau} \& {Lifshitz}}{{Landau} \&
  {Lifshitz}}{1959}]{landau_lifshitz}
{Landau} L.~D.,  {Lifshitz} E.~M.,  1959, {Fluid mechanics}

\bibitem[\protect\citeauthoryear{{Lazarian}}{{Lazarian}}{2016}]{lazarian_2016}
{Lazarian} A.,  2016, \mn@doi [\apj] {10.3847/1538-4357/833/2/131}, \href
  {https://ui.adsabs.harvard.edu/abs/2016ApJ...833..131L} {833, 131}

\bibitem[\protect\citeauthoryear{{Lazarian} \& {Xu}}{{Lazarian} \&
  {Xu}}{2021}]{lazarian_xu_mirror}
{Lazarian} A.,  {Xu} S.,  2021, \mn@doi [\apj] {10.3847/1538-4357/ac2de9},
  \href {https://ui.adsabs.harvard.edu/abs/2021ApJ...923...53L} {923, 53}

\bibitem[\protect\citeauthoryear{{Lee} \& {V{\"o}lk}}{{Lee} \&
  {V{\"o}lk}}{1973}]{lee_volk_1973}
{Lee} M.~A.,  {V{\"o}lk} H.~J.,  1973, \mn@doi [\apss] {10.1007/BF00648673},
  \href {https://ui.adsabs.harvard.edu/abs/1973Ap&SS..24...31L} {24, 31}

\bibitem[\protect\citeauthoryear{{Linden}, {Profumo}  \& {Anderson}}{{Linden}
  et~al.}{2010}]{Linden2010}
{Linden} T.,  {Profumo} S.,   {Anderson} B.,  2010, \mn@doi [\prd]
  {10.1103/PhysRevD.82.063529}, \href
  {https://ui.adsabs.harvard.edu/abs/2010PhRvD..82f3529L} {82, 063529}

\bibitem[\protect\citeauthoryear{{Liping}, {Hui}, {Fan}, {Xiaocan}, {Shengtai},
  {jiansen}, {Lei}  \& {Xueshang}}{{Liping} et~al.}{2020}]{yang_2020}
{Liping} Y.,  {Hui} L.,  {Fan} G.,  {Xiaocan} L.,  {Shengtai} L.,  {jiansen}
  H.,  {Lei} Z.,   {Xueshang} F.,  2020, arXiv e-prints, \href
  {https://ui.adsabs.harvard.edu/abs/2020arXiv200906253L} {p. arXiv:2009.06253}

\bibitem[\protect\citeauthoryear{{Loureiro} \& {Boldyrev}}{{Loureiro} \&
  {Boldyrev}}{2016}]{loureiro_boldyrev_2016}
{Loureiro} N.~F.,  {Boldyrev} S.,  2016, arXiv e-prints, \href
  {https://ui.adsabs.harvard.edu/abs/2016arXiv161207266L} {p. arXiv:1612.07266}

\bibitem[\protect\citeauthoryear{{Makwana} \& {Yan}}{{Makwana} \&
  {Yan}}{2020}]{makwana_yan_2020}
{Makwana} K.~D.,  {Yan} H.,  2020, \mn@doi [Physical Review X]
  {10.1103/PhysRevX.10.031021}, \href
  {https://ui.adsabs.harvard.edu/abs/2020PhRvX..10c1021M} {10, 031021}

\bibitem[\protect\citeauthoryear{{Mallet}, {Schekochihin}  \&
  {Chandran}}{{Mallet} et~al.}{2017a}]{mallet_2017_collisionless}
{Mallet} A.,  {Schekochihin} A.~A.,   {Chandran} B. D.~G.,  2017a, \mn@doi
  [Journal of Plasma Physics] {10.1017/S0022377817000812}, \href
  {https://ui.adsabs.harvard.edu/abs/2017JPlPh..83f9009M} {83, 905830609}

\bibitem[\protect\citeauthoryear{{Mallet}, {Schekochihin}  \&
  {Chandran}}{{Mallet} et~al.}{2017b}]{mallet_2017_colless_rec}
{Mallet} A.,  {Schekochihin} A.~A.,   {Chandran} B. D.~G.,  2017b, \mn@doi
  [Journal of Plasma Physics] {10.1017/S0022377817000812}, \href
  {https://ui.adsabs.harvard.edu/abs/2017JPlPh..83f9009M} {83, 905830609}

\bibitem[\protect\citeauthoryear{{Mallet}, {Schekochihin}  \&
  {Chandran}}{{Mallet} et~al.}{2017c}]{mallet_2017_tearing}
{Mallet} A.,  {Schekochihin} A.~A.,   {Chandran} B.~D.~G.,  2017c, \mn@doi
  [\mnras] {10.1093/mnras/stx670}, \href
  {https://ui.adsabs.harvard.edu/abs/2017MNRAS.468.4862M} {468, 4862}

\bibitem[\protect\citeauthoryear{{McKee} \& {Ostriker}}{{McKee} \&
  {Ostriker}}{1977}]{mckee_ostriker_1977}
{McKee} C.~F.,  {Ostriker} J.~P.,  1977, \mn@doi [\apj] {10.1086/155667}, \href
  {https://ui.adsabs.harvard.edu/abs/1977ApJ...218..148M} {218, 148}

\bibitem[\protect\citeauthoryear{{Meyrand}, {Kanekar}, {Dorland}  \&
  {Schekochihin}}{{Meyrand} et~al.}{2019}]{meyrand_2019}
{Meyrand} R.,  {Kanekar} A.,  {Dorland} W.,   {Schekochihin} A.~A.,  2019,
  \mn@doi [Proceedings of the National Academy of Science]
  {10.1073/pnas.1813913116}, \href
  {https://ui.adsabs.harvard.edu/abs/2019PNAS..116.1185M} {116, 1185}

\bibitem[\protect\citeauthoryear{{Miville-Desch{\^e}nes}, {Ysard}, {Lavabre},
  {Ponthieu}, {Mac{\'\i}as-P{\'e}rez}, {Aumont}  \&
  {Bernard}}{{Miville-Desch{\^e}nes} et~al.}{2008}]{miville_deschenes_2008}
{Miville-Desch{\^e}nes} M.~A.,  {Ysard} N.,  {Lavabre} A.,  {Ponthieu} N.,
  {Mac{\'\i}as-P{\'e}rez} J.~F.,  {Aumont} J.,   {Bernard} J.~P.,  2008,
  \mn@doi [\aap] {10.1051/0004-6361:200809484}, \href
  {https://ui.adsabs.harvard.edu/abs/2008A&A...490.1093M} {490, 1093}

\bibitem[\protect\citeauthoryear{{Ptuskin}, {Voelk}, {Zirakashvili}  \&
  {Breitschwerdt}}{{Ptuskin} et~al.}{1997}]{ptuskin_1997}
{Ptuskin} V.~S.,  {Voelk} H.~J.,  {Zirakashvili} V.~N.,   {Breitschwerdt} D.,
  1997, \aap, \href {https://ui.adsabs.harvard.edu/abs/1997A&A...321..434P}
  {321, 434}

\bibitem[\protect\citeauthoryear{{Quataert}, {Thompson}  \& {Jiang}}{{Quataert}
  et~al.}{2021a}]{qtj_2021_diff}
{Quataert} E.,  {Thompson} T.~A.,   {Jiang} Y.-F.,  2021a, arXiv e-prints,
  \href {https://ui.adsabs.harvard.edu/abs/2021arXiv210205696Q} {p.
  arXiv:2102.05696}

\bibitem[\protect\citeauthoryear{{Quataert}, {Jiang}  \& {Thompson}}{{Quataert}
  et~al.}{2021b}]{qtj_2021_streaming}
{Quataert} E.,  {Jiang} Y.-F.,   {Thompson} T.~A.,  2021b, arXiv e-prints,
  \href {https://ui.adsabs.harvard.edu/abs/2021arXiv210608404Q} {p.
  arXiv:2106.08404}

\bibitem[\protect\citeauthoryear{{Recchia}, {Blasi}  \& {Morlino}}{{Recchia}
  et~al.}{2016}]{recchia_2016}
{Recchia} S.,  {Blasi} P.,   {Morlino} G.,  2016, \mn@doi [\mnras]
  {10.1093/mnras/stw1966}, \href
  {https://ui.adsabs.harvard.edu/abs/2016MNRAS.462.4227R} {462, 4227}

\bibitem[\protect\citeauthoryear{{Ruszkowski}, {Yang}  \&
  {Zweibel}}{{Ruszkowski} et~al.}{2017}]{ruszkowski17}
{Ruszkowski} M.,  {Yang} H. Y.~K.,   {Zweibel} E.,  2017, \mn@doi [\apj]
  {10.3847/1538-4357/834/2/208}, \href
  {https://ui.adsabs.harvard.edu/abs/2017ApJ...834..208R} {834, 208}

\bibitem[\protect\citeauthoryear{{Schekochihin}}{{Schekochihin}}{2020}]{schekochihin_mhd}
{Schekochihin} A.~A.,  2020, arXiv e-prints, \href
  {https://ui.adsabs.harvard.edu/abs/2020arXiv201000699S} {p. arXiv:2010.00699}

\bibitem[\protect\citeauthoryear{{Schekochihin}, {Parker}, {Highcock},
  {Dellar}, {Dorland}  \& {Hammett}}{{Schekochihin}
  et~al.}{2016}]{scheko_2016_echo}
{Schekochihin} A.~A.,  {Parker} J.~T.,  {Highcock} E.~G.,  {Dellar} P.~J.,
  {Dorland} W.,   {Hammett} G.~W.,  2016, \mn@doi [Journal of Plasma Physics]
  {10.1017/S0022377816000374}, \href
  {https://ui.adsabs.harvard.edu/abs/2016JPlPh..82b9012S} {82, 905820212}

\bibitem[\protect\citeauthoryear{{Schlickeiser}}{{Schlickeiser}}{2002}]{schlickeiser2002}
{Schlickeiser} R.,  2002, {Cosmic Ray Astrophysics}

\bibitem[\protect\citeauthoryear{{Skilling}}{{Skilling}}{1971}]{skilling71}
{Skilling} J.,  1971, \mn@doi [\apj] {10.1086/151210}, 170, 265

\bibitem[\protect\citeauthoryear{{Skilling}}{{Skilling}}{1975}]{skilling_1975}
{Skilling} J.,  1975, \mn@doi [\mnras] {10.1093/mnras/172.3.557}, \href
  {https://ui.adsabs.harvard.edu/abs/1975MNRAS.172..557S} {172, 557}

\bibitem[\protect\citeauthoryear{{Squire} \& {Hopkins}}{{Squire} \&
  {Hopkins}}{2018}]{squire_hopkins_2018_rdi}
{Squire} J.,  {Hopkins} P.~F.,  2018, \mn@doi [\apjl]
  {10.3847/2041-8213/aab54d}, \href
  {https://ui.adsabs.harvard.edu/abs/2018ApJ...856L..15S} {856, L15}

\bibitem[\protect\citeauthoryear{{Squire}, {Schekochihin}  \&
  {Quataert}}{{Squire} et~al.}{2017}]{squire_et_al_2017}
{Squire} J.,  {Schekochihin} A.~A.,   {Quataert} E.,  2017, \mn@doi [New
  Journal of Physics] {10.1088/1367-2630/aa6bb1}, \href
  {https://ui.adsabs.harvard.edu/abs/2017NJPh...19e5005S} {19, 055005}

\bibitem[\protect\citeauthoryear{{Squire}, {Hopkins}, {Quataert}  \&
  {Kempski}}{{Squire} et~al.}{2021}]{squire_dust_2021}
{Squire} J.,  {Hopkins} P.~F.,  {Quataert} E.,   {Kempski} P.,  2021, \mn@doi
  [\mnras] {10.1093/mnras/stab179}, \href
  {https://ui.adsabs.harvard.edu/abs/2021MNRAS.502.2630S} {502, 2630}

\bibitem[\protect\citeauthoryear{{Stone}, {Cummings}, {McDonald}, {Heikkila},
  {Lal}  \& {Webber}}{{Stone} et~al.}{2013}]{stone_voyager}
{Stone} E.~C.,  {Cummings} A.~C.,  {McDonald} F.~B.,  {Heikkila} B.~C.,  {Lal}
  N.,   {Webber} W.~R.,  2013, \mn@doi [Science] {10.1126/science.1236408},
  \href {https://ui.adsabs.harvard.edu/abs/2013Sci...341..150S} {341, 150}

\bibitem[\protect\citeauthoryear{{Strong}, {Porter}, {Digel},
  {J{\'o}hannesson}, {Martin}, {Moskalenko}, {Murphy}  \& {Orlando}}{{Strong}
  et~al.}{2010}]{strong_2010}
{Strong} A.~W.,  {Porter} T.~A.,  {Digel} S.~W.,  {J{\'o}hannesson} G.,
  {Martin} P.,  {Moskalenko} I.~V.,  {Murphy} E.~J.,   {Orlando} E.,  2010,
  \mn@doi [\apjl] {10.1088/2041-8205/722/1/L58}, \href
  {https://ui.adsabs.harvard.edu/abs/2010ApJ...722L..58S} {722, L58}

\bibitem[\protect\citeauthoryear{{Tomassetti}}{{Tomassetti}}{2012}]{tomassetti_2012}
{Tomassetti} N.,  2012, \mn@doi [\apjl] {10.1088/2041-8205/752/1/L13}, \href
  {https://ui.adsabs.harvard.edu/abs/2012ApJ...752L..13T} {752, L13}

\bibitem[\protect\citeauthoryear{{Trotta}, {J{\'o}hannesson}, {Moskalenko},
  {Porter}, {Ruiz de Austri}  \& {Strong}}{{Trotta} et~al.}{2011}]{trotta_2011}
{Trotta} R.,  {J{\'o}hannesson} G.,  {Moskalenko} I.~V.,  {Porter} T.~A.,
  {Ruiz de Austri} R.,   {Strong} A.~W.,  2011, \mn@doi [\apj]
  {10.1088/0004-637X/729/2/106}, \href
  {https://ui.adsabs.harvard.edu/abs/2011ApJ...729..106T} {729, 106}

\bibitem[\protect\citeauthoryear{{Verscharen}, {Chen}  \& {Wicks}}{{Verscharen}
  et~al.}{2017}]{verscharen_2017}
{Verscharen} D.,  {Chen} C. H.~K.,   {Wicks} R.~T.,  2017, \mn@doi [\apj]
  {10.3847/1538-4357/aa6a56}, \href
  {https://ui.adsabs.harvard.edu/abs/2017ApJ...840..106V} {840, 106}

\bibitem[\protect\citeauthoryear{{Voelk}}{{Voelk}}{1975}]{voelk_1975}
{Voelk} H.~J.,  1975, \mn@doi [Reviews of Geophysics and Space Physics]
  {10.1029/RG013i004p00547}, \href
  {https://ui.adsabs.harvard.edu/abs/1975RvGSP..13..547V} {13, 547}

\bibitem[\protect\citeauthoryear{{V{\"o}lk}}{{V{\"o}lk}}{1973}]{voelk_1973}
{V{\"o}lk} H.~J.,  1973, \mn@doi [\apss] {10.1007/BF00649186}, \href
  {https://ui.adsabs.harvard.edu/abs/1973Ap&SS..25..471V} {25, 471}

\bibitem[\protect\citeauthoryear{{Wiener}, {Oh}  \& {Guo}}{{Wiener}
  et~al.}{2013}]{wiener2013}
{Wiener} J.,  {Oh} S.~P.,   {Guo} F.,  2013, \mn@doi [\mnras]
  {10.1093/mnras/stt1163}, \href
  {https://ui.adsabs.harvard.edu/abs/2013MNRAS.434.2209W} {434, 2209}

\bibitem[\protect\citeauthoryear{{Xu} \& {Lazarian}}{{Xu} \&
  {Lazarian}}{2018}]{xu_lazarian_ttd}
{Xu} S.,  {Lazarian} A.,  2018, \mn@doi [\apj] {10.3847/1538-4357/aae840},
  \href {https://ui.adsabs.harvard.edu/abs/2018ApJ...868...36X} {868, 36}

\bibitem[\protect\citeauthoryear{{Xu}, {Yan}  \& {Lazarian}}{{Xu}
  et~al.}{2016}]{xu_2016}
{Xu} S.,  {Yan} H.,   {Lazarian} A.,  2016, \mn@doi [\apj]
  {10.3847/0004-637X/826/2/166}, \href
  {https://ui.adsabs.harvard.edu/abs/2016ApJ...826..166X} {826, 166}

\bibitem[\protect\citeauthoryear{{Yan} \& {Lazarian}}{{Yan} \&
  {Lazarian}}{2004}]{yan_lazarian_2004}
{Yan} H.,  {Lazarian} A.,  2004, \mn@doi [\apj] {10.1086/423733}, \href
  {https://ui.adsabs.harvard.edu/abs/2004ApJ...614..757Y} {614, 757}

\bibitem[\protect\citeauthoryear{{Yan} \& {Lazarian}}{{Yan} \&
  {Lazarian}}{2008}]{yan_lazarian_2008}
{Yan} H.,  {Lazarian} A.,  2008, \mn@doi [\apj] {10.1086/524771}, \href
  {https://ui.adsabs.harvard.edu/abs/2008ApJ...673..942Y} {673, 942}

\bibitem[\protect\citeauthoryear{{Yang}, {Aharonian}  \& {Evoli}}{{Yang}
  et~al.}{2016}]{yang_2016}
{Yang} R.,  {Aharonian} F.,   {Evoli} C.,  2016, \mn@doi [\prd]
  {10.1103/PhysRevD.93.123007}, \href
  {https://ui.adsabs.harvard.edu/abs/2016PhRvD..93l3007Y} {93, 123007}

\bibitem[\protect\citeauthoryear{{Zakharov} \& {Sagdeev}}{{Zakharov} \&
  {Sagdeev}}{1970}]{zakharov_sagdeev_1970}
{Zakharov} V.~E.,  {Sagdeev} R.~Z.,  1970, Soviet Physics Doklady, \href
  {https://ui.adsabs.harvard.edu/abs/1970SPhD...15..439Z} {15, 439}

\bibitem[\protect\citeauthoryear{Zweibel}{Zweibel}{2017}]{zweibel2017_wind}
Zweibel E.~G.,  2017, \mn@doi [Physics of Plasmas] {10.1063/1.4984017}, 24,
  055402

\makeatother
\end{thebibliography}


\appendix

\section{CR diffusion coefficient in weak fast-mode turbulence }
\label{app:diff_fast_mode}
To calculate the CR diffusion coefficient in isotropic weak fast-mode turbulence, we use the results of \cite{yan_lazarian_2008}. We use the same notation as \cite{yan_lazarian_2008} and define the cosine of the wave pitch angle $\xi = \cos \theta$, the dimensionless wavenumber $x=kL$, the dimensionless rigidity $R={\rm v}/(L \Omega)$ and the perpendicular wavenumber normalised by the CR gyroradius, $w=k_\perp r_{L} = x_\perp  R (1-\mu^2)^{1/2}$, where ${\rm v \approx c}$ is the speed of the CR particle, $L$ is the outer scale of the turbulence and $\Omega$ is the relativistic gyrofrequency. The CR pitch-angle diffusion coefficient due to gyroresonant scattering is then (\citealt{yan_lazarian_2008}),

\begin{equation}\label{eq:Dmumu_G}
  \begin{aligned}
    D^G_{\mu \mu} = {\rm Ma}^2 \frac{{\rm v} \pi^{1/2}(1-\mu^2)}{2 L R^2}  \int_0^1 d \xi & \int_1^{k_{\rm max}(\xi) L}  dx \frac{x^{-5/2} \xi}{\Delta \mu}  [J'_1(w)]^2 \\ & \exp\Big(- \frac{ (\mu - (x\xi R)^{-1})^2 }{\Delta \mu^2} \Big)
  \end{aligned}
\end{equation}
where $J_1$ is the Bessel function of the first kind and the prime indicates a derivative, $\Delta \mu^2 = {\rm Ma} (1-\mu^2)$ and the exponential reflects the broadening of the gyro-resonance condition $k_\parallel {\rm v_\parallel} \approx \Omega$ due to large-scale magnetosonic fluctuations (\citealt{voelk_1973}, \citealt{voelk_1975}, \citealt{yan_lazarian_2008}). The CR pitch angle diffusion coefficient due to nonresonant scattering by transit-time damping (TTD) is,
\begin{equation}\label{eq:Dmumu_T}
  \begin{aligned}
    D^T_{\mu \mu} = {\rm Ma}^2 \frac{{\rm v} \pi^{1/2}(1-\mu^2)}{2 L R^2}  \int_0^1 d \xi & \int_1^{k_{\rm max}(\xi) L}  dx \frac{x^{-5/2} \xi}{\Delta \mu}  [J_1(w)]^2 \\ & \exp\Big(- \frac{ (\mu - {\rm v_A} / ({\rm v}\xi) )^2 }{\Delta \mu^2} \Big).
  \end{aligned}
\end{equation}
The spatial diffusion coefficient due to gyroresonant and TTD scattering is (see, e.g., \citealt{zweibel2017_wind}),
\begin{equation} \label{eq:kappa_dmumu}
    \kappa = \frac{{\rm v^2}}{4} \int_0^1 d \mu \frac{(1-\mu^2)^2}{D^G_{\mu \mu} + D^T_{\mu \mu} } .
\end{equation}
Because nonresonant scattering by TTD is dominated by large-scale modes,  $D^T_{\mu \mu}$ is essentially energy independent. Thus, energy-dependence in the CR diffusion coefficient comes from gyroresonant scattering, i.e. $D^G_{\mu \mu}$. The impact of wave damping on the cascade is reflected by the upper bound $=k_{\rm max}(\xi) L$ in the $x$ integral in equations \ref{eq:Dmumu_G} and \ref{eq:Dmumu_T}. In particular, the angle-dependent cutoff $k_{\rm max}(\xi) L$  corresponds to the scale $k(\xi=\cos\theta_c)$ where the cascade timescale is equal to the wave damping timescale. As in \citealt{yan_lazarian_2004} and \citealt{yan_lazarian_2008}, we here assume that fast modes with $\theta < \theta_c$  continue cascading to smaller scales unaffected by the damping, while the remaining modes are fully damped. However, we stress again that this assumption is quite uncertain. For example, non-linearities might transfer energy from small $\theta$ to larger $\theta$ where energy is dissipated, thus removing energy from the otherwise weakly damped parallel propagating modes.

For a fully ionized plasma, the main damping mechanisms are viscous damping on scales larger than the thermal ion mean free path and collisionless damping on scales smaller than the mean free path. $k_{\rm max}(\xi) L$ is then found by equating the cascade rate,
\begin{equation} \label{eq:Gamma_casc}
    \tau_{\rm casc}^{-1} \sim \Big(\frac{k}{L} \Big)^{1/2} \frac{\delta V^2}{V_{\rm ph}}
\end{equation}
($\delta V$ is the amplitude at the injection scale) and the relevant damping rate. For $\beta \ll 1$ in the collisional limit, i.e. $k_\parallel l_{\rm mfp} {\rm v_A / v_{th}} \ll 1$, the damping rate is (\citealt{br65})
\begin{equation}\label{eq:Gamma_brag}
  \begin{aligned}
    \Gamma (k_\parallel l_{\rm mfp} \frac{\rm v_A} {\rm  v_{th}} \ll 1) &= \frac{\nu_{\rm B} k^2}{6} (1-\xi^2 ) \\ & \sim \frac{\nu_{\rm B} k^2}{6} \theta^2 \qquad \theta \ll 1,
  \end{aligned}
\end{equation}
where $\nu_{\rm B} \sim l_{\rm mfp} v_{\rm th}$ is the anisotropic Braginskii viscosity and in the second line we took the small-angle limit. For $\beta \ll 1$ and $k_\parallel l_{\rm mfp} {\rm v_A / v_{th}} \gg 1$, the collisionless damping rate is (\citealt{ginzburg1961})
\begin{equation}\label{eq:Gamma_ginz}
  \begin{aligned}
    \Gamma (k_\parallel l_{\rm mfp} \frac{\rm v_A} {\rm  v_{th}} \gg 1) &= \frac{(\pi \beta)^{1/2} \sin^2\theta  }{4 \cos \theta} k {\rm v_A}   \Big[   \Big(\frac{m_e}{m_i} \Big)^{1/2} \exp \Big( - \frac{m_e}{\beta m_i \cos^2\theta} \Big) \\ &+ 5 \exp \Big(-\frac{1}{\beta \cos^2 \theta} \Big) \Big] \\ & \sim \frac{\sqrt{\pi \beta} \theta^2  }{4} \Big( \frac{m_e}{m_i} \Big)^{1/2} k {\rm v_A} \qquad \theta \ll 1,
  \end{aligned}
\end{equation}
where in the last step we again took the small-angle limit. We note that both damping mechanisms $\rightarrow 0$ for modes propagating parallel to the local magnetic field ($\theta \approx 0$) We calculate $k_{\rm max}(\xi) L$ by equating the cascade rate in \eqref{eq:Gamma_casc} with the damping rates in \eqref{eq:Gamma_brag} and \eqref{eq:Gamma_ginz}, and then evaluate the integrals in \eqref{eq:Dmumu_G} and \eqref{eq:Dmumu_T} numerically. To evaluate the cascade and damping rates in \eqref{eq:Gamma_casc}--\eqref{eq:Gamma_ginz}, we take the average of the cascade/damping rate in the interval $(\theta- \delta \theta, \theta + \delta \theta)$, where $\delta \theta$ is the spread in mode pitch angle experienced by a fast mode during one cascade time due to turbulent magnetic-field-line wandering. The field-line wandering due to ambient Alfv\'enic turbulence with ${\rm Ma_{Alf}} \sim 1$ experienced by fast modes with wavenumber $k$ and pitch angle $\theta$ can be roughly approximated as,
    \begin{equation} \label{eq:dtheta_alf_iso_app}
    \frac{\delta B}{B} \sim \Big[\frac{ \cos \theta}{{\rm Ma}^2 (kL)^{1/2}} + \Big( \frac{ \sin \theta}{{\rm Ma}^2 (kL)^{1/2}} \Big)^{2/3}  \Big]^{1/2}.
\end{equation}  
We stress that we include collisionless damping only on scales $<l_{\rm mfp} {\rm v_A / v_{th}}$ and collisional damping only on scales $> l_{\rm mfp} {\rm v_A / v_{th}}$. For example, if equating \eqref{eq:Gamma_casc} and \eqref{eq:Gamma_ginz} yields $k_{\rm max, \parallel}(\xi) {\rm v_A / v_{th}} < l_{\rm mfp}^{-1}$ for a $\xi$ that was not damped in the viscous regime, we correct it by setting $k_{\rm max, \parallel} = l_{\rm mfp}^{-1} {\rm v_{th} / v_A}$, as collisionless damping is not the appropriate damping on scales $k_\parallel l_{\rm mfp} {\rm v_A / v_{th}} < 1$. Figure \ref{fig:kappa_turb} shows the results of this calculation.

\section{Simultaneous scattering by self-excited waves and weak fast-mode turbulence} \label{app:sc_et_same_phase}

In some phenomenological models of CR transport it is assumed that the self-excited waves cascade (diffuse in $k$-space) due to the ambient Kolmogorov-like turbulence, which produces a spectrum of waves slightly different from Kolmogorov in the wavenumber range affected by the streaming instability (e.g. \citealt{blasi12}; \citealt{aloisio_blasi_2013}; \citealt{aloisio_2015}). The self-excited Alfv\'en waves can in principle interact with either the Alfv\'enic component of the turbulence or the fast-mode component.  While three-wave interactions between high-frequency self-excited Alfv\'en waves and turbulent fast modes can in principle occur at low $\beta$ (as the modes can satisfy frequency-matching conditions; \citealt{chandran_2005}), the fast-mode cascade is significantly affected by damping in the wavelength regime where self-excitation is important (Figure \ref{fig:kappa_turb}). As a result, the fast-mode cascade differs significantly from Kolmogorov (Figure \ref{fig:kappa_turb}; it may in fact be completely suppressed on small scales, see Figure \ref{fig:kappa_turb_anisotropic}).  Moreover, even in the absence of fast-mode damping, self-excited Alfv\'en waves are more likely to be sheared away to high $k_\perp$ by the background Alfv\'enic turbulence (\citealt{farmer_goldreich}; \citealt{lazarian_2016}). To see this, we compare the cascade time of the self-excited Alfv\'en waves due to fast modes, ${\tau_{\rm casc}}$, to the shearing time set by background Alfv\'enic turbulence, $\tau_{\rm shear}$. At low $\beta$ and for turbulence injected with ${\rm Ma_{Alf}} < 1$, 
\begin{equation} \label{eq:timescale_ratio_Ma<1}
    \frac{\tau_{\rm casc}^{-1}}{\tau_{\rm shear}^{-1}} \sim \Big(\frac{\rm Ma}{\rm Ma_{Alf}} \Big)^{2} ,
\end{equation}
while for ${\rm Ma_{Alf}}>1$,
\begin{equation}\label{eq:timescale_ratio_Ma>1}
      \frac{\tau_{\rm casc}^{-1}}{\tau_{\rm shear}^{-1}} \sim \Big(\frac{\rm Ma}{\rm Ma_{Alf}} \Big)^{3/2} {\rm Ma}^{1/2} ,  
\end{equation}
Most of the turbulent energy is expected to reside in the Alfv\'enic branch, ${\rm Ma_{Alf}} > {\rm Ma}$, and so $\tau_{\rm shear} < \tau_{\rm casc}$, i.e., the cascading of self-excited Alfv\'en waves to higher $k_\parallel$ by fast-mode turbulence is likely ineffective. We reiterate that damping reduces the power in fast-mode turbulence on small scales, further increasing $\tau_{\rm casc} / \tau_{\rm shear}$ and reinforcing our conclusion.

This implies that self-excited Alfv\'en waves and weak-fast mode turbulence act as independent scatterers of CRs. Their amplitudes and thus CR scattering rates are set by very different physics (different driving, cascading and damping mechanisms). Combined with the non-uniform scaling with energy of $\kappa_{\rm turb}$ in Figure \ref{fig:kappa_turb}, this implies that generating CR spectra with a roughly constant power-law slope over a wide range of energies using a combination of these scattering mechanisms requires significant fine-tuning of plasma conditions, regardless of whether CR propagation is single- or multi-phase. This is in contrast to phenomenological models based on an undamped Kolmogorov cascade, which do not face such significant fine-tuning issues.

\section{Impact of ionisation losses on the B/C spectrum} \label{app:bc_loss}

As we pointed out in Section \ref{sec:sc}, the energy loss term from eq. \ref{eq:adv_diff_turb} is more correctly written as a flux in momentum space. Using the correct form turns out to be particularly important for B nuclei at low energies, as we now show. In the loss regime, B nuclei satisfy,
\begin{equation} \label{eq:E_loss_eq}
   - \frac{1}{p^2} \frac{\partial }{\partial p} \Big(p^2 \frac{p}{\tau_{\rm loss }(p)} f_B \Big) =\frac{f_c}{\tau_{\rm spall}}= n \sigma_{\rm spall} {\rm v} f_C, 
\end{equation}
where the energy loss rate of a nucleus with mass $m_B$ and charge $Z$ due to ionisation of ISM material is  (\citealt{schlickeiser2002}),
\begin{gather}\label{eq:tau_ion}
\begin{aligned}
    \tau_{\rm loss}^{-1} = \tau_{\rm ion}^{-1} & \approx   2 \times 10^{-16} \ s^{-1} \ Z^2 \frac{n}{1 {\rm \ cm^{-3}}} \Big(\frac{\rm v}{c}\Big)^{-2} \Big(\frac{p}{\rm GeV / c} \Big)^{-1} \\ & \approx L \Big(\frac{p}{p_0}\Big)^{-3}, \quad p \ll p_0 = m_{B} c,
\end{aligned}
\end{gather}
where $L$ in the last step is a constant that depends on the density of the medium. \eqref{eq:E_loss_eq}  and \eqref{eq:tau_ion} imply that in the energy-loss dominated regime, the proton spectrum is $f_p \sim Q_p \tau_{\rm loss} \propto p^{-\gamma_{\rm inj} + 3}$ and similarly the carbon spectrum is  $f_C \sim Q_C \tau_{\rm loss} \propto p^{-\gamma_{\rm inj}+3}$. One might further guess that the boron spectrum is given by, $f_B \sim f_C \tau_{\rm loss}/\tau_{\rm spall} \sim Q_C \tau_{\rm loss}^2 /\tau_{\rm spall}  \sim p^{-\gamma_{\rm inj} + 7 }$. However, this turns out to not be a valid solution of \eqref{eq:E_loss_eq}. In particular, the absolute value of the energy flux due to ionisation losses associated with this spectrum decreases with decreasing momentum (the LHS of eq. \ref{eq:E_loss_eq} is negative becasue $\partial f_B / \partial p >0$). Equivalently, a given momentum shell is populated from higher momenta at a rate that is faster than the rate at which it loses particles to lower momenta. The correct solution of equation \ref{eq:E_loss_eq} is instead given by,
\begin{equation} \label{eq:fB_loss}
f_B =  A   -  \frac{1}{ 7-\gamma_{\rm inj}} \Big(\frac{p}{p_0} \Big)^3 L^{-1} \tau_{\rm spall}^{-1}  f_C \approx A
\end{equation}
where $A$ is a constant. In the last step we used the requirement $f_B \geq 0$, which implies that the B spectrum is set by $A$ at most momenta in the energy-loss dominated regime, due to the strong momentum dependence of the second term. The constant $A$ can be found by imposing continuity between the loss-dominated and escape-dominated regimes. The loss-dominated B/C spectrum is then,
\begin{equation} \label{eq:BC_loss}
\frac{f_B}{f_C} =  \frac{A}{f_C}   -  \frac{1}{ 7-\gamma_{\rm inj}} \Big(\frac{p}{p_0} \Big)^3 L^{-1} \tau_{\rm spall}^{-1} \approx \frac{A}{f_C} \propto p^{\gamma_{\rm inj} -3},
\end{equation}
which is $\propto p^{1.3}$ for $\gamma_{\rm inj} = 4.3$.

\bsp	
\label{lastpage}
\end{document}